\documentclass[aps]{revtex4}
\usepackage{eurosym}
\usepackage{amsfonts}
\usepackage{amsmath}
\usepackage{amssymb,epsf}
\usepackage{color}
\usepackage{graphicx}
\usepackage{natbib}
\usepackage{float}
\usepackage{caption}
\usepackage{subfig}
\usepackage{epstopdf}

\begin{document}
\title{Geometrical thermodynamics and P-V criticality of \\ charged accelerating AdS black holes}
 \author{Kh. Jafarzade$^{1,2}$\footnote{email address: khadije.jafarzade@gmail.com},
 J. Sadeghi$^{1,2}$\footnote{email address: pouriya@ipm.ir},
 B. Eslam Panah$^{1,3,2,4}$\footnote{email address: eslampanah@umz.ac.ir (corresponding author)} and
 S. H. Hendi$^{5,6}$\footnote{email address: hendi@shirazu.ac.ir}}
 \affiliation{$^1$ Department of Physics, Sciences Faculty, University of Mazandaran, P. O. Box 47415-416, Babolsar, Iran\\
$^2$  ICRANet-Mazandaran, University of Mazandaran, P. O. Box 47415-416, Babolsar, Iran\\
 $^3$ Research Institute for Astronomy and Astrophysics of Maragha (RIAAM), P.O. Box 55134-441, Maragha, Iran \\
 $^4$ ICRANet, Piazza della Repubblica 10, I-65122 Pescara, Italy \\
 $^5$ Department of Physics, School of Science, Shiraz University, Shiraz 71454, Iran\\
 $^6$ Biruni Observatory, School of Science, Shiraz University, Shiraz 71454, Iran}

\begin{abstract}

The unusual asymptotic structure of the accelerating black
holes led to ambiguity in their geometric characteristics and
thermodynamic behavior. Motivated by the interesting properties of
such black holes and the significant role of electric charge and
string tension on their structure, we study the thermodynamic
behavior of these black holes by two methods and examine the changes
of free parameters on the thermal behavior of the black holes.
First, we investigate phase transition and thermal stability of
the system through the use of heat capacity in the non-extended
phase space. We examine the effects of electric charge, string
tension and the cosmological constant on the phase transition and
stability of the system. We also find that to have a phase
transition, we have to apply some constraints on the free
parameters. Then, we employ the geometrical thermodynamic (GT)
method to study phase transition and compare the obtained results
with those of the heat capacity. Next, we work in the extended
phase space by considering the cosmological constant as a
dynamical pressure and evaluate the existence of van der Waals
like phase transition. We obtain critical quantities and study the
effective role of electric charge and string tension on these
quantities. Finally, we make use of the GT method in the extended
phase space and find that the results of the GT method, heat
capacity and $P-V$ diagram lead to a consistent conclusion.
\end{abstract}

\maketitle

\section{Introduction}

Black hole thermodynamics is one of the exciting and
challenging subjects in theoretical physics. The historical study
of the black hole as a thermodynamic system backs to the
pioneering works of Bardeen, Carter, Hawking \cite{Bardeen,Hawking1a} and Bekenstein
\cite{Bekenstein1a,Bekenstein1b}. They clarified the laws of
black hole mechanics and showed that these laws are corresponding
to ordinary thermodynamics by appropriate identification of the
related quantities, such as temperature, entropy, energy and so
on. The investigation of black hole thermodynamics in anti-de
Sitter (AdS) spacetime has been aroused the attention of
researchers. Studying thermodynamic properties of AdS black holes
is preliminary steps of investigating the quantum nature of
gravity, one of the most important theoretical subjects in
physical communities. Besides, according to the AdS/CFT
correspondence, the gravity side on asymptotically AdS spacetime
is equivalent to the conformal field theory in one fewer dimension
\cite{2,3,4,5,6}.

Among the various topics in black hole thermodynamics, the
phase transition is of particular importance. At first, Hawking
and Page \cite{7} demonstrated the existence of a certain phase
transition (so-called Hawking-Page) between thermal AdS and
Schwarzschild-AdS black hole which corresponds to the
confinement/deconfinement phase transition in the dual strongly
coupled gauge theory. Afterward, some efforts were conducted in
the context of phase transition for more complicated backgrounds
\cite{8,9}. Considering the phase transition of charged AdS black
holes showed that critical behavior of Reissner-Nordstr\"{o}m-AdS
black holes is superficially analogous to the van der Waals
liquid-gas phase transition \cite{10,11}. It has been proposed
that such similarity will be more precise by considering the
cosmological constant as a dynamical pressure in the extended
phase space \cite{12,13}. The consideration of cosmological
constant as a thermodynamical pressure and its conjugate quantity
as a thermodynamical volume gives us a better insight into
understanding van der Waals like behavior, the first and
second order phase transitions of black holes. In this regard,
some efforts have been made in the context of $P-V$ criticality of
black holes in modified theories of gravitation, such as
Gauss-Bonnet gravity \cite{GB,GBI,GBII,GBIII}, dilaton gravity
\cite{Dilaton,DilatonII}, Lovelock gravity
\cite{LoveI,LoveII,LoveIII}, Horava-Lifshitz gravity
\cite{HL0,HLI,HLII,HLIII,HLIV}, massive gravity
\cite{massiveI,massiveII,massiveIII,massiveIV,massiveV,massiveVI,massiveVII,massiveVIII,massiveIX,massiveX},
$F(R)$ gravity \cite{F(R)I,F(R)II}, gravity's rainbow
\cite{RainI,RainII}, and massive gravity's rainbow \cite{MassRain}. Moreover, the effects of linear and nonlinear
electrodynamics on the $P-V$ criticality of black holes have been
studied before \cite{NED,NEDI,NEDIII,NEDIV,NEDVI,NEDVII,NEDVIII,NEDIX,NEDX,NEDXI,NEDXII}.
Indeed, the variety of gravitating systems enriches our
knowledge of the phase structure of different black holes.

Among different methods to investigate phase transition, studying
the heat capacity of a thermodynamical system is more common.
Using the heat capacity, one can investigate two distinctive
points as bound and phase transition points. The bound point is
where the sign of temperature is changed. Since the positive
(negative) temperature is representing a physical (non-physical)
solution, this point is a limitation point, distinguishing the
physical black hole from the non-physical one. The existence of a
direct relationship between the heat capacity and the temperature
helps us to find these two quantities usually share the same
roots. So, one concludes that the bound point is where the
numerator of heat capacity (temperature) vanishes. On the other
hand, the phase transition point is related to divergency of the
heat capacity. Considering the heat capacity also provides a
mechanism to study thermal stability of the system. Indeed, the
positivity of heat capacity guarantees thermal stability of the
system, while its negativity indicates that the system under
consideration is in an unstable state
\cite{HeatI,HeatII,HeatIII,HeatIV,HeatV}. Since
investigating thermal stability/instability of the system is
possible through the heat capacity, we study the behavior of this
thermodynamic quantity for charged accelerating black holes and
inspect the effects of acceleration (string tension) and electric
charge in order to have a stable system. We also obtain a relation
between the free parameters of the model for having phase
transition. We explore this special condition in the non-extended
phase space via studying the heat capacity.

Another formalism for the investigation of black hole phase
transition is geometrical thermodynamics (GTs). This formalism is
based on the construction of a thermodynamical metric by using
thermodynamical potential (internal energy or entropy) and its
derivative with respect to the extensive parameters of system.
In GTs, a metric is introduced on the equilibrium
thermodynamical phase space. In fact, a sort of Riemannian metric
is defined as the Hessian of the thermodynamical potential, where
the derivatives are taken with respect to the extensive
thermodynamic variables. Depending on the choice of the
thermodynamical potential, the components of the phase space
differ in their structures (this is because each thermodynamical
quantity has its specific extensive parameters). The divergencies
of Ricci scalar in the constructed metric provide information
related to the phase transition. Studying GTs was first began by
Weinhold \cite{WeinI,WeinII}. He introduced a metric on the space
of equilibrium states where its components are given as the second
derivatives of internal energy with respect to entropy and other
extensive quantities. Then, Ruppeiner introduced another metric
which is defined as the negative second derivatives of entropy
with respect to the internal energy and other extensive quantities
\cite{RupI,RupII}. It is shown that there is a conformally
relationship between Weinhold and Ruppeiner metrics where the
conformal factor is the inverse of temperature \cite{Salamon}.
These metrics encounter with some problems. Indeed, the obtained
curvature scalars from Weinhold and Ruppeiner metrics are known to
diverge at the critical points of usual systems which indicated by
a singularity in the specific heat. But for black holes, there are
some contradictions. For example, the curvature scalar of the
Weinhold metric is singular for Reissner-Nordstr\"{o}m black hole,
which its singularity is not consistent with the singularity of
specific heat. The mentioned problem comes from the fact that both
the Weinhold and Ruppeiner metrics are not Legendre invariant and
so they are not suitable to describe thermodynamic properties of
various black holes \cite{QueI,QueII}. In this regard, Quevedo
proposed a new type of thermodynamical metric which was invariant
under Legendre transformation \cite{QueI,QueII}. The importance of
Legendre invariance lies in the thermodynamics itself, meaning
that once a representation is chosen to describe a system, its
Legendre transform contains the same information as the original
representation. In fact, Legendre invariance is an essential
ingredient of a geometric construction \cite{QueIII}. Another
thermodynamical metric is also introduced by Mansoori et al.
\cite{MansooriI,MansooriII}. But, these thermodynamic metrics had
some shortcomings. Hence, Hendi et al. proposed a new metric that
eliminated the problems of previous thermodynamical metrics (see
the Refs. \cite{HPEMI,HPEMII,HPEMIII,HPEMV,HPEMVI}, for more
details on various black holes). It is worthwhile to mention that
this new thermodynamical metric is defined the same as Quevedo's
metric with a different functional form of the conformal factor.
Recently, two new metrics were introduced by geometric
interpretation of criticality conditions where the divergency of
the Ricci scalar of these metrics is representing critical point
\cite{50,51}. It is worth mentioning that the basic
motivation of GTs is to give an independent picture regarding
thermodynamical aspects of systems. In addition, GTs provides
information regarding bound points, phase transitions, their types
and stability conditions. Furthermore, it can give microscopic
information about a thermodynamic system. In other words, by using
the sign of thermodynamical Ricci scalar, one can study the kind
of intermolecular interaction along the transition curve.
Positivity (negativity) of $R$ refers to the dominance of
repulsive (attractive) interaction in the thermodynamic system,
whereas $R=0$ indicates that there is no interaction in such a
system \cite{RupI}. Since Ruppeiner geometry is obtained from the
thermodynamic fluctuation theory \cite{RupII}, it is usual to
employ such geometry to probe the microstructure of a
thermodynamic system. Pineda et al. have recently displayed a
direct connection between Legendre invariant metrics and
fluctuation theory \cite{Pineda}. In Ref. \cite{QueIII}, Quevedo
and Tapias used the formalism of GTs to describe chemical
reactions in the context of equilibrium thermodynamics and showed
that the curvature of the equilibrium manifold with a Legendre
invariant metric reflects the thermodynamical interaction. They
conducted their study in the context of an ideal gas and a van der
Waals fluid. In this paper, we employ the mentioned thermodynamic
metrics and show which one can provide a suitable picture of phase
transition of charged accelerating black holes in both extended
and non-extended phase space. According to the critical
conditions, we also illustrate that the introduced metrics are a
suitable candidate for only spherical symmetric black holes, not
for accelerating ones. Besides, we employ the GT idea to
investigate the micro-structure of charged accelerating black holes
that have van der Waals like behavior and show that the
microscopic properties of such black holes are similar to those of
van der Waals fluid.

In recent years, the accelerating black holes attract much attention
\cite{AccI,AccII,AccIII,AccIV,AccV,AccVI,AccVII,AccVIII,AccIX,AccX,AccXI,AccXII,AccXIII,AccXIV}.
These black holes are described by the C-metric \cite{CMI,CMII,CMIII,CMIV}, interpreting as two uniformly
accelerating black holes. One of the main properties of this
metric is the existence of a string-like singularity along one
polar axis attached to the black hole. This conical singularity
can be imagined as a cosmic string with a tension providing the
force driving the acceleration. It is noteworthy that these conical singularities can be eliminated after imposing appropriate restrictions on physical parameters and then embedding the solutions into $D = 11$ supergravity. Besides, an interesting feature of
the accelerating black holes is their uncommon asymptotic behavior
where the curvature of background is not constant at spatial
infinity. Moreover, depending on the range of the parameters, an
acceleration horizon can be appeared, causing a complicated
structure. The existence of such a horizon raises the problem of
thermodynamic equilibrium. Considering a negative cosmological
constant, one can eliminate this problem \cite{AccI}. The
mentioned black hole is called a slowly accelerating and is
displaced from the center of AdS spacetime via a cosmic string
ending on the black hole horizon \cite{52}. The thermodynamics of
such black holes was investigated by Appels et al in Ref.
\cite{AccI}. They showed that the first law of these black holes
can be expressed in the standard form under the satisfaction of
two certain conditions. Another interesting feature is related to the fact that the circular orbits of the photons deviate from the
equatorial plane and the property of the black hole shadow changes due to the acceleration. Indeed, the latitude of the circular orbit increases by increasing the acceleration \cite{Zhang1kj}.

Within general relativity (GR), the C-metric has been used to investigate radiation at infinity \cite{Ashtekar1abc,Podolsky2bc}. However, the application of C-metric is not limited to GR. It can be used to describe the production of black hole pairs in strong background fields \cite{Horowitz3abc}, the construction of the black ring solution in five dimensions \cite{Emparan4abc}, the generation of black holes in an electric or magnetic field and the splitting of cosmic strings \cite{Dowker1ab,Eardley1ab}. Since the oppositely accelerating black holes are causally disconnected and the metric can be expressed in appropriate coordinates to cover just one of the “moving” black holes, C-metric can be regarded as a reasonable candidate to describe boosted black holes \cite{Ross1mn}. Besides, this metric has been generalized to include rotation, cosmological constant and a Newman-Unti-Tamburino (NUT) parameter \cite{CMII}. The gravitational lensing \cite{Zhang1kj,Zhang2kjl,Frost1kjl}, quasinormal modes \cite{Destounis2kn}, holographic complexity \cite{Shun1sd}, stability of the Cauchy horizon \cite{Destounis2sd}, gravitational entropy \cite{Guha2sd}, spinning spindles \cite{Ferrero1kj} and thermodynamics \cite{Tavakoli2bkj,Huang2br,Ball2bkj,Ball2kj} of accelerating black holes have been investigated before. Moreover, the effects of $f(R)$ gravity \cite{Zhang2akj,Belhaj2fd,Pourhassan} and NUT parameter \cite{Podolsky2lkb} on the properties of accelerating black holes have been reported in the litrature.

It was shown that depending on the gravity under consideration and employed matter fields, the critical behavior of the system may be modified and some conditions regarding the existence/absence of van der Waals like behavior may appear. So, it would be interesting to investigate how the acceleration of the black hole affects its phase transition. In the present paper, we also study the van der Waals like behavior, critical properties and conditions for observing the critical behavior of these black holes.

We organize this paper as follows: in Sec. \ref{FE}, we briefly
review the charged accelerating AdS black holes. In Sec.
\ref{Thermo-nonE}, we investigate the phase transition of these
black holes by using the heat capacity and GT method in
non-extended phase space and study thermal stability conditions.
Section \ref{Thermo-E} devoted to employ the analogy between the
cosmological constant and thermodynamical pressure and extend
phase space. We investigate van der Waals like phase transition
and obtain the critical point by two different methods and show
that their results are identical. Then, we study the critical
behavior of the system via heat capacity and GT method and
indicate that these approaches lead to the same results. In the
last section, we present our conclusions.

\section{Charged accelerating AdS black holes \label{FE}}

In this section, we review some basic properties of the charged accelerating
AdS black holes. The metric governing the charged accelerating black holes
is \cite{AccI}
\begin{equation}
ds^{2}=\frac{1}{\Omega ^{2}}\left[ f(r)dt^{2}-\frac{dr^{2}}{f(r)}%
-r^{2}\left( \frac{d\theta ^{2}}{g(\theta )}+g(\theta )sin^{2}\theta \frac{%
d\phi ^{2}}{K^{2}}\right) \right] ,  \label{Eq1}
\end{equation}%
where
\begin{eqnarray}
f(r) &=&\left( 1-A^{2}r^{2}\right) \left( 1-\frac{2m}{r}+\frac{e^{2}}{r^{2}}%
\right) +\frac{r^{2}}{\ell ^{2}},  \notag \\
&&  \notag \\
g(\theta ) &=&1+2mAcos\theta +e^{2}A^{2}cos^{2}\theta ,  \label{Eq2}
\end{eqnarray}%
also $\Omega $ is conformal factor which is given by
\begin{equation}
\Omega =1+Ar~cos\theta .  \label{Eq3}
\end{equation}

This factor determines the conformal infinity or boundary of the AdS
spacetime. The parameters $A$ ($A>0$) and $\ell =\sqrt{\frac{-3}{\Lambda }}$
($\Lambda $\ is cosmological constant) are the acceleration parameter and
AdS radius, respectively. Also, the parameters $m$ and $e$ are related to
the mass and electric charge of the black hole as \cite{AccVIII,Ashtekar,Das}
\begin{eqnarray}
M &=&\left[ 1-\left( 1+A^{2}e^{2}\right) A^{2}\ell ^{2}\right] \frac{m}{%
K\alpha },  \notag \\
&&  \notag \\
Q &=&\frac{1}{4\pi }\int_{\Omega =0}\ast F=\frac{e}{K},  \label{Eq7}
\end{eqnarray}%
where  the parameter $ K $ encodes information about the
conical deficit on the south and north poles, so that $\phi \in
[0, 2\pi]$.  In addition, the parameter $\alpha$ is a
normalization factor, re-scaling the time coordinate, which is
necessary to have a well-defined thermodynamic behavior
\cite{AccV,AccVIII,AccIX,Gibbons}. Considering the normalization
factor, the proper time coordinate of an asymptotic observer is
$\tau =\alpha t$, in which $\alpha
=\sqrt{(1+e^{2}A^{2})(1-A^{2}\ell ^{2}(1+e^{2}A^{2}))}$ (more
details are given in appendix A).  The electromagnetic field
two-form $F$ is related to gauge potential one form $B$ as
\cite{AccIX}
\begin{equation}
F=dB,~\ ~\ \&~~~~B=-\frac{e}{\alpha }\left( \frac{1}{r}-\frac{1}{r_{+}}%
\right) dt.  \label{Eq8}
\end{equation}

It is to be noted that the conical singularity in Eq. (\ref{Eq1})
is removed by adjusting $m=e=0$, the resultant spacetime is a pure
AdS metric in Rindler-type coordinates. We will introduce a proper
coordinate transformation in appendix B to remove this problem.

Taking a close look at the angular part of metric and the behavior of $%
g(\theta )$ at poles $\theta _{+}=0$ ($\cos 0=1$) and $\theta _{-}=\pi $ ($%
\cos \pi =-1$), one can find the presence of cosmic string. The regularity
of metric at a pole leads to
\begin{equation}
K_{\pm }=g\left( \theta _{\pm }\right) =1\pm 2mA+e^{2}A^{2}.  \label{Eq4}
\end{equation}

As it is seen in Eq. (\ref{Eq4}), for $mA\neq 0$, one cannot fix the
parameter of $K$ and therefore, there exists irregularity at both poles. The
irregularity at an axis causes the conical singularity. Thus, $K$ is chosen
to regularize one pole, leaving either a conical deficit or a conical excess
along the other pole. The conical deficits on the north pole ($\theta _{+}=0$%
) and the south pole ($\theta _{-}=\pi $) are given by
\begin{equation}
\delta _{\pm }=2\pi \left( 1-\frac{g\left( \theta _{\pm }\right) }{K}\right)
,  \label{Eq5}
\end{equation}%
which corresponding to a cosmic string with tension \cite{AccV,AccII}
\begin{equation}
\mu _{\pm }=\frac{\delta _{\pm }}{8\pi }=\frac{1}{4}-\frac{1\pm
2mA+e^{2}A^{2}}{4K}.  \label{Eq6}
\end{equation}

By setting $K=K_{+}=1+2mA+e^{2}A^{2}$ (i. e., $\mu _{+}=0$ ), one can remove
the conical singularity on the north pole $(\theta =0)$. So, there is a
conical deficit on the south pole with the string tension $\mu =\mu _{-}=%
\frac{mA}{K}$. The $mA$-term should be restricted to $mA<\frac{1}{2}$ to
preserve the metric signature \cite{AccV}. For simplicity, we consider $%
\mathcal{B}=mA$ as a constant parameter throughout this paper.

In order to have a well-defined metric (\ref{Eq1}), corresponding physically
to a slowly accelerated black hole in the bulk, certain conditions should be
imposed in the range of different parameters as follows:

I) The function $g(\theta )$ should be a positive definite function for $%
\theta $ in $[0,\pi ]$.

II) Due to the requirement of slow acceleration, the function $f(r)$ should
has no root on the boundary $\left( f\left( -\frac{1}{A\cos \theta }\right)
>0\right) $.

III) The spacetime has to admit a black hole in the bulk.

\begin{figure*}[tbh]
\centering
\includegraphics[width=0.3\linewidth]{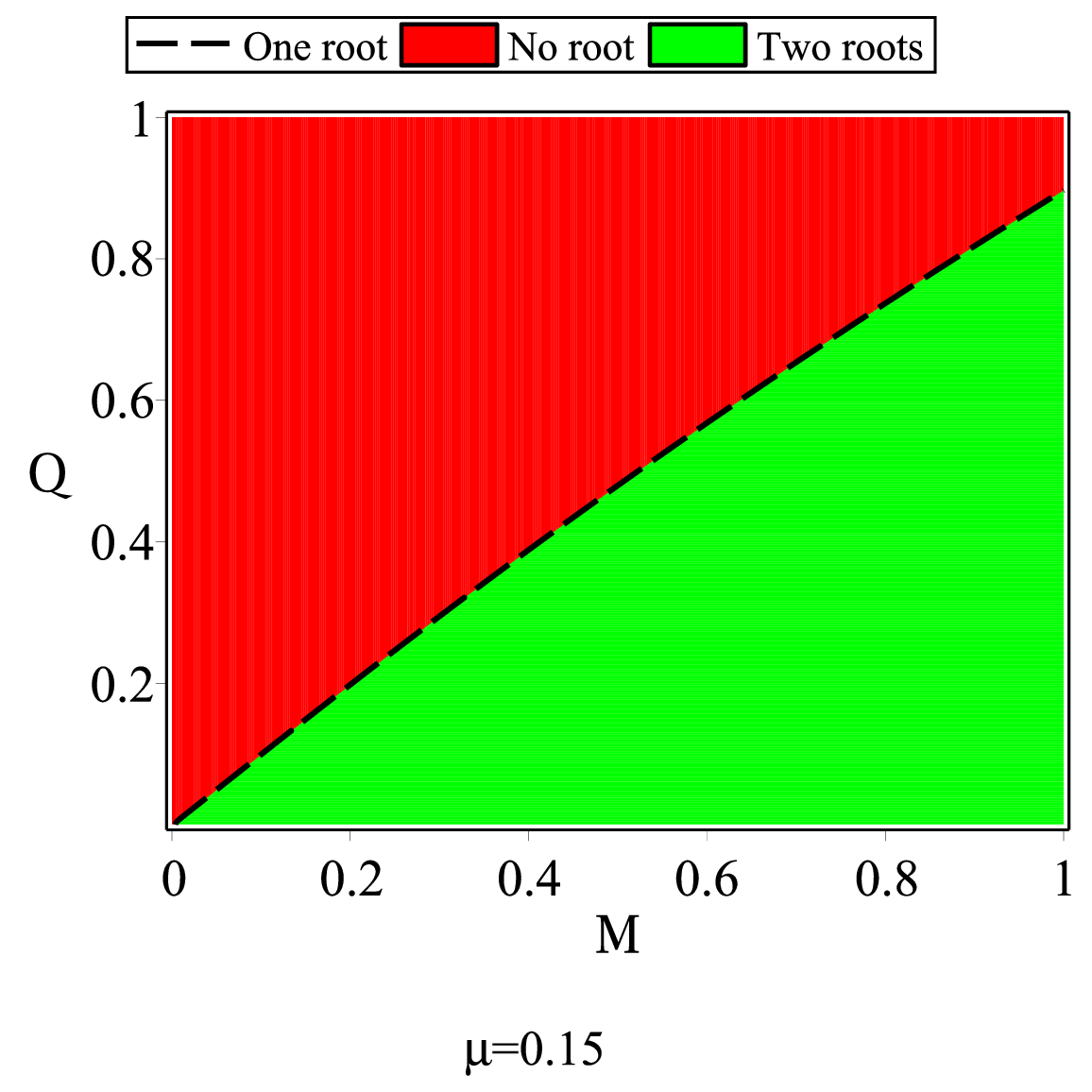}\hfil
\includegraphics[width=0.3\linewidth]{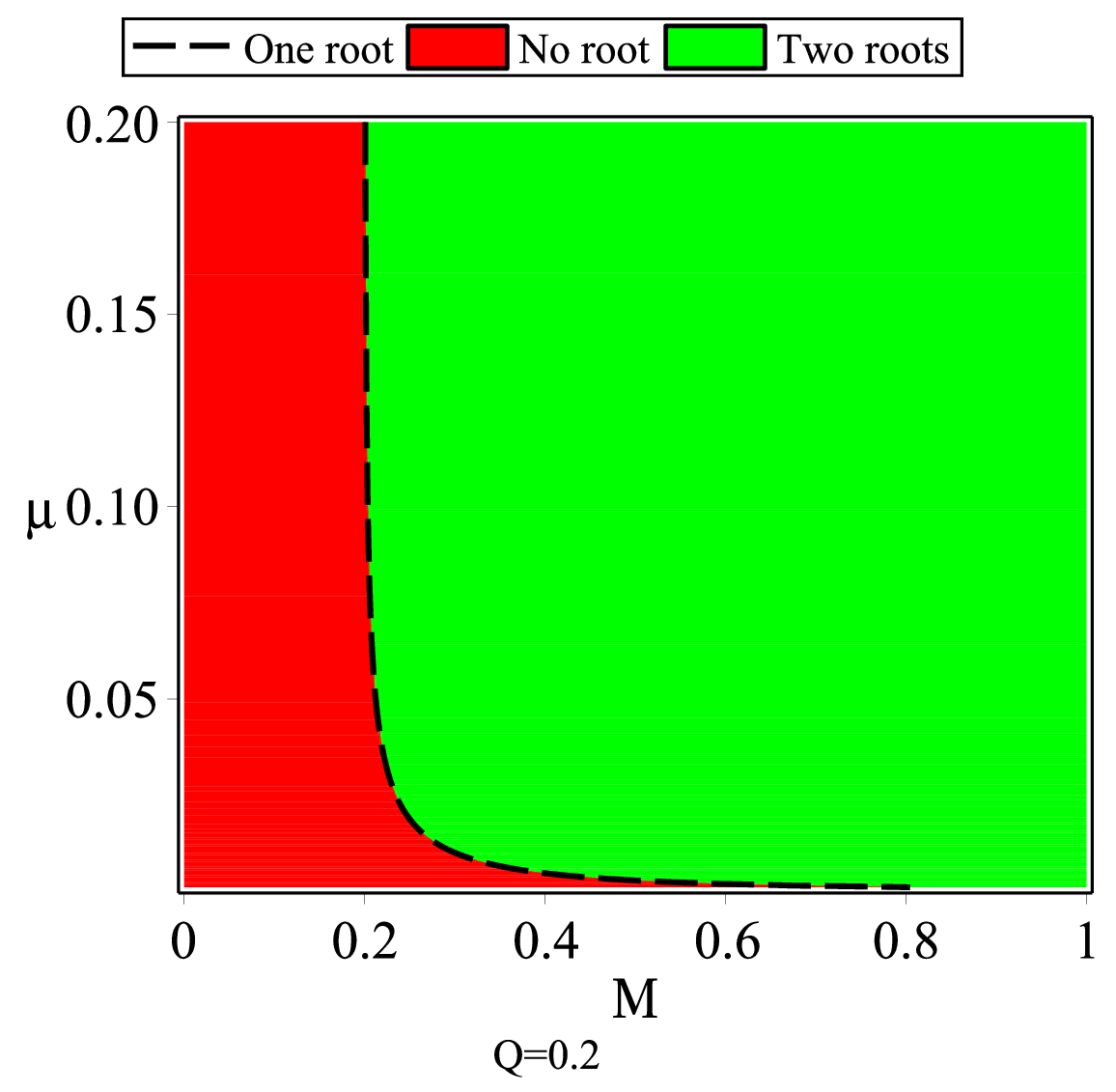}\hfil
\includegraphics[width=0.3\linewidth]{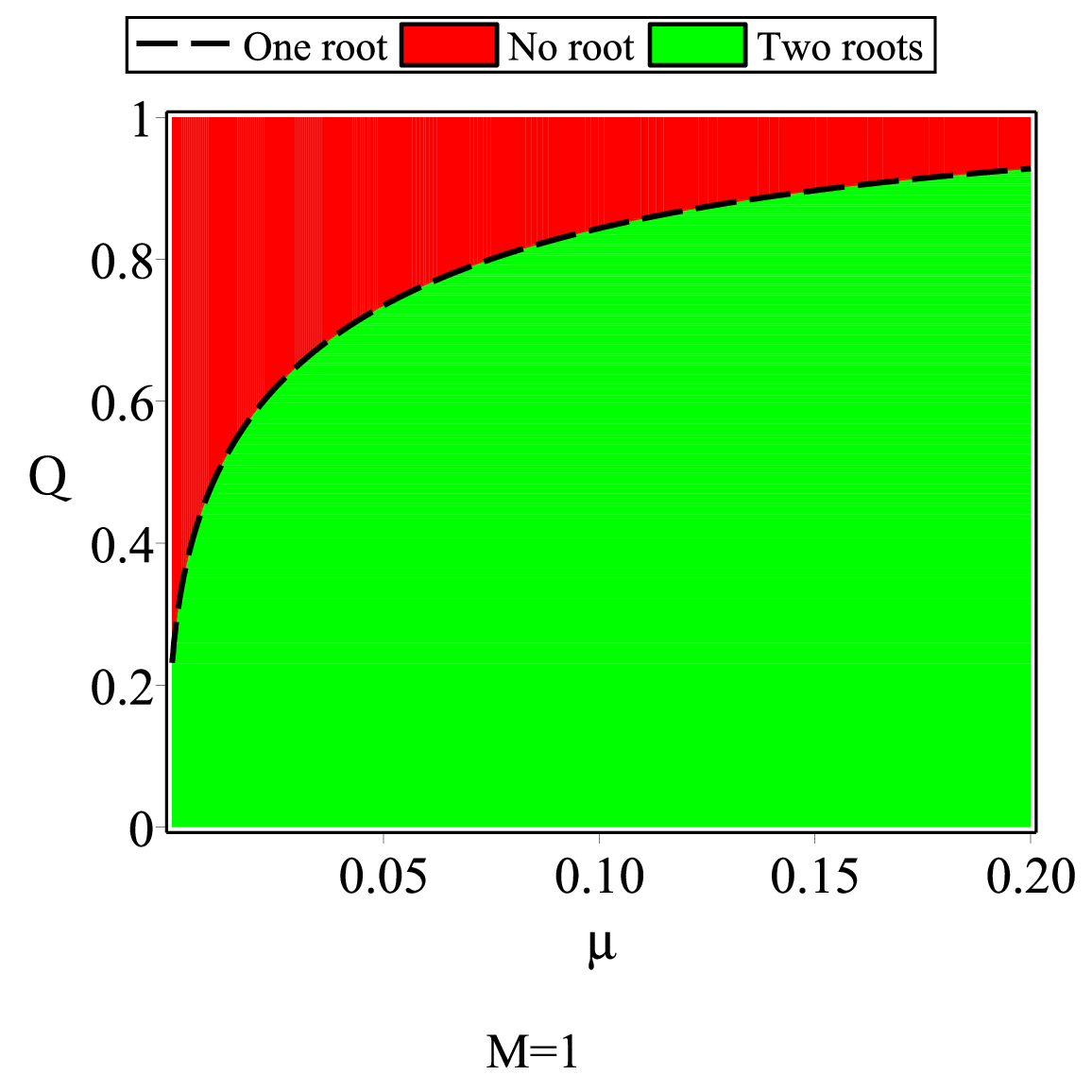}
\caption{Variation of $f(r)$ as a function of different parameters for $%
\mathcal{B}=0.2$, $\protect\beta =0.04$, $\ell =2$ and $A=0.02$.}
\label{Fig1}
\end{figure*}

We study the behavior of metric function in Fig. \ref{Fig1}. One can observe
three different cases. a) Two real roots which are Cauchy and event
horizons. b) One real root (extreme horizon). c) Absence of real root (naked
singularity). Fig. \ref{Fig1}, shows regions in which the metric function
has one or two real roots.

Our analysis shows that in order to have black hole solutions, one
has to consider the small values for the electric charge ($Q$)
and the large values for string tension ($\mu $), mass ($M$).
\section{Thermodynamical structure in non-extended phase space \label{Thermo-nonE}}

In this section, we study phase transition and thermal stability
of the charged accelerating AdS black holes in the context of the
canonical ensemble by calculating the heat capacity. We also
investigate the effects of electric charge, the AdS radius (or the
cosmological constant) and string tension on phase transition and
stability of the system and show that a certain relation between
these parameters should be satisfied in order to have phase
transition. Then, we employ thermodynamical metrics to study phase
transition and compare obtained results with those of heat
capacity.

\subsection{Temperature and (non)physical black hole solutions}

In order to study the physical black hole solutions, we have to
evaluate the behavior of temperature. According to the pioneering
work of Hawking, the temperature of black holes is related to the
surface gravity $\kappa$ by the relation $ T=\frac{\kappa}{2\pi} $
\cite{Hawking1a}. Replacing the geometrical mass $m$ which is
obtained by evaluating the metric function on horizon $(f(r=r_{+}) =0)$ as \cite{Dolana1008}
\begin{equation}
m=\frac{\left( r_{+}^{2}+e^{2}\right) \left( 1-A^{2}r_{+}^{2}\right) +\frac{%
r_{+}^{4}}{\ell ^{2}}}{2r_{+}\left( 1-A^{2}r_{+}^{2}\right) }, \label{Eq12}
\end{equation}
in the temperature relation, one can find Hawking temperature as
\begin{eqnarray}
T=\frac{f^{\prime }\left( r_{+}\right) }{4\pi \alpha }=\frac{3r_{+}^{2}+\ell ^{2}\left( 1-\frac{Q^{2}\mathcal{B}^{2}}{\mu
^{2}r_{+}^{2}}\right) -\beta^{2}r_{+}^{2}\left(\ell ^{-2} r_{+}^{2}+ \left(
2-A^{2}r_{+}^{2}\right) \left( 1-%
\frac{Q^{2}\mathcal{B}^{2}}{\mu ^{2}r_{+}^{2}}\right) \right) }{4\pi \alpha r_{+}\left(
\ell ^{2}-\beta^{2}r_{+}^{2}\right) },  \label{Eq13}
\end{eqnarray}
where $\mathcal{B}=mA$ and $\beta =A\ell $, are considered as
fixed parameters throughout the paper. We should note that to
preserve the metric signature and remove the acceleration horizon,
we consider, respectively, $2\mathcal{B}<1$ and $\beta <1$
\cite{AccV}. According to the numerator of temperature (Eq.
(\ref{Eq13})), one can find that it has at least one root. More
investigations confirm that there are three roots for the
temperature which only real one is
\begin{equation}
r_{+}|_{T=0}=\sqrt{2\sqrt{\frac{-\rho }{3}}\sin \left(
\frac{1}{3}\sin ^{-1}\left( \frac{3\sqrt{3}g}{2(\sqrt{-\rho
})^{3}}\right) \right)+\frac{(3-2\beta ^{2})}{3A^{2}}},
\label{rootT}
\end{equation}
where
\begin{eqnarray}
\rho &=&-\frac{1}{3A^{4}}\left[ 9\left( 1-\beta ^{2}\right) +2\beta
^{4}\left( 2+\frac{3Q^{2}\mathcal{B}^{2}}{\mu ^{2}\ell ^{2}}\right) \right] ,
\notag \\
&&  \notag \\
g &=&-\frac{1}{27A^{6}}\left[ 54-81\beta ^{2}+27\beta ^{4}\left( 2+\frac{%
Q^{2}\mathcal{B}^{2}}{\mu ^{2}\ell ^{2}}\right) +4\beta ^{6}\left( 4+\frac{%
9Q^{2}\mathcal{B}^{2}}{\mu ^{2}\ell ^{2}}\right) \right] .
\end{eqnarray}%

\begin{figure*}[tbh]
\centering
\includegraphics[width=0.35\linewidth]{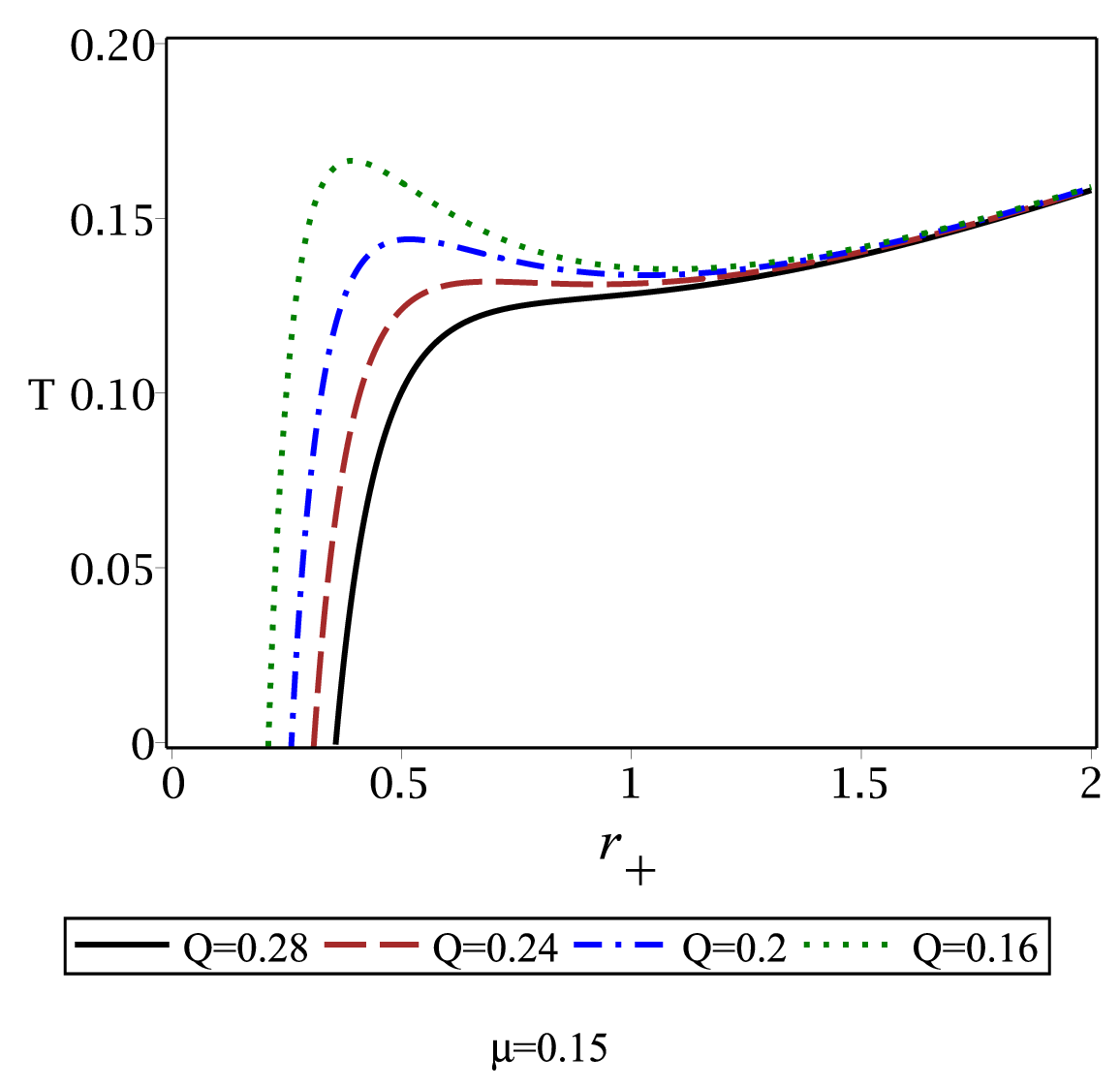} \includegraphics[width=0.35%
\linewidth]{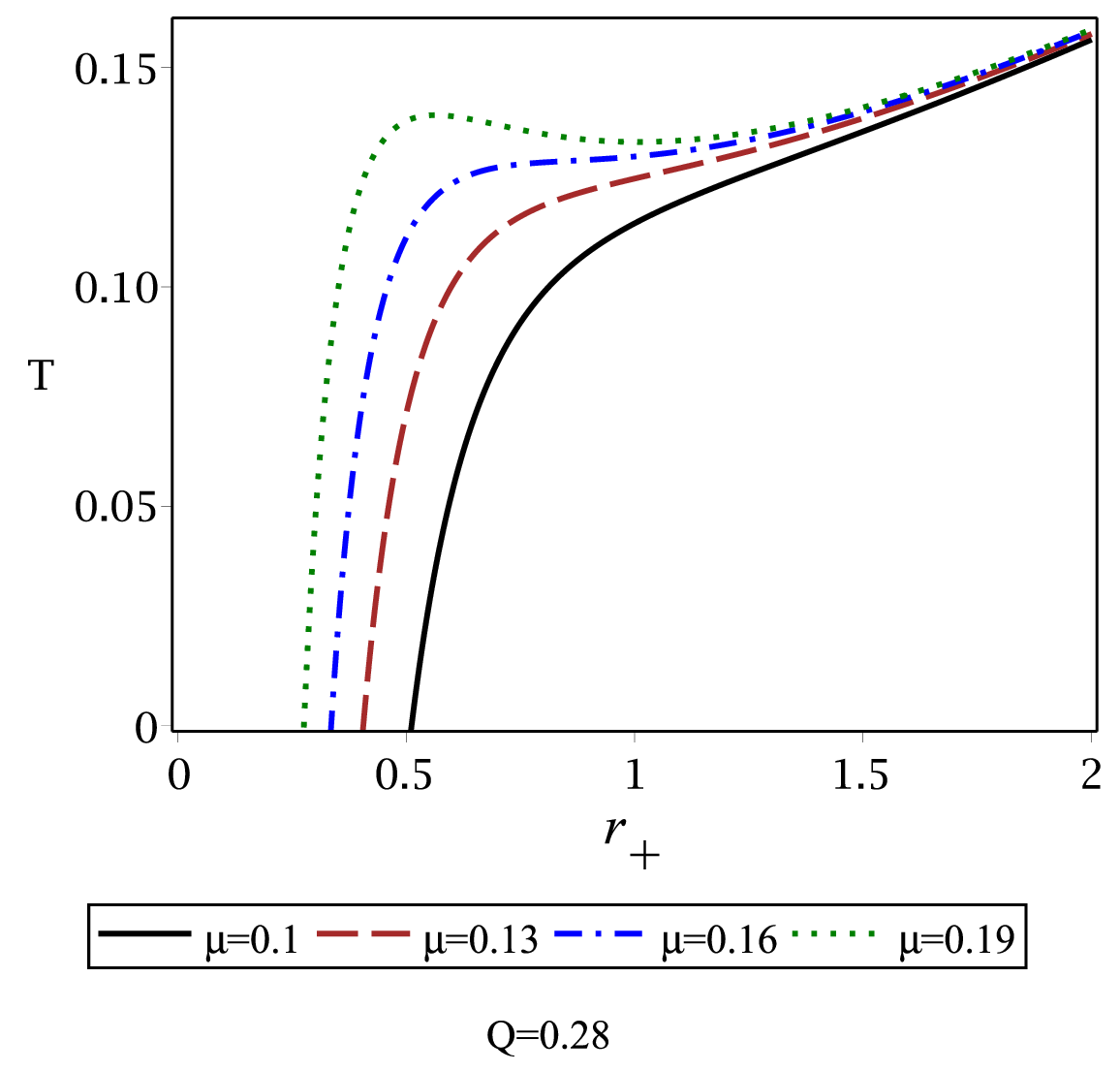}
\caption{$T$ versus $r_{+}$ for $\mathcal{B}=0.2$, $\protect\beta =0.04$, $%
\ell =2$ and $A=0.02$. Left panel for different values of the electric
charge. Right panel for different values of the string tension.}
\label{Fig2}
\end{figure*}

Taking a look at Eq. (\ref{Eq13}), one can see that the temperature diverges at $Ar_{+}=1$. To investigate the effects of black hole parameters on the temperature, we have plotted Fig. \ref{Fig2}.

Analyzing the temperature for small and large values of the horizon radius, gives us interesting information regarding these black holes. Using the series expansion of $T$ for vanishing horizon radius (or very small charged accelerating black holes) results into
\begin{equation}
\lim_{r_{+}\longrightarrow 0}T\propto -\frac{Q^{2}\mathcal{B}^{2}}{4\pi
\alpha \mu ^{2}r_{+}^{3}}+\frac{1}{4\pi \alpha r_{+}}\left( 1+\frac{%
A^{2}Q^{2}\mathcal{B}^{2}}{\mu ^{2}}\right) +O(r_{+}),  \label{EqST}
\end{equation}%
which confirms the following important points:

I) The only quantity which does not have direct effect on
the high energy limit of the temperature is the AdS
radius.

II) The dominant term in the high energy limit of the
temperature which includes the electric charge and string tension
is always negative. It means that the temperature diverges at
$r_{+}\longrightarrow 0$ for non-vanishing $B$, $Q$ and $\mu$.
Since negative temperature corresponds to a non-physical solution,
there is no physical solution for a very small charged
accelerating black hole. Indeed, by decreasing the event horizon
radius, one can find that the temperature vanishes before it
negatively diverges.

III) In the absence of electric charge, the temperature is
always positive, indicating that a very small uncharged
accelerating black hole has a physical solution, unlike its
charged counterpart.

The asymptotic behavior of the temperature is given by
\begin{equation}
\lim_{r_{+}\longrightarrow \infty }T\propto \frac{\left( 1-\beta ^{2}\right)
r_{+}}{4\pi \alpha \ell ^{2}}+O\left( \frac{1}{r_{+}}\right) ,  \label{EqLT}
\end{equation}%
which shows that for large black holes, the temperature is
governed by the AdS radius. As it was mentioned, an accelerating
black hole has an acceleration horizon. To resolve such a problem,
one should consider $\beta <1 $ \cite{AccV}. So, existence of such
a constraint ensures physical solution for large black holes.

Now, we focus on the behavior of the temperature and the effects of electric
charge and string tension on it. As one can see from Fig. \ref{Fig2}, the
temperature is zero for certain values of the horizon radius. At this point $(r_{+}=r_{0})$, the sign of temperature changes from negative to positive.
Where for $(r_{+}<r_{0})$, the temperature is always negative which is
representing a non-physical solution. But for $r_{+}>r_{0}$, the temperature is positive and the black hole has a physical solution.

The first parameter that we study its effect on the temperature is
the electric charge ($Q$). The left panel of Fig. \ref{Fig2},
shows that $r_{0}$ increases by increasing $Q$. Therefore, the
region of the non-physical solution increases. Here, the
interesting issue is that as $Q$ decreases, the temperature
acquires one or two extrema. Evidently, in extremum, the
denominator of the heat capacity is zero. In other words, the heat
capacity diverges at this point. Then, one can say that extrema
are the same points in which the black hole undergoes a phase
transition. Increasing $Q$, these extrema disappear and only a
bound point will be observed. So, by using the temperature
diagram, one can study both bound and phase transition points. On
the other hand, by employing this diagram, one can achieve a
better insight in understanding single phase regions which are
small/large black holes. The region $r_{0}<r_{+}<r_{max}$ is
related to small black holes and the region $r_{+}>r_{min}$ is denoted by large black holes ($r_{max}$ and $r_{min}$ are the first and second extremum, respectively). Medium black holes are located between these two extrema (see Refs. \cite{HeatIV,HendiarXiv}, for more details).

The next parameter which affects the temperature is the string
tension. Taking a closer look at the right panel of Fig.
\ref{Fig2}, one can find that the role of $\mu $ is opposite of
that of electric charge. In other words, by increasing $\mu $, the
extrema are formed while for its small values, the only bound
point is observed. Also, one can see that as the string tension
increases the bound point shifts to a smaller horizon radius.


\subsection{Heat capacity and phase transition}

Heat capacity is one of the interesting thermodynamical quantities which describes two important issues of a thermodynamical system as thermal stability and phase transition. The positivity of heat capacity indicates thermal stability of the system while its negativity is representing instability. One can consider two cases for an unstable black hole. It may go to a stable state by a phase transition or it has a non-physical solution which in this case no phase transition takes place. So, we can extract two characteristic points by using the heat capacity: bound and phase transition points. As it was already mentioned \cite{HeatIV}, the bound point is related to the root of heat capacity. Whereas, the phase transition point(s) is(are) where heat capacity diverges. The heat capacity is given by

\begin{equation}
C_{Q,\mu}=T\left( \frac{\partial S}{\partial T}\right) _{Q}=\frac{\left( \frac{%
\partial M}{\partial S}\right) _{Q}}{\left( \frac{\partial ^{2}M}{\partial
S^{2}}\right) _{Q}}.  \label{Eq10}
\end{equation}

The entropy $S$ of the black hole is identified with a quarter of the horizon area as
\begin{equation}
S=\frac{\mathcal{A}}{4}=\frac{\pi r_{+}^{2}}{K\left( 1-A^{2}r_{+}^{2}\right)
}.  \label{Eq9}
\end{equation}

It is evident that the entropy diverges at $Ar_{+}=1$. To avoid divergency
and negatively, one should consider $A<\frac{1}{r_{+}}$.

As it was pointed out, in the canonical ensemble, the phase transition
points are detected as divergence points of the heat capacity. By employing Eqs. (\ref{Eq9}), (\ref{Eq10}) and (\ref{Eq13}), one finds
\begin{equation}
C_{Q,\mu}=\frac{2\pi \mu r_{+}^{2}\left( 3r_{+}^{2}\left( 1-\frac{A^{2}r_{+}^{2}%
}{3}\right) +\ell ^{2}\left( 1-\frac{Q^{2}\mathcal{B}^{2}}{\mu ^{2}r_{+}^{2}}%
\right) \left[ 1-2A^{2}r_{+}^{2}\left( 1-\frac{A^{2}r_{+}^{2}}{2}\right) %
\right] \right) }{(1-A^{2}r_{+}^{2})\left( 3r_{+}^{2}\left( 1+\frac{%
A^{2}\beta ^{2}X}{3\mu ^{2}}\right) +\ell ^{2}\left( 1-\frac{7Q^{2}\mathcal{B%
}^{2}}{\mu ^{2}r_{+}^{2}}\right) \left( \beta ^{2}r_{+}^{2}-1\right) -\frac{%
4\ell ^{2}Q^{2}\mathcal{B}^{2}}{\mu ^{2}r_{+}^{2}}\right) \mathcal{B}},
\label{Eq14}
\end{equation}%
where
\begin{equation*}
X=\mu ^{2}r_{+}^{2}\left( 1-A^{2}r_{+}^{2}\right) \left( 1+\frac{Q^{2}%
\mathcal{B}^{2}}{\mu ^{2}r_{+}^{2}}\right) +4Q^{2}\mathcal{B}^{2}+\frac{\mu
^{2}r_{+}^{4}}{\ell ^{2}}.
\end{equation*}

To understand the heat capacity in more details, we investigate the its
limiting behaviors as
\begin{equation}
C_{Q,\mu}~\Rightarrow \left\{
\begin{array}{cc}
\lim_{r_{+}\longrightarrow 0}C_{Q,\mu}\propto -\frac{2\pi \mu r_{+}^{2}}{3\mathcal{B}%
}+O(r_{+}^{4}) & \text{small black holes} \\
&  \\
\lim_{r_{+}\longrightarrow \infty }C_{Q,\mu}\propto \frac{2\pi \mu }{r_{+}^{2}A^{4}%
\mathcal{B}}+O(\frac{1}{r_{+}^{4}}), & \text{large black holes}%
\end{array}%
\right. .
\end{equation}

For small black holes: as it was already mentioned, there is a physical
solution in this limit just in the absence of electric charge (see Eq. \ref{EqST}). But as we see, they are not thermally stable due to the negativity
of heat capacity.

For large black holes: the heat capacity is highly governed by the
acceleration parameter and string tension and due to the positivity of heat capacity in this limit, these black holes are thermally stable.

\begin{figure}[!htb]
\centering
\includegraphics[width=0.33\textwidth]{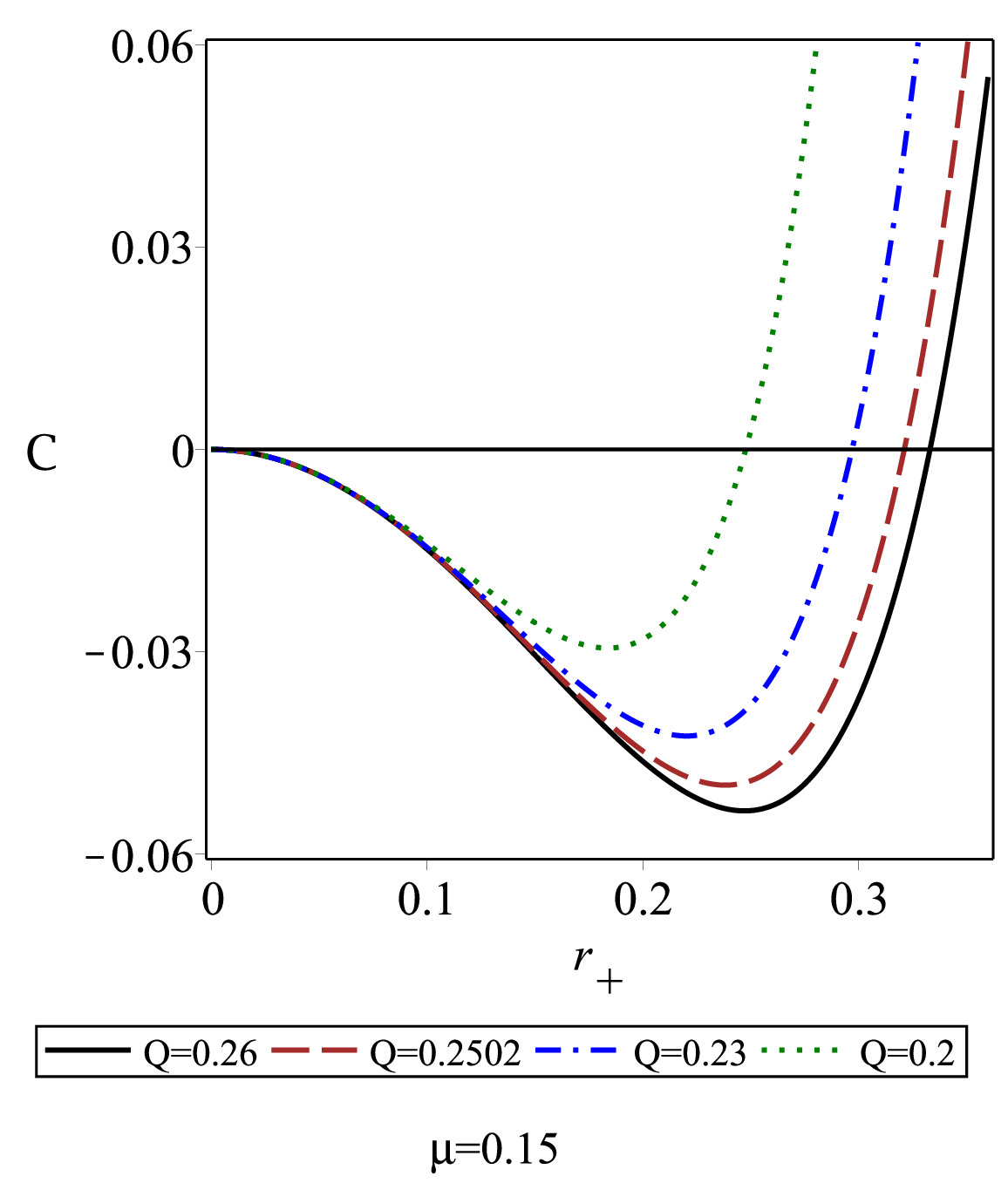} \includegraphics[width=0.33%
\textwidth]{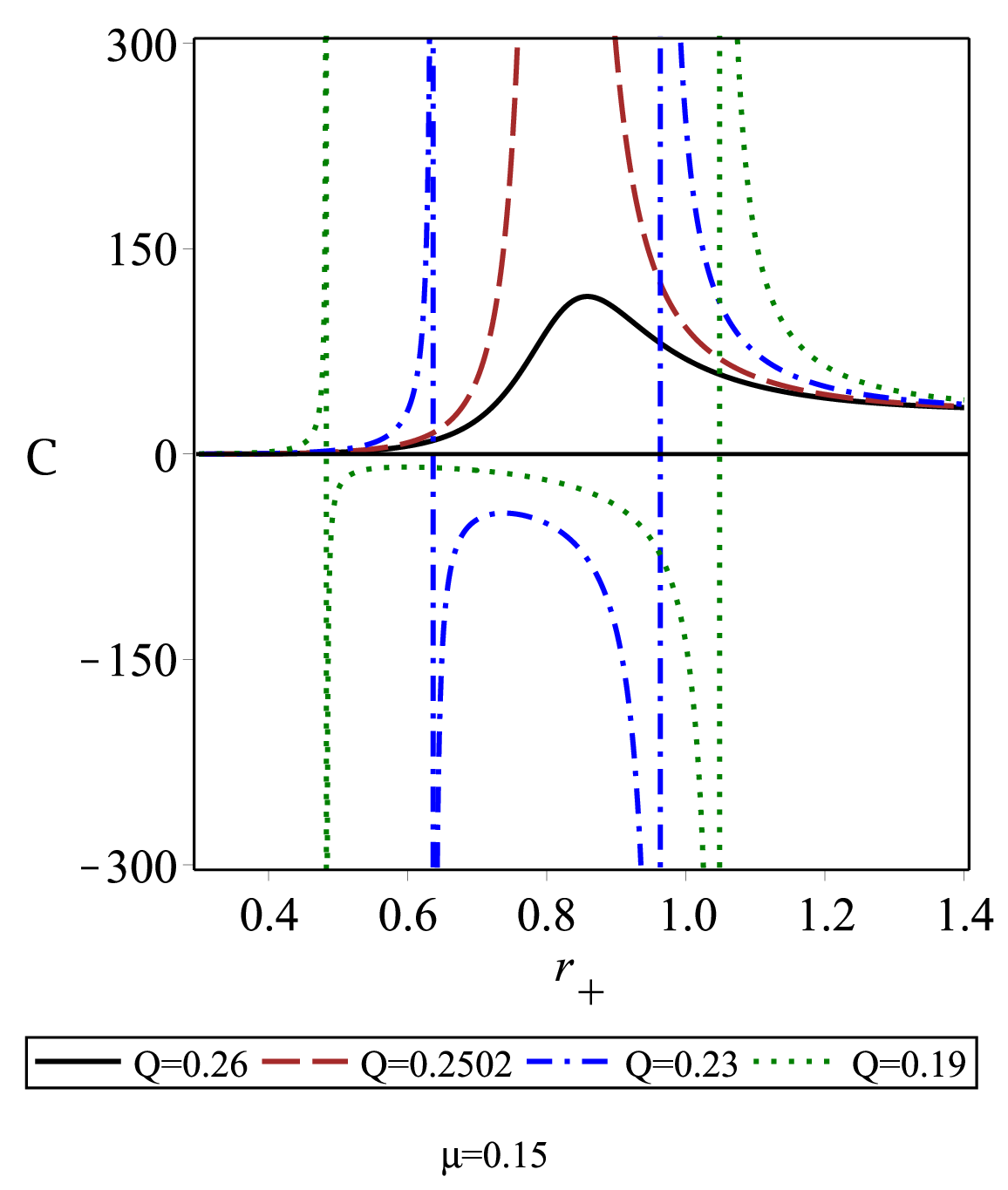} \newline
\includegraphics[width=0.33\textwidth]{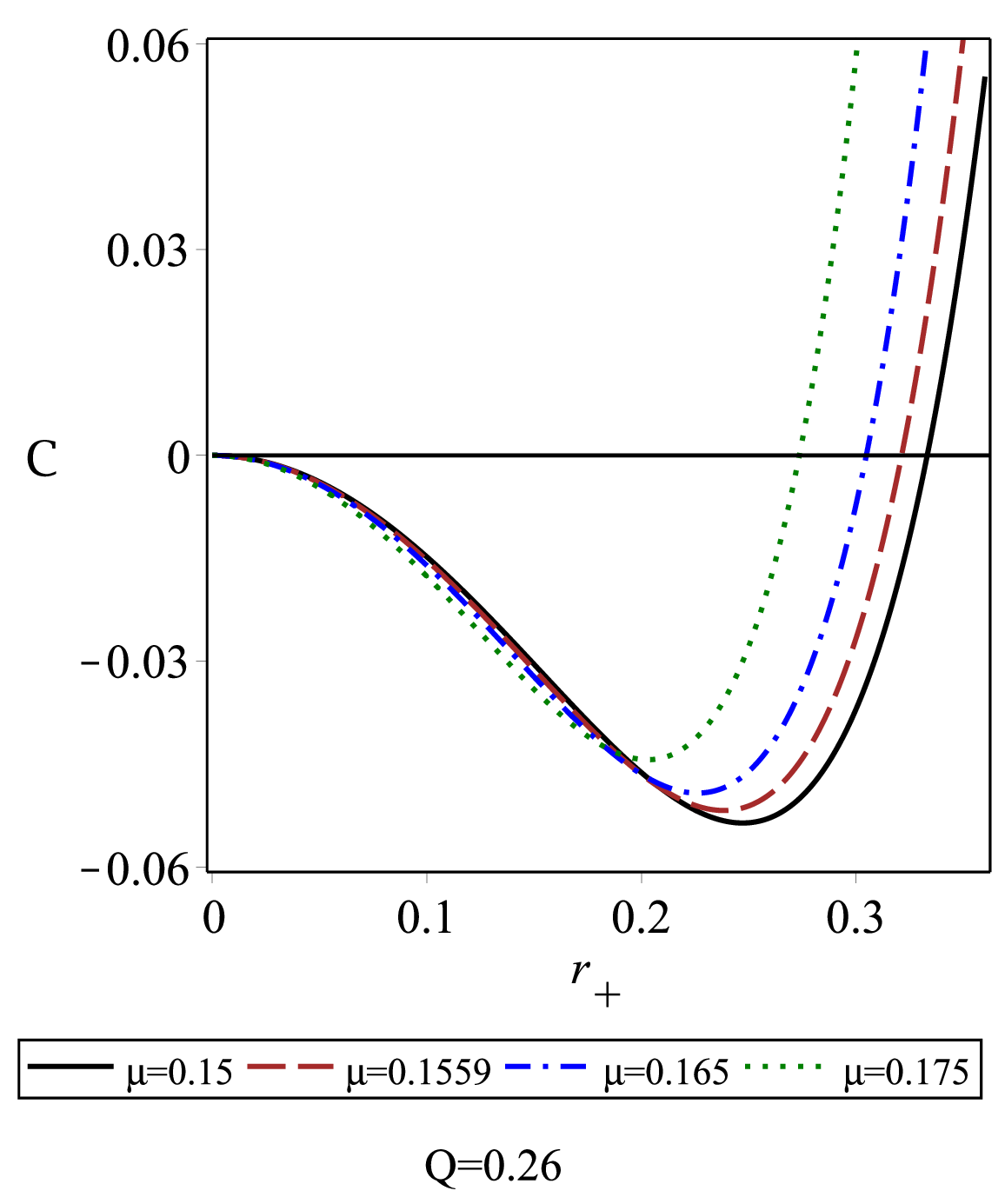} \includegraphics[width=0.33%
\textwidth]{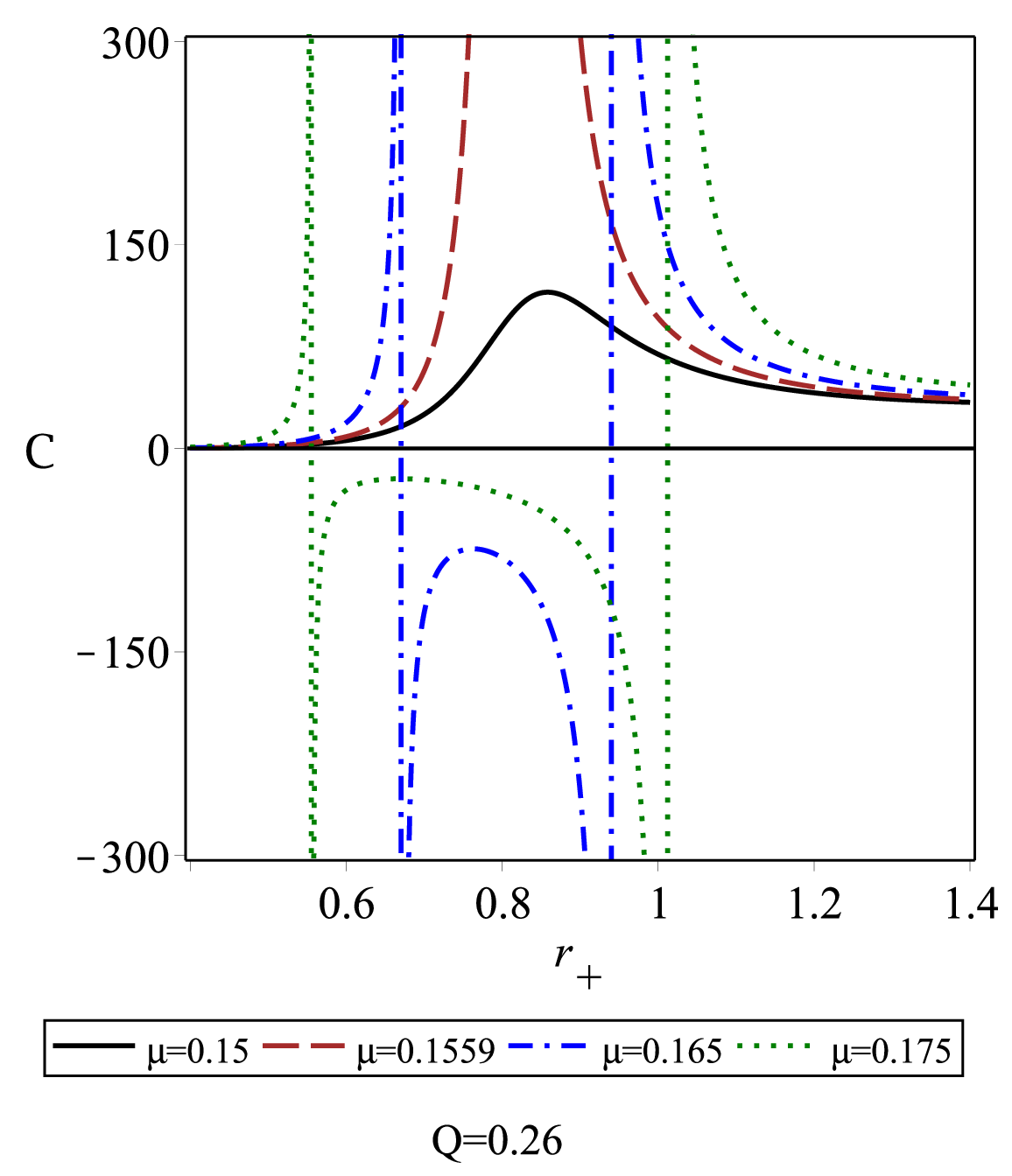} \newline
\caption{$C_{Q,\mu}$ versus $r_{+}$ for $\mathcal{B} =0.2 $, $\protect\beta=0.04
$, $\ell=2 $ and $A=0.02 $. Up panels for different values of the electric
charge (for different scales). Down panels for different values of the
string tension (for different scales).}
\label{Fig3}
\end{figure}

\begin{figure*}[tbh]
\centering
\includegraphics[width=0.3\linewidth]{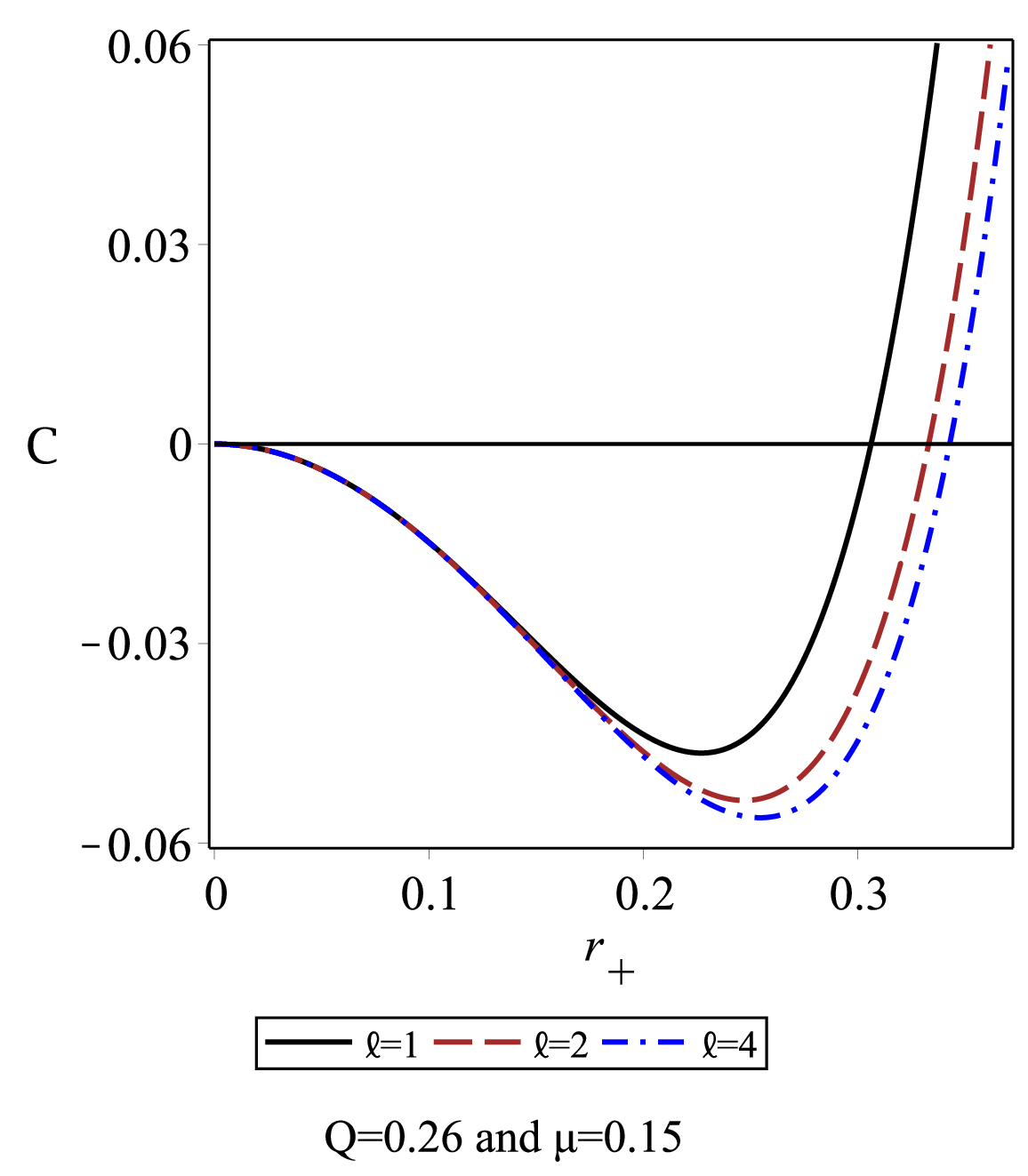}\hfil
\includegraphics[width=0.3\linewidth]{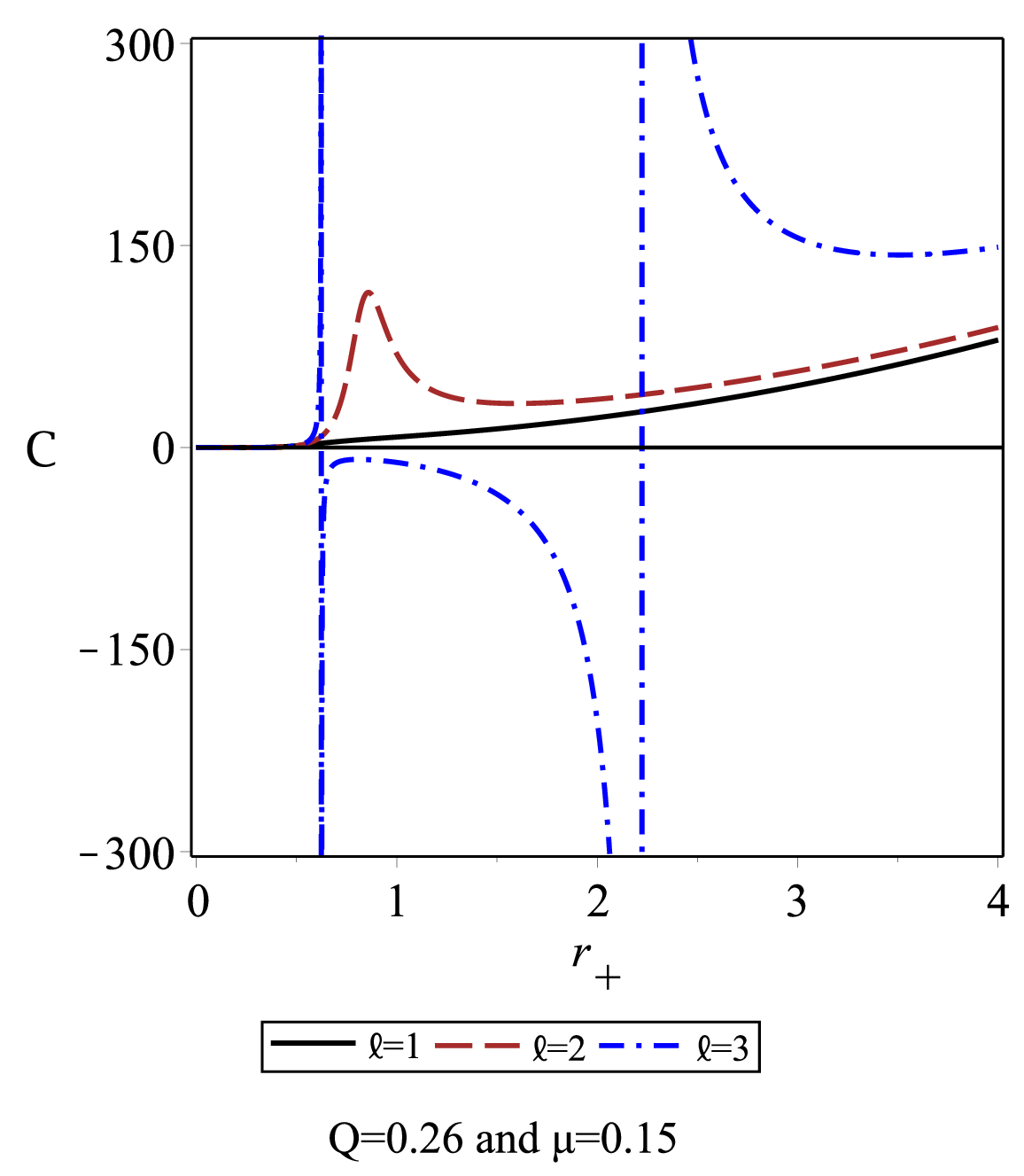}\hfil
\includegraphics[width=0.32\linewidth]{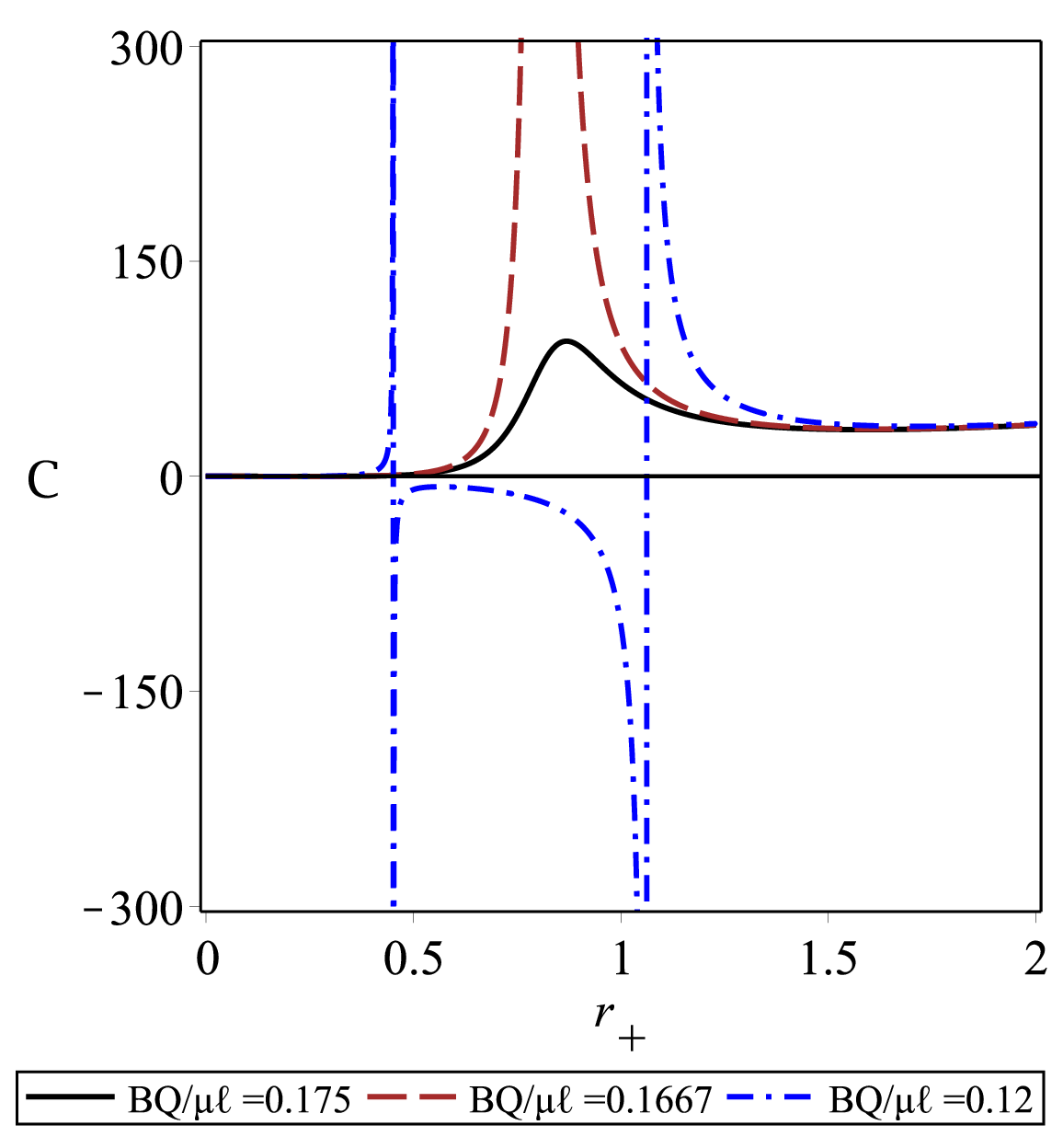}
\caption{$C_{Q,\mu}$ versus $r_{+}$ for $\mathcal{B}=0.2$. Two left panels for $\protect\beta =0.04$, $A=0.04$ (continuous line), $A=0.02$ (dashed line), $%
A=0.01$ (dash-dotted line) and different values of the AdS radius (for
different scales). Right panel: the relation between black hole parameters
for having phase transition.}
\label{Fig4}
\end{figure*}

To study the behavior of heat capacity and the effects of different
parameters on this quantity, we have plotted some diagrams in Figs. \ref%
{Fig3}, and \ref{Fig4}. Evidently, the heat capacity has only one
root for $Q>0.2502$ at $r_{+}=r_{0}$ which is a bound point (see
the up panels of Fig. \ref{Fig3}). For $r_{+}<r_{0}$, both
temperature and heat capacity are negative and so there is no
physical solution. In contrast, for $r_{+}>r_{0}$, thermally
stable phase can be observed for such black holes. Up panels of
Fig. \ref{Fig3}, show that the heat capacity has a root and two
divergencies ($r_{Div1}$ and $r_{Div2}$ where $r_{Div1}<r_{Div2}$)
for small electric charges. Between root and the smaller divergency,
the heat capacity has a positive value and black holes which are
placed in this region are in a stable phase. The region between
two divergencies is related to medium black holes that due to the
negativity of heat capacity, they are thermally unstable. For
region after the larger divergency, the heat capacity is positive
and large black holes are in a stable phase. As it was pointed out,
medium black holes are in an unstable phase and undergo a phase
transition.

Now, we would like to study the crucial role of string tension on
the phase transition of the system. In the bottom panels of Fig.
\ref{Fig3}, we have considered $Q$ as a fixed parameter and
investigated the behavior of heat capacity for different values of $\mu$. From up panels of Fig. \ref{Fig3}, we saw that only charged accelerating black holes with small electric charge enjoy the existence of phase transition. But down panels of Fig. \ref{Fig3}, show that such black holes undergo phase transition by increasing the string tension.

As it was mentioned \cite{AccV}, accelerating black holes
have well-defined thermodynamics just under satisfaction of the
certain condition $A\ell <1$. The cosmological constant which is
proportional to AdS radius is representing the natural curvature
of the spacetime. In other words, the curvature is an increasing/a
decreasing function of $\ell $ \cite{54}. Now, we are interested
to investigate how the AdS radius affects the phase transition of
the system. In left and middle panels of Fig. \ref{Fig4}, we
studied the role of $\ell $ on the heat capacity with fixed $\mu $
and $Q$. As we see, black holes with low acceleration (small
string tension) and large electric charge can experience phase
transition in a high curvature background.

As a result, one can find that charged accelerating black
holes with small electric charge undergo phase transition easily.
But black holes with large electric charge should be pulled by
stronger strings or they should be located in a higher curvature
background for having a phase transition. Therefore, one can find
that there is an implicit relationship between the electric
charge, string tension and AdS radius for observing a phase
transition. According to the right panel of Fig. \ref{Fig4}, we
find that for the mentioned parameters, the heat capacity diverges for $\frac{Q%
\mathcal{B}}{\mu \ell }<0.1667$. In contrast, for $\frac{Q\mathcal{B}}{\mu
\ell }>0.1667$, the heat capacity has no phase transition point. For the
case of $\frac{Q\mathcal{B}}{\mu \ell }=0.1667$, the heat capacity has only one divergence point where two phases of medium and large black holes are in equilibrium. This indicates that such a condition $\left( \frac{Q\mathcal{B}%
}{\mu \ell }<0.166\right) $ should be satisfied to have phase transition for charged accelerating black holes.

Now, we focus on the effects of electric charge and string tension on
thermal stability of the system. Fig. \ref{Fig3}, shows that as the electric charge (string tension) increases (decreases), bound point shifts to the larger horizon radius. So, the physical region is a decreasing (an increasing) function of $Q$ ($\mu $). Taking a close look at this figure, one can find that as the string tension (electric charge) increases (decreases), $r_{Div1}$ decreases whereas $r_{Div2}$ increases. So, the unstable region increases by increasing (decreasing) $\mu $ ($Q$), see also Fig. 5 for more details.

\begin{figure*}[!htb]
\centering
\includegraphics[width=0.38\linewidth]{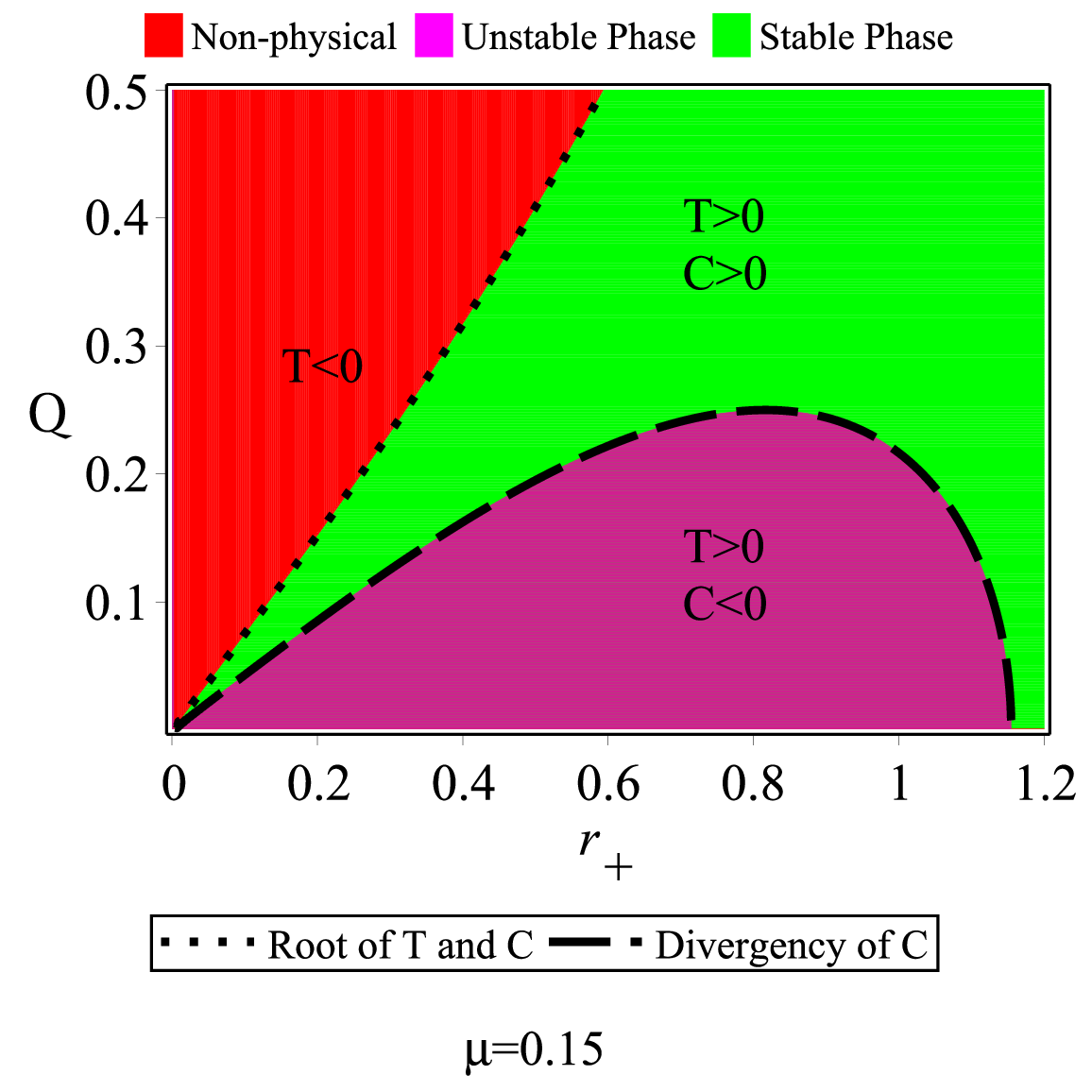} \includegraphics[width=0.38%
\linewidth]{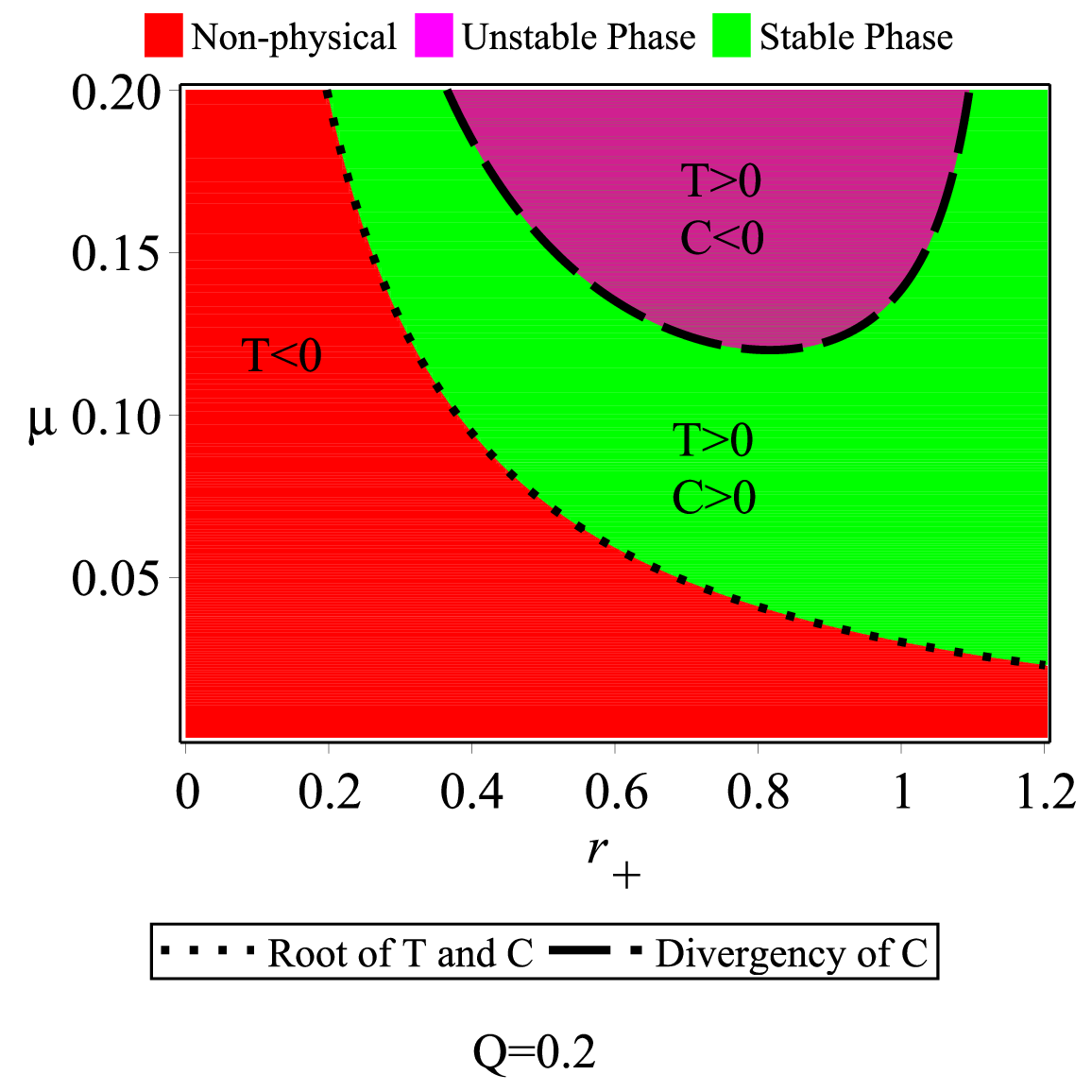}
\caption{Thermally stable and/or unstable regions of the black holes for $%
\mathcal{B}=0.2 $, $\protect\beta=0.04 $, $\ell=2 $ and $A=0.02 $.}
\label{Fig5}
\end{figure*}
\subsection{Geometrical thermodynamics and phase transition}

Another approach to study the thermodynamical behavior of a system
is geometrical thermodynamics. In this method, a thermodynamical
phase space (metric) is constructed by considering one of the
thermodynamical quantities as thermodynamical potential and other
quantities as extensive parameters. By calculating the Ricci
scalar of such a thermodynamical metric and determining its
divergence points, one can obtain phase transition points of the
system. It was shown that obtained results are matched with the
results of the heat capacity. It means that divergence points of
the Ricci scalar and divergence points (and root) of the heat
capacity are coincident (see Refs. \cite{HPEMI,HPEMII,HPEMIII,HPEMV,HPEMVI}, for more details). There
are several methods to build thermodynamical metrics known as
Weinhold, Ruppeiner, Quevedo and HPEM metrics. Here, we would like
to investigate phase transition of the charged accelerating AdS
black holes in the non-extended phase space via the GT method. To
introduce a suitable metric, we consider the total mass as a
thermodynamical potential with the entropy and the electric charge
as extensive parameters. The thermodynamical metrics in the
context of GTs are given by
\begin{figure*}[!htb]
\centering
\subfloat[]{
        \includegraphics[width=0.28\textwidth]{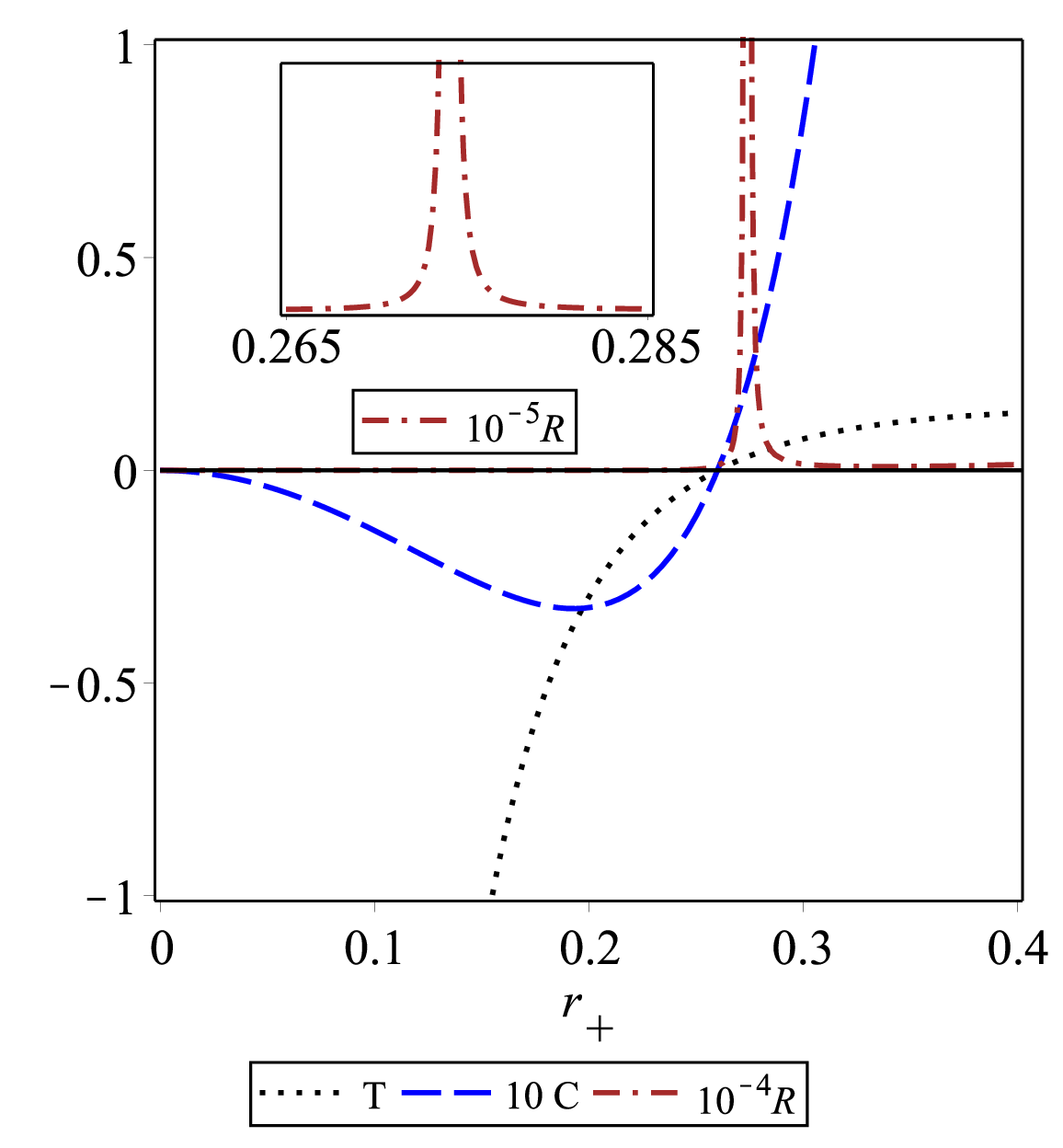}}
\subfloat[]{
        \includegraphics[width=0.27\textwidth]{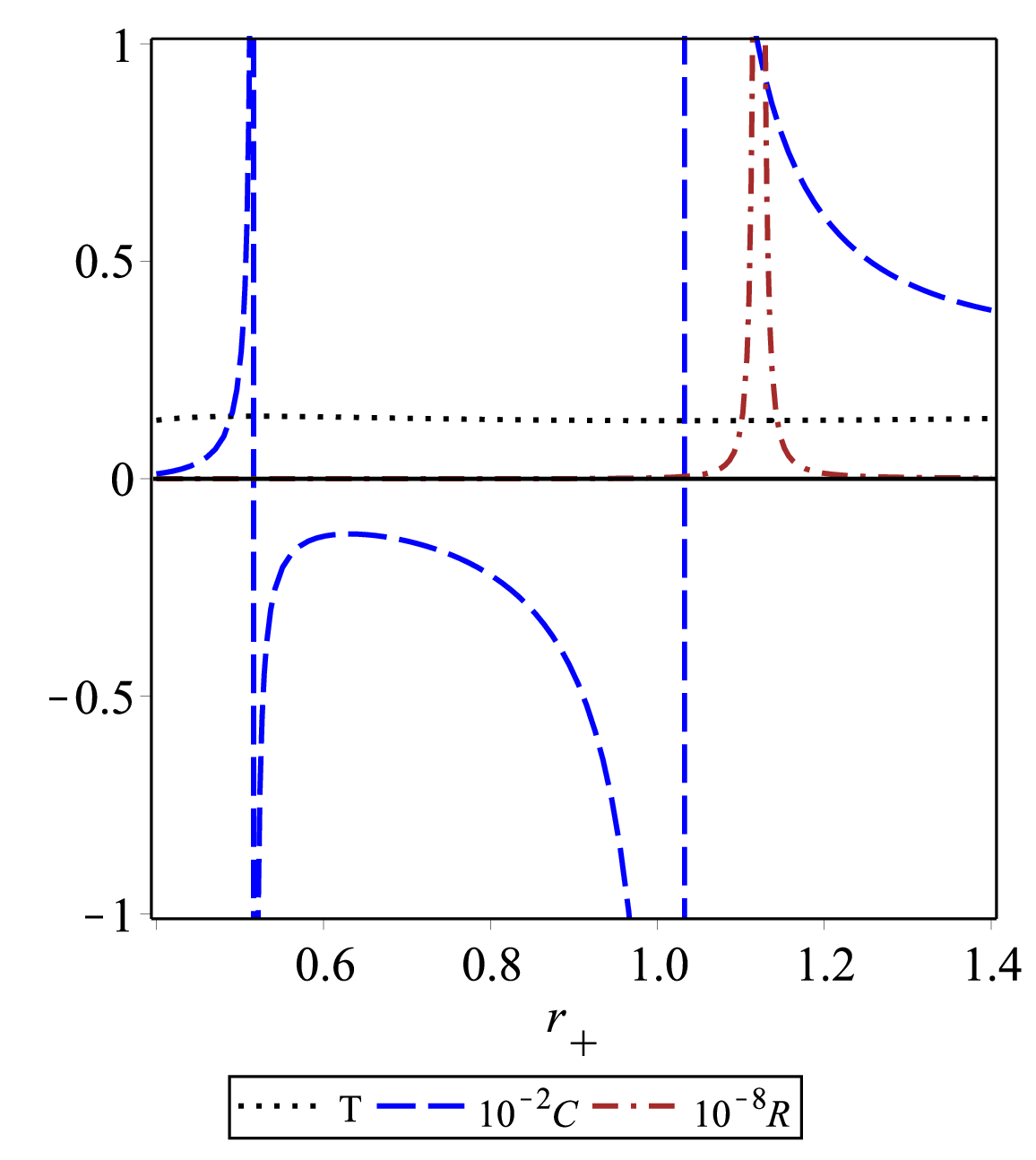}} \newline
\subfloat[]{
        \includegraphics[width=0.27\textwidth]{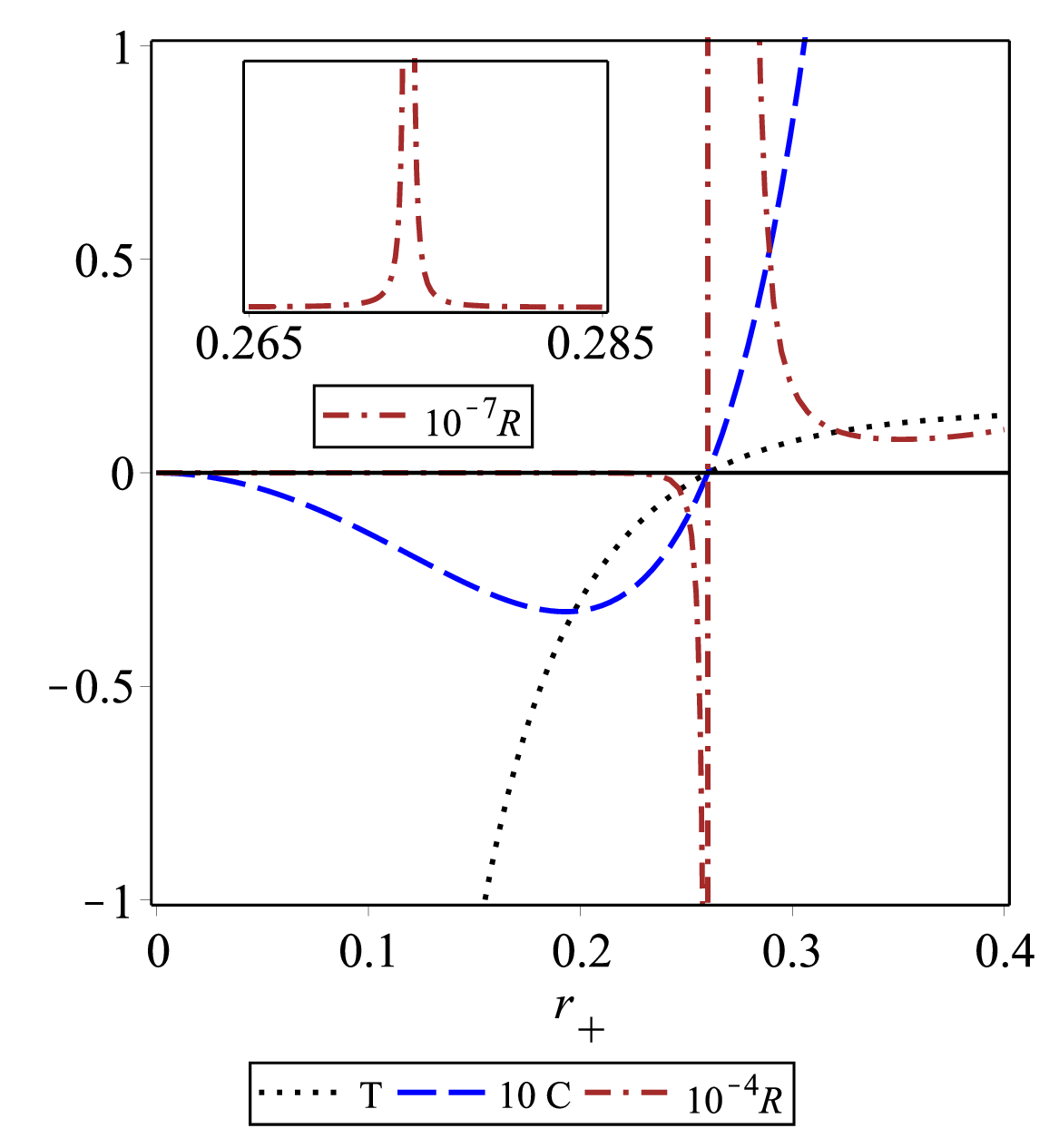}}
\subfloat[]{
        \includegraphics[width=0.27\textwidth]{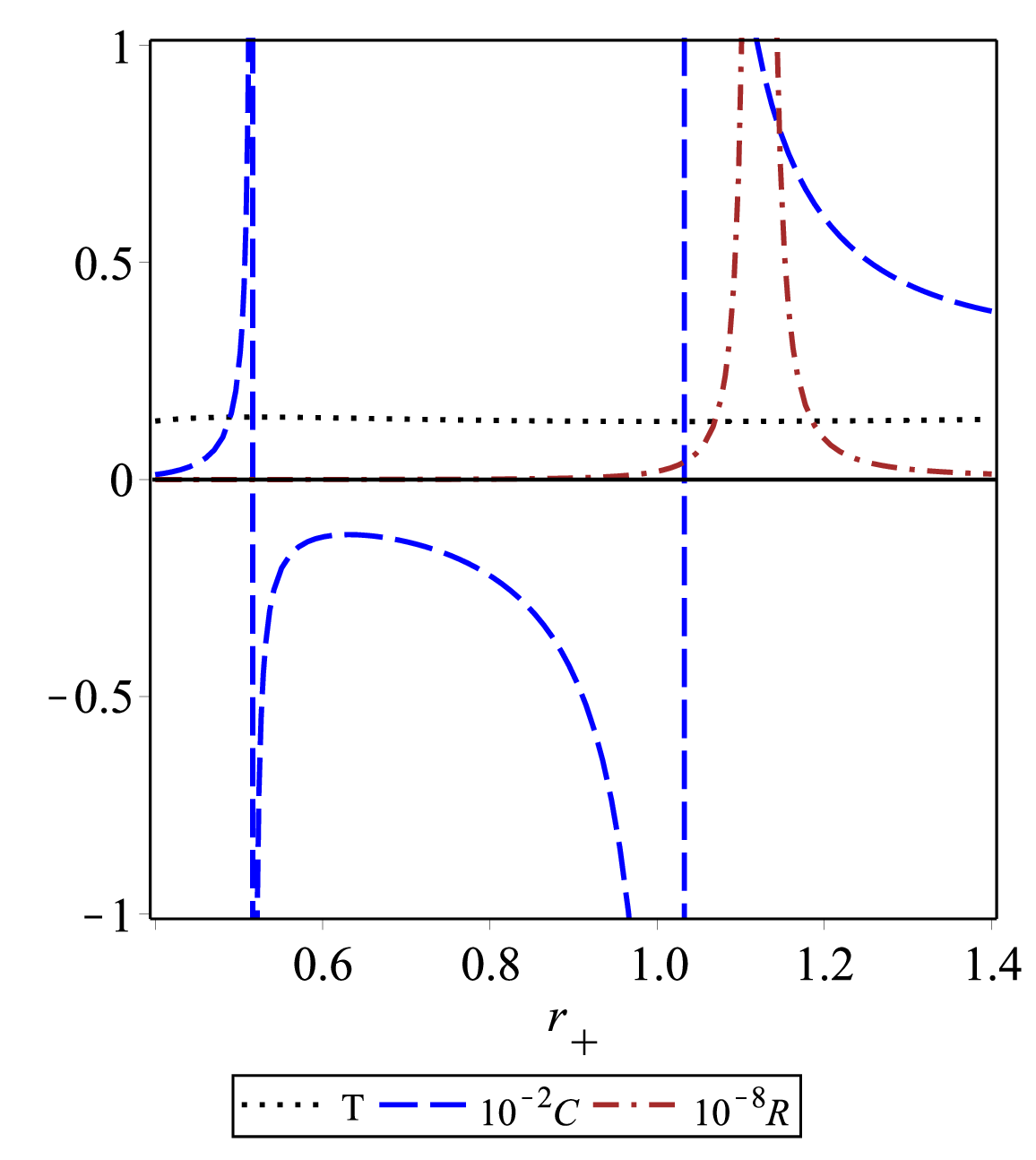}} \newline
\subfloat[]{
        \includegraphics[width=0.28\textwidth]{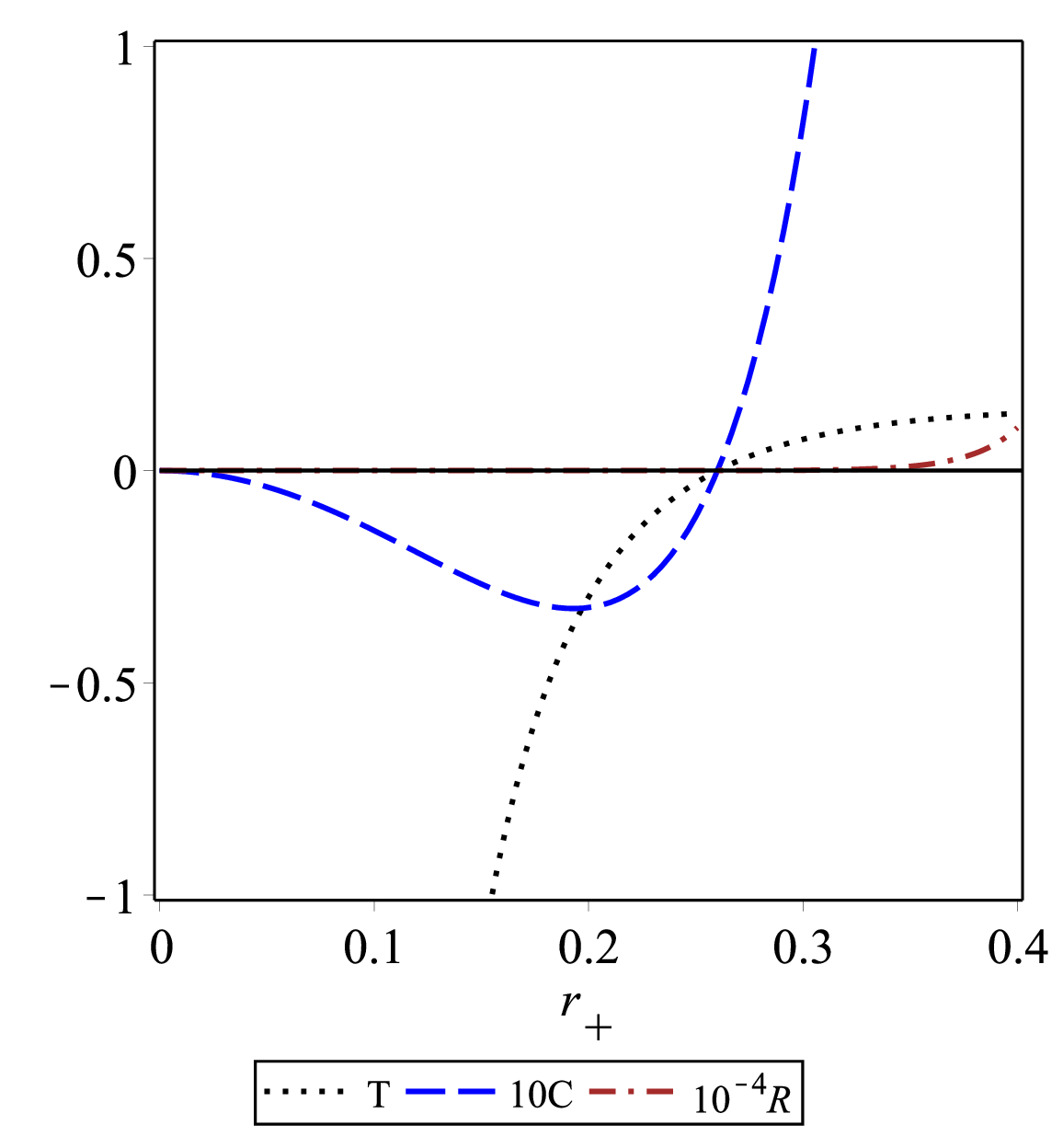}}
\subfloat[]{
        \includegraphics[width=0.27\textwidth]{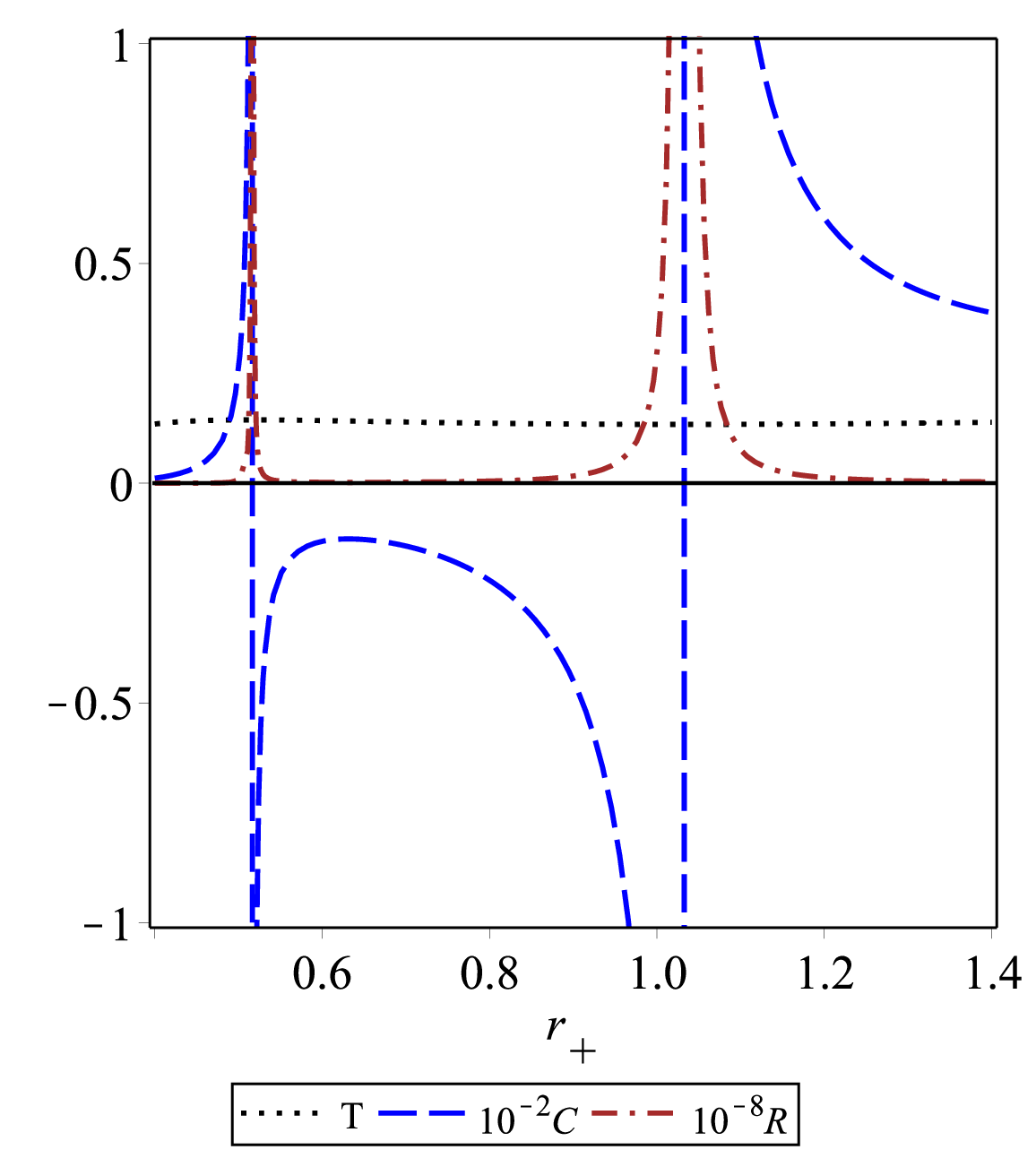}} \newline
\caption{$T$ (dotted line), $C_{Q,\mu}$ (dashed line) and
Ricci scalar ( dash-dotted line) versus $r_{+}$ for $Q=0.2$, $\protect\mu =0.15$, $\mathcal{B}=0.2$, $\protect\beta =0.04$, $\ell =2$ and $A=0.02$. Weinhold's Ricci scalar (up panels), Ruppeiner's Ricci scalar (middle panels) and Quevedo's Ricci scalar
(down panels).}
\label{Fig6}
\end{figure*}
\begin{equation}
ds^{2}=\left\{
\begin{array}{cc}
Mg_{ab}^{W}dX^{a}dX^{b}, & \text{Weinhold} \\
&  \\
-T^{-1}Mg_{ab}^{W}dX^{a}dX^{b}, & \text{Ruppeiner} \\
&  \\
(SM_{S}+QM_{Q})(-M_{SS}dS^{2}+M_{QQ}dQ^{2}), & \text{Quevedo} \\
&  \\
S\frac{M_{S}}{M_{QQ}^{3}}(-M_{SS}dS^{2}+M_{QQ}dQ^{2}), & \text{HPEM}%
\end{array}%
\right. ,  \label{Eq15}
\end{equation}
where $ M_{k}=\frac{\partial M}{\partial k} $ and $
M_{kj}= \frac{\partial^{2} M}{\partial k \partial j} $. Since we
are looking for the divergence points of the Ricci scalar and due
to the fact that its numerator is a smooth finite function, we
focus on its denominator. The denominators of the Ricci scalars are
\begin{equation}
denom(\mathcal{R})=\left\{
\begin{array}{cc}
(M_{SS}M_{QQ}-M_{SQ}^{2})^{2}M^{2}, & \text{Weinhold} \\
&  \\
(M_{SS}M_{QQ}-M_{SQ}^{2})^{2}TM^{2}, & \text{Ruppeiner} \\
&  \\
M_{SS}^{2}M_{QQ}^{2}(SM_{S}+QM_{Q})^{3}, & \text{Quevedo} \\
&  \\
S^{3}M_{S}^{3}M_{SS}^{2}, & \text{HPEM}%
\end{array}%
\right. .  \label{Eq16}
\end{equation}

\begin{figure}[!htb]
\centering
\includegraphics[width=0.33\textwidth]{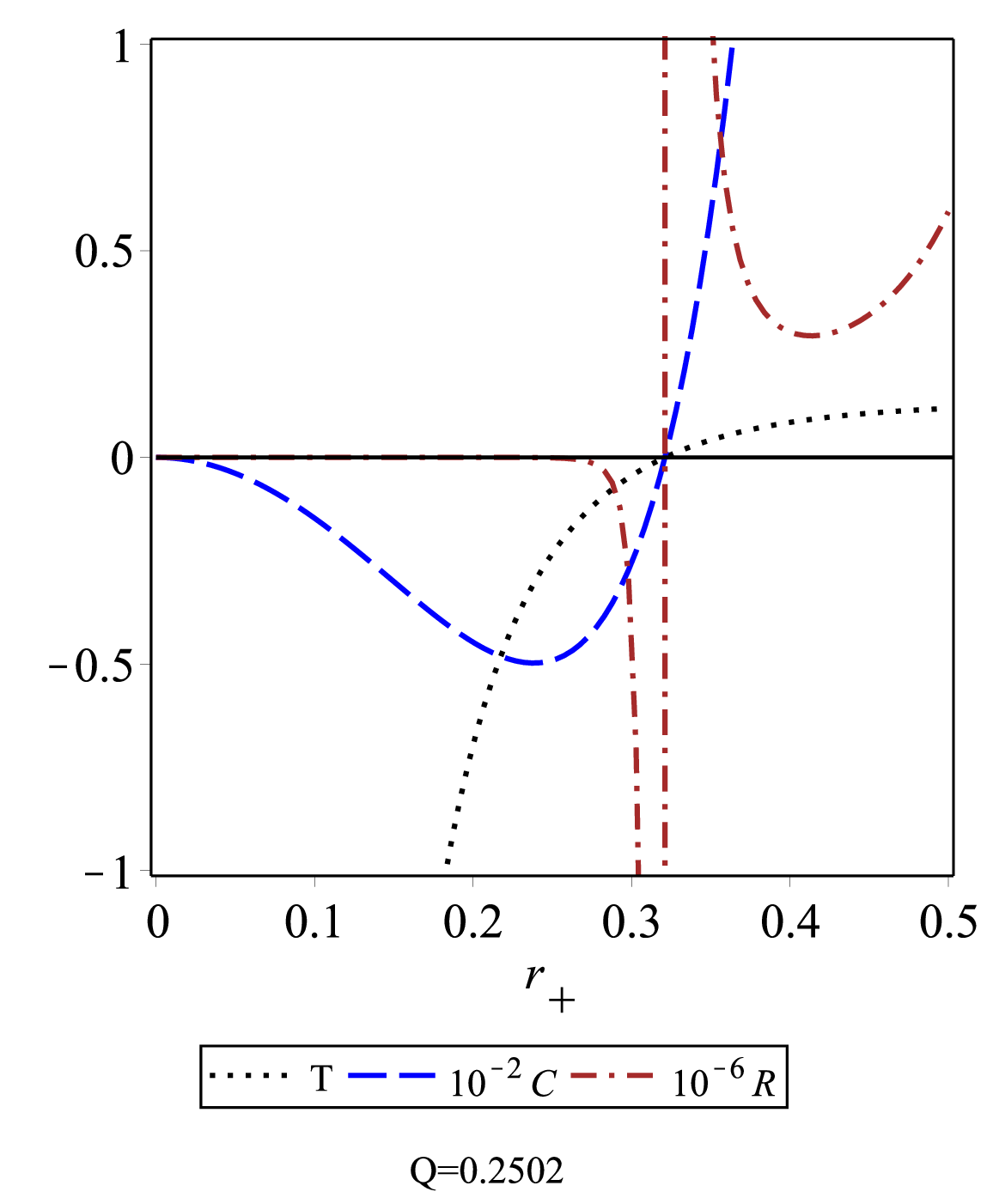} \includegraphics[width=0.33%
\textwidth]{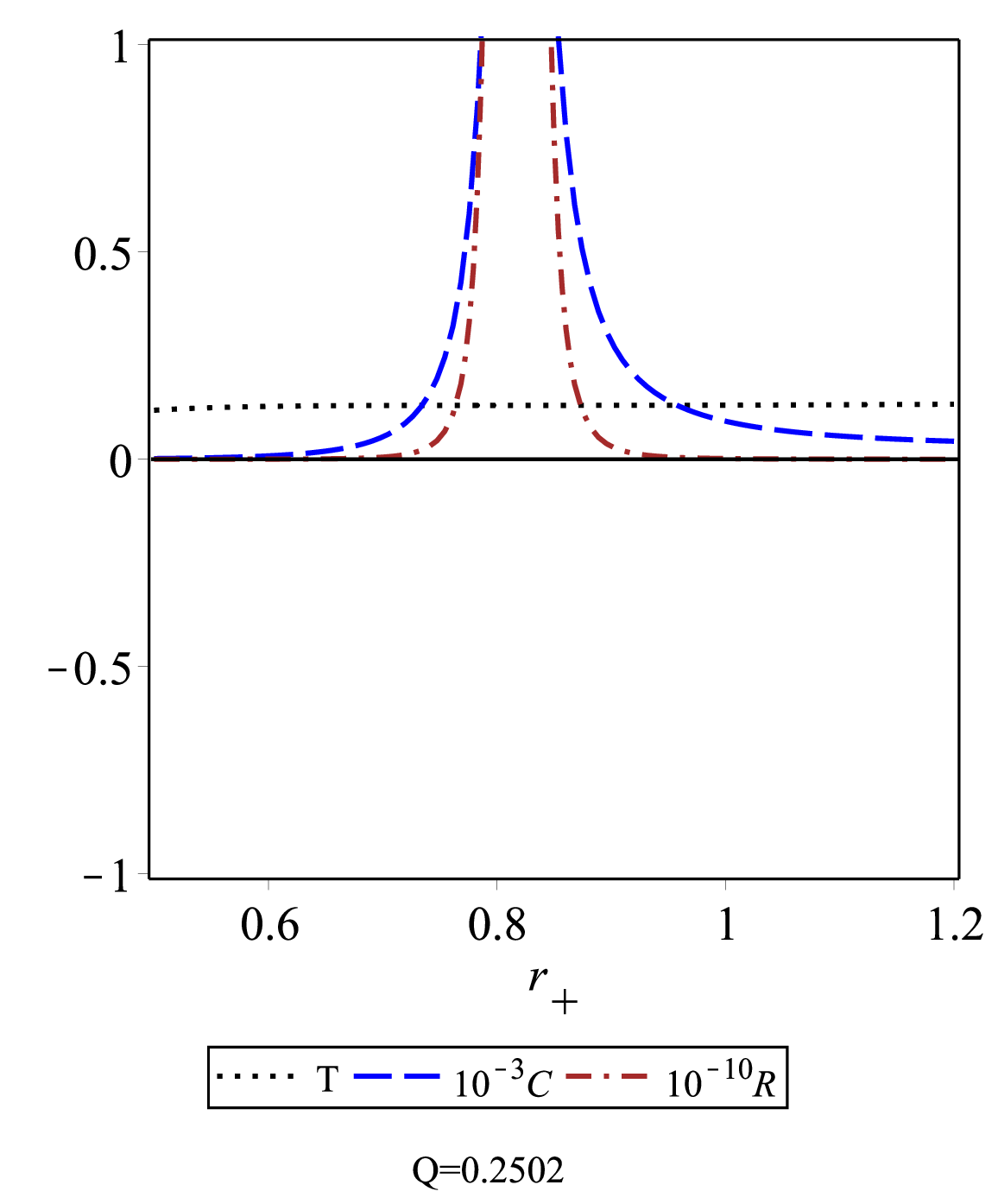} \newline
\includegraphics[width=0.33\textwidth]{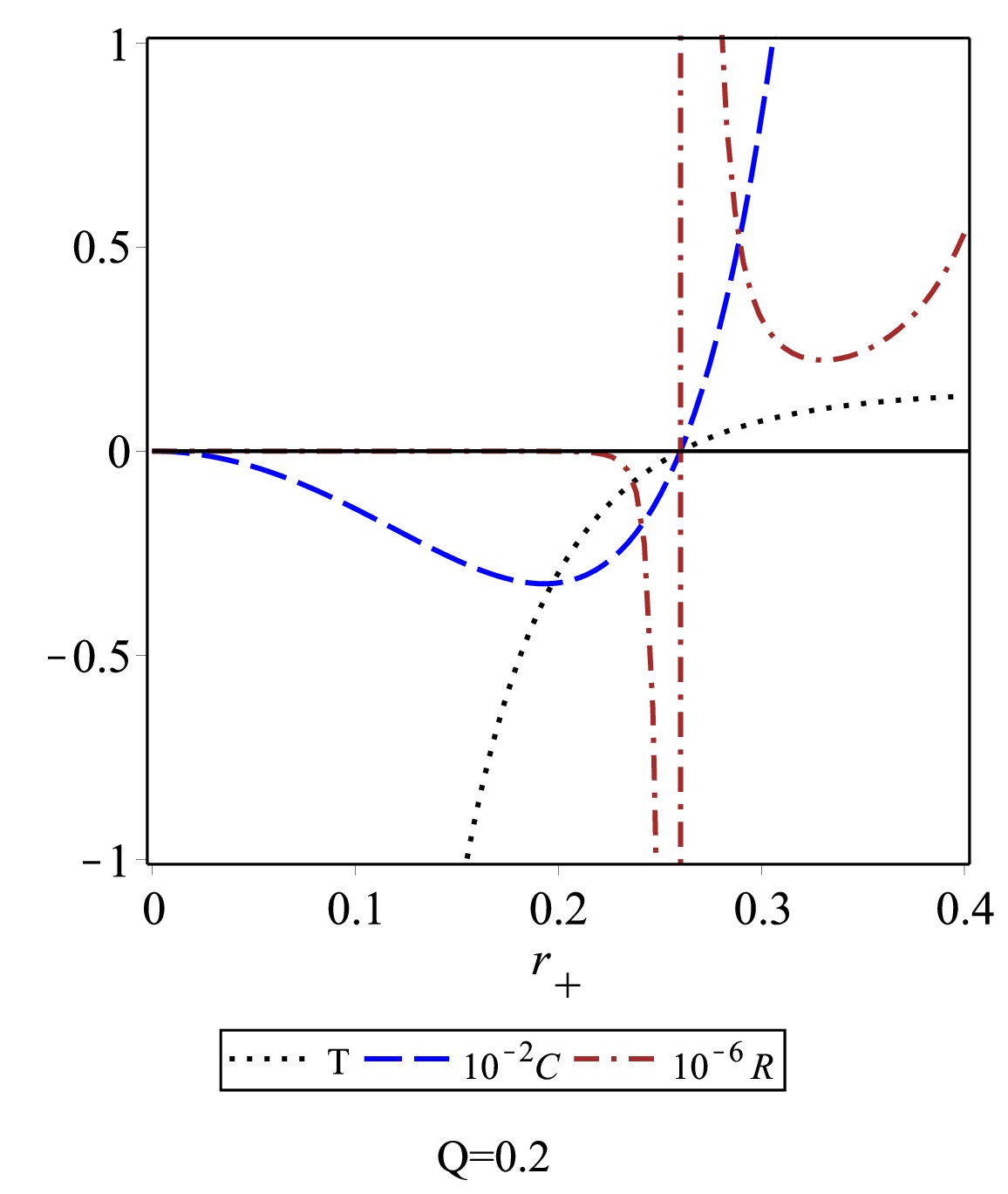} \includegraphics[width=0.33%
\textwidth]{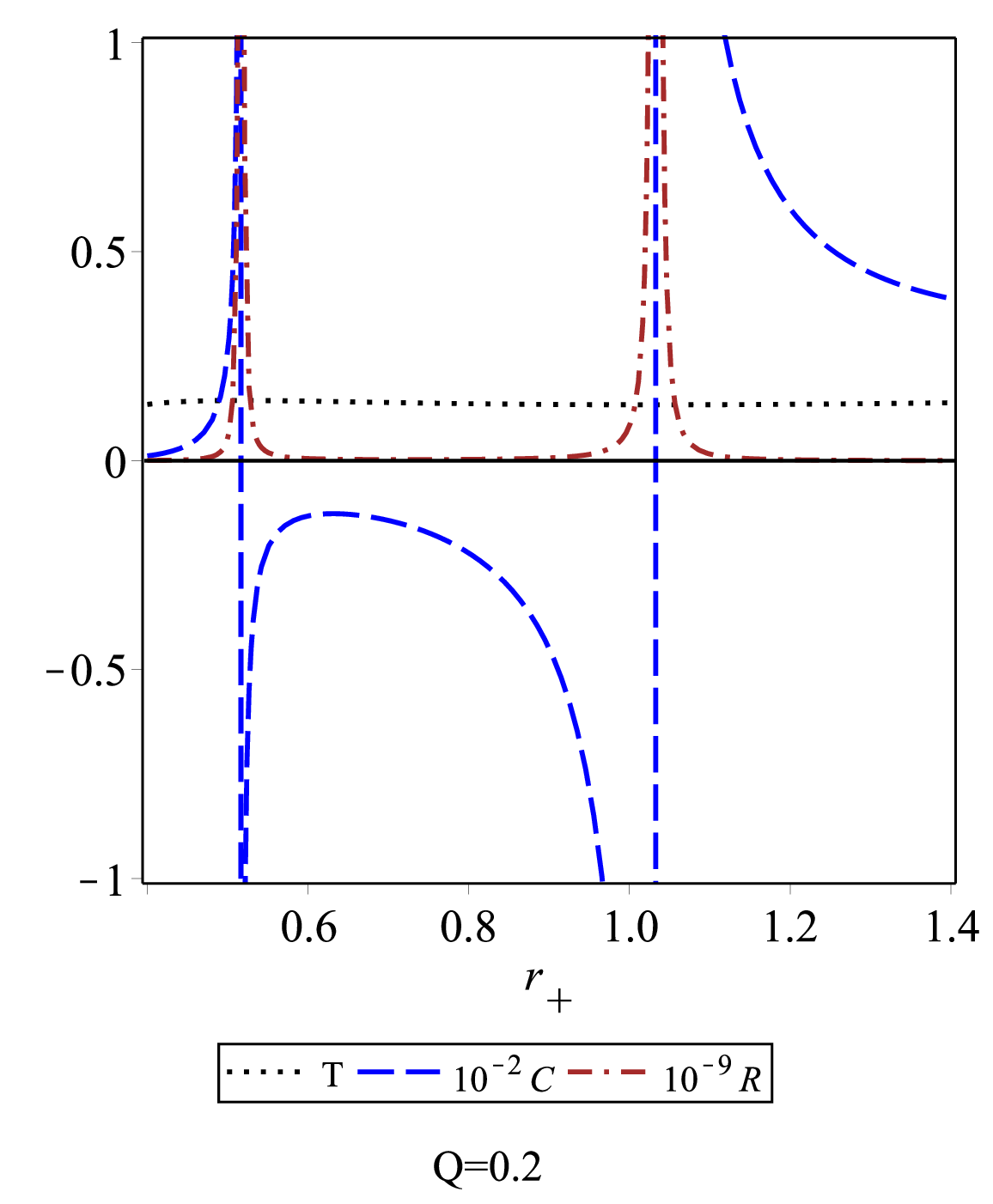} \newline
\caption{$T$ (dotted line), $C_{Q,\mu}$ (dashed line) and $\mathcal{R}_{HPEM}$
(dash-dotted line) versus $r_{+}$ for $Q=0.2$, $\protect\mu =0.15$, $%
\mathcal{B}=0.2$, $\protect\beta =0.04$, $\ell =2$, and $A=0.02$.}
\label{Fig7}
\end{figure}

Taking Eq. (\ref{Eq16}) into account, we are in a position to find
that among the mentioned metrics, which one is an efficient tool
to describe phase transitions of the charged accelerating black
holes. To have a proper geometrical approach for studying phase
transitions, the thermodynamic Ricci scalar should diverge at
points we will mention: bound and phase transition
points. Regarding Eq. (\ref{Eq16}), it is evident that only in a special case $M_{SQ}=0$ and nonzero $M_{QQ}$, the divergence points of the heat
capacity coincide with divergencies of the Weinhold and Ruppeiner
Ricci scalars. Due to the presence of the temperature in the
denominator of Ruppeiner's Ricci scalar, this metric can describe
a bound point, while Weinhold's metric just under the satisfaction
of condition $M_{SS}M_{QQ}-M_{SQ}^{2}=M_{S}$, is able to describe
this point. For the case of $M_{SS}=\frac{M_{SQ}^{2}}{M_{QQ}}$,
there are extra divergencies for $R_{W}$ (Weinhold's Ricci scalar)
and $R_{R}$ (Ruppeiner's Ricci scalar) which are not related to
any phase transition of the heat capacity. Regarding Quevedo's
metric, the existence of $M_{SS}$ in the denominator of Ricci
scalar guarantees that the divergencies of the heat capacity and
Quevedo's Ricci scalar coincide. However, considering the presence
of $M_{QQ}^{2}$, one may find an extra singular point for
vanishing $M_{QQ}$. It is worthwhile to mention that another
divergence point may be appeared from zero of $SM_{S}+QM_{Q}=0$ as
well. Regarding the bound point, as we see from the denominator of
$R_{Q}$ (Quevedo's Ricci scalar), coincidence with this point
takes place only for vanishing $M_{Q}$ (which is in general a
nonzero function). We continue our investigation by employing
HPEM's metric. Due to the structure of denominator of $R_{H}$
(HPEM's Ricci scalar), the divergence points and root of the heat
capacity coincide with divergencies of the HPEM's Ricci scalar.
So, this metric provides a successful mechanism for investigating
bound and phase transition points of such black holes. In
order to have a more precise picture, we have plotted some diagrams in Figs. \ref{Fig6} and \ref{Fig7} which confirm our analysis. A remarkable issue regarding the HPEM's metric is that, in addition to describe bound and phase transition points, one can distinguish these points from each other. It means that the behavior of HPEM's Ricci scalar differs around root and divergencies of $C_{Q,\mu}$ (compare the left panels of Fig. \ref{Fig7}, with its right panels). Considering the root of heat capacity, the sign of Ricci scalar near divergence point changes from $-\infty $ to $+\infty $. Whereas, for the case of divergencies of heat capacity the sign of Ricci scalar does not change. So, one can recognize bound point and phase transition points by studying the behavior of the HPEM's Ricci scalar around its divergencies.

A significant point about geometrical methods is that they are based on the grand canonical ensemble foundation. Here, we connect these methods to the heat capacity only for the sake of comparison and check the validity of the results.
\section{Thermodynamical structure in extended phase space \label{Thermo-E}}

In this section, we extend the phase space by considering
the cosmological constant as a thermodynamical quantity and
investigate the phase structure of the system. Recently, the idea
of variable $\Lambda$ has been attracted a lot of attention
\cite{12,HLIV}. Considering the cosmological constant as a
thermodynamical pressure and its conjugate quantity as a
thermodynamical volume leads to a new insight into thermodynamical
structure and phase transition of the black holes. Here, we study
the phase structure of the system through three approaches including
heat capacity, thermodynamical geometry and van der Waals like
behavior and show that these approaches yield consistent results.

\subsection{$P - V$ criticality of the charged accelerating black holes}

Regarding the normalization-free of the time coordinate, the van der Waals
like behavior of such black holes was investigated in Ref. \cite{53}. They
considered $Ar_{+}$ as a constant parameter and studied phase transition of the black hole in the extended phase space. They showed that in order to have a phase transition, one has to consider $Ar_{+}=cte$. In other words, only under a certain condition, there is a small/large black hole phase transition for accelerating black holes. Here, we relax this condition and investigate van der Waals like behavior by re-scaling the time coordinate. We determine critical values by using a new method which was introduced in Ref. \cite{54}. To start, we review some basic thermodynamical properties of charged accelerating AdS black holes. The pressure associated with the cosmological constant is given by
\begin{equation}
P=-\frac{\Lambda }{8\pi }=\frac{3}{8\pi \ell ^{2}}.  \label{Eq17}
\end{equation}

Using Eqs. (\ref{Eq7}), (\ref{Eq12}) and (\ref{Eq17}), one can obtain
thermodynamical volume which is the conjugate quantity of pressure as
\begin{equation}
V=\left( \frac{dM}{dP}\right) _{Q,\mu }=\frac{4\pi }{3K\alpha }\left(
r_{+}^{3}+2A^{2}r_{+}^{5}+\frac{3A^{2}r_{+}^{3}}{16\pi P}+\frac{9A^{2}r_{+}}{%
128\pi ^{2}P^{2}}\left( 1+\frac{Q^{2}\mathcal{B}^{2}}{\mu ^{2}r_{+}^{2}}%
\right) \right)
\label{EqV17}.
\end{equation}

The Helmholtz free energy which is employed to extract information regarding the phase transitions and chemical equilibrium is expressed as \cite{NEDXII,AccX,AccXI}
\begin{equation}
F=M-TS,  \label{Eq19}
\end{equation}%
the thermodynamic equilibrium corresponds to the global minimum of $F$.

To study the van der Waals like phase transition, obtaining the equation of state is necessary. Inserting Eq. (\ref{Eq17}) into Eq. (\ref{Eq13}), one can calculate the equation of state as
\begin{equation}
P=\frac{3\left( 1-A^{2}r_{+}^{2}\right) \left[ 2\pi Tr_{+}\left( 2-\beta
^{2}+A^{2}Q^{2}\right) +\left( 1-\frac{\mathcal{B}^{2}Q^{2}}{\mu
^{2}r_{+}^{2}}\right) \left( A^{2}r_{+}^{2}-1\right) \right] }{8\pi
r_{+}^{2}\left( 3-A^{2}r_{+}^{2}\right) }.  \label{Eq20}
\end{equation}

To have a better understanding of pressure's properties, we find
its limiting behaviors as
\begin{equation}\label{limitP}
P~\Rightarrow \left\{
\begin{array}{cc}
\lim_{r_{+}\longrightarrow 0}P\propto \frac{Q^{2}\mathcal{B}^{2}}{8\pi \mu
^{2}r_{+}^{4}}-\frac{5A^{2}Q^{2}\mathcal{B}^{2}}{24\pi \mu ^{2}r_{+}^{2}}-%
\frac{1}{8\pi r_{+}^{2}}+O(r_{+}), & \text{small black holes} \\
&  \\
\lim_{r_{+}\longrightarrow \infty }P\propto \frac{3A^{2}}{8\pi }+\frac{3T}{%
2r_{+}}-\frac{3T\beta ^{2}}{4r_{+}}+\frac{3TA^{2}Q^{2}\mathcal{B}^{2}}{4\mu
^{2}r_{+}}+O(\frac{1}{r_{+}^{2}}), & \text{large black holes}%
\end{array}%
\right. .
\end{equation}

Considering Eq. (\ref{limitP}), we find that in the
absence of electric charge, the dominant term of the pressure is
only a function of the horizon radius that is a negative term.
Since the negative pressure is not physically acceptable and it
contradicts with the positive definite definition of the pressure
in Eq. (\ref{Eq17}), we conclude that although very small
uncharged accelerating black holes have a physical temperature,
they are impermissible from the thermodynamic point of view in the
extended phase space. While for large black holes, one finds that
the acceleration parameter is the governing factor in pressure and
depending on the values of different parameters, it can be
positive or negative.

Now, we are going to investigate the existence of van der Waals like phase
transition for such black holes. To do so, we obtain the specific volume
which is related to the horizon radius as
\begin{equation}
\upsilon =2\left( \frac{3V}{4\pi }\right) ^{\frac{1}{3}}=\frac{2r_{+}}{K^{%
\frac{1}{3}}}\left[ 1+\frac{\beta ^{2}}{6}+\frac{A^{2}}{16\pi P}+\frac{%
A^{2}r_{+}^{2}}{6}\left( 4-\frac{Q^{2}\mathcal{B}^{2}}{\mu ^{2}r_{+}^{2}}%
\right) +\frac{3A^{2}}{128\pi ^{2}P^{2}r_{+}^{2}}\left( 1+\frac{Q^{2}%
\mathcal{B}^{2}}{\mu ^{2}r_{+}^{2}}\right) \right] ,  \label{Eq21}
\end{equation}%
and the equation of state (\ref{Eq20}) can be arranged as

\begin{eqnarray}
\label{EqPv7}
P&=&\frac{T\mu^{\frac{1}{3}}}{\mathcal{B}^{\frac{1}{3}}\upsilon}-\frac{T\mu^{\frac{1}{3}}\beta^{2}}{3\mathcal{B}^{\frac{1}{3}}\upsilon}+\frac{TA^{2}Q^{2}\mathcal{B}^{\frac{5}{3}}}{3\mu^{\frac{5}{3}}\upsilon }-\frac{\mu^{\frac{2}{3}}}{2\pi \mathcal{B}^{\frac{2}{3}}\upsilon^{2}}-\frac{\beta^{2}\mu^{\frac{2}{3}}}{6\pi \mathcal{B}^{\frac{2}{3}}\upsilon^{2}}+ \frac{2A^{2}\mathcal{B}^{\frac{4}{3}}Q^{2}}{3\pi\mu^{\frac{4}{3}}\upsilon^{2} }+\frac{2\mathcal{B}^{\frac{2}{3}}Q^{2}}{\pi \mu^{\frac{2}{3}}\upsilon^{4}}+\frac{4Q^{2}\beta^{2}\mathcal{B}^{\frac{2}{3}}}{3\pi \mu^{\frac{2}{3}}\upsilon^{4}}\\ \nonumber
&+&\frac{A^{2}}{24\pi} -\frac{4A^{2}Q^{4}\mathcal{B}^{\frac{8}{3}}}{3\pi \mu^{\frac{8}{3}}\upsilon^{4}}+\frac{A^{2}Q^{2}\mathcal{B}^{2}}{\pi \chi}+\frac{\mathcal{B}\mu TA^{2}\upsilon^{3}}{8\chi}-\frac{\mathcal{B}^{\frac{2}{3}}\mu^{\frac{4}{3}} A^{2}\upsilon^{2}}{8\pi\chi}-\frac{3\mathcal{B}^{\frac{4}{3}}\mu^{\frac{8}{3}} A^{2}\upsilon^{4}}{8\pi\chi^{2}}+\frac{3 T\mathcal{B}^{\frac{5}{3}}\mu^{\frac{7}{3}} A^{2}\upsilon^{5}}{8\chi^{2}}\\ \nonumber
&+&\frac{3 \mathcal{B}^{\frac{8}{3}}\mu^{\frac{4}{3}} A^{2}Q^{2}\upsilon^{2}}{2\pi\chi^{2}}+\frac{3 T \mu A^{2}Q^{2}\mathcal{B}^{3} \upsilon^{3}}{2\chi^{2}}+\frac{12 A^{2}Q^{4}\mathcal{B}^{4}}{2\pi\chi^{2}}
,  
\end{eqnarray}%
where
\begin{equation*}
\chi =4Q^{2}\mathcal{B}^{2}+2\pi T\mu \upsilon ^{3}\mathcal{B}-\mu ^{\frac{4%
}{3}}\upsilon ^{2}\mathcal{B}^{\frac{2}{3}}.
\end{equation*}

As we know, the van der Waals liquid$-$gas system goes under a first order phase
transition for temperatures smaller than the critical temperature ($T<T_{c}$) whereas, its phase transition is a second order one at the
critical temperature \cite{12,55}. Fig. \ref{Fig8}, confirms van der Waals
like behavior for the charged accelerating AdS black holes. Formation of the swallow-tail shape in $F-T$ diagram (continuous line of Fig. \ref{Fig8}) indicates the existence of a first-order small-large black hole transition for $P<P_{c}$. The critical point of the system which coincides with the inflection point of $P-\upsilon $ diagram is obtained as
\begin{equation}
\frac{\partial P}{\partial \upsilon }\bigg|_{\upsilon =\upsilon
_{c},T=T_{c}}=0~~~\&~~~\frac{\partial ^{2}P}{\partial \upsilon ^{2}}\bigg|%
_{\upsilon =\upsilon _{c},T=T_{c}}=0.  \label{Eq23}
\end{equation}

Eq. (\ref{EqPv7}) is much complicated to determine critical quantities
analytically by usual method. But, one can obtain approximate critical
values for very small $Q$ and $A$ as follows
\begin{eqnarray}
\upsilon _{c} &=&\frac{2\sqrt{6}Q\mathcal{B}^{\frac{2}{3}}\sqrt{3\left( 1+%
\frac{2\beta ^{2}}{3}\right) -\frac{2A^{2}Q^{2}\mathcal{B}^{2}}{\mu ^{2}}}}{%
\mu ^{\frac{2}{3}}\sqrt{3\left( 1+\frac{\beta ^{2}}{3}\right) -\frac{%
4A^{2}Q^{2}\mathcal{B}^{2}}{\mu ^{2}}}},  \notag \\
&&  \notag \\
T_{c} &=&\frac{8\left[ \frac{2A^{2}Q^{2}\mathcal{B}^{2}}{\mu ^{2}}-3\left( 1+%
\frac{2\beta ^{2}}{3}\right) \right] Q^{2}\mathcal{B}^{2}+\mu ^{\frac{4}{3}%
}\upsilon _{c}^{2}\left( 3\left( 1+\frac{\beta ^{2}}{3}\right) -\frac{%
4A^{2}Q^{2}\mathcal{B}^{2}}{\mu ^{2}}\right) \mathcal{B}^{\frac{2}{3}}}{\pi
\mu \upsilon _{c}^{3}\left( 3\left( 1-\frac{\beta ^{2}}{3}\right) +\frac{%
A^{2}Q^{2}\mathcal{B}^{2}}{\mu ^{2}}\right) \mathcal{B}},  \notag \\
&&  \notag \\
P_{c} &=&\frac{12\mu^{2}\upsilon _{c}^{2}\left( 1+\frac{\beta ^{2}}{3}\right)+A^{2}\upsilon _{c}^{4}\mathcal{B}^{\frac{2}{3}}\mu ^{\frac{4}{3}}-16 Q^{2}\mathcal{B}^{\frac{4}{3}}\left( 3\mu ^{\frac{2}{3}}\left( 3+2\beta ^{2}\right) +\upsilon
_{c}^{2}A^{2}\mathcal{B}^{\frac{2}{3}}-\frac{6A^{2}Q^{2}\mathcal{B}^{2}}{\mu
^{\frac{4}{3}}}\right) }{24\pi \upsilon_{c}^{4}\mathcal{B}^{\frac{2}{3}}\mu ^{\frac{4}{3}}}.
\label{Eq24}
\end{eqnarray}%

In tables \ref{tab2} and \ref{tab3}, we show that how critical
quantities and universal critical ratio ($\frac{P_{c}\upsilon
_{c}}{T_{c}}$) change under variation of black hole parameters. In order to show the effects of electric charge and string
tension on the critical values of phase transition, we have
plotted Fig. \ref{Fig9}, and presented tables \ref{tab2}, and
\ref{tab3}. Fig. \ref{Fig9}, and table \ref{tab2}, indicate that
as $Q$ increases, the critical pressure and temperature decrease,
whereas the critical volume and the universal critical ratio
increase. As for the effects of string tension, one can see that
the critical volume is a decreasing function of this parameter,
whereas the critical pressure, temperature and universal critical
ratio are an increasing function of this parameter (see Fig.
\ref{Fig9}, and table \ref{tab3}).

\begin{table}[tbp]
\caption{Critical values for $\mathcal{B}=0.2 $, $\protect\beta=0.04 $, $%
A=0.02 $ and $\protect\mu=0.15 $.}
\label{tab2}
\begin{center}
\begin{tabular}{||c|c|c|c|c||}
\hline\hline
$Q$ & $\upsilon_{c}$ & $T_{c}$ & $P_{c}$ & $\frac{P_{c}\upsilon_{c}}{T_{c}}$
\\ \hline\hline
$0.16$ & $0.9511$ & $0.2041$ & $0.0728$ & $0.339601$ \\ \hline
$0.18$ & $1.0700$ & $0.1814$ & $0.0575$ & $0.339608$ \\ \hline
$0.20$ & $1.1889$ & $0.1633$ & $0.0466$ & $0.339616$ \\ \hline
$0.22$ & $1.3078$ & $0.1484$ & $0.0385$ & $0.339625$ \\ \hline\hline
\end{tabular}%
\end{center}
\end{table}

\begin{table}[tbp]
\caption{Critical values for $\mathcal{B}=0.2 $, $\protect\beta=0.04 $, $%
A=0.02 $ and $Q=0.2$.}
\label{tab3}
\begin{center}
\begin{tabular}{||c|c|c|c|c||}
\hline\hline
$\mu $ & $\upsilon_{c}$ & $T_{c}$ & $P_{c}$ & $\frac{P_{c}\upsilon_{c}}{T_{c}%
}$ \\ \hline\hline
$0.14$ & $1.2448$ & $0.1524$ & $0.0406$ & $0.3319$ \\ \hline
$0.15$ & $1.1889$ & $0.1633$ & $0.0466$ & $0.3396$ \\ \hline
$0.16$ & $1.1388$ & $0.1742$ & $0.0530$ & $0.3469$ \\ \hline
$0.17$ & $1.0937$ & $0.1851$ & $0.0599$ & $0.3540$ \\ \hline\hline
\end{tabular}%
\end{center}
\end{table}

\subsection{Critical points of the charged accelerating black holes via new
prescription}

As it was mentioned, one cannot obtain the critical values
analytically via the usual method due to the complexity of Eq.
(\ref{EqPv7}). Since this method is not practical for black holes
with the non-spherical horizon in most gravitational theories
\cite{HendiarXiv}, we employ an alternative approach for obtaining
the critical values. This method is based on the denominator of
heat capacity \cite{54}. Using the analogy between pressure and
the cosmological constant and solving denominator with respect to
pressure, one can obtain a new relation for pressure. It is
worthwhile to mention that this new pressure is not the same
pressure that was obtained in Eq. (\ref{EqPv7}). Substituting Eqs. (\ref%
{Eq17}) and (\ref{Eq21}) in Eq. (\ref{Eq14}), the new pressure is obtained
as follows
\begin{equation}
P_{new}=-\frac{\chi_{2}}{3}+\left( -\frac{w}{2}+\frac{\sqrt{12\varrho^{3}+81w^{2}}%
}{18}\right) ^{\frac{1}{3}}+\left( -\frac{w}{2}-\frac{\sqrt{12\varrho^{3}+81w^{2}}%
}{18}\right) ^{\frac{1}{3}},  \label{Eq27}
\end{equation}%
where
\begin{eqnarray*}
\varrho &=&-\frac{\chi_{2}^{2}}{3}-\frac{\mu ^{\frac{2}{3}}A^{2}\left( 1+\frac{%
\upsilon ^{2}A^{2}\mathcal{B}^{\frac{2}{3}}}{2\mu ^{\frac{2}{3}}}\right)
^{-1}}{16\pi ^{2}\upsilon ^{2}\mathcal{B}^{\frac{2}{3}}}, \\
&& \\
w &=&\frac{2\chi_{2}^{3}}{27}+\frac{\mu ^{\frac{2}{3}}\chi_{2} A^{2}\left(
1+\frac{\upsilon ^{2}A^{2}\mathcal{B}^{\frac{2}{3}}}{2\mu ^{\frac{2}{3}}}%
\right) ^{-1}}{48\pi ^{2}\upsilon ^{2}\mathcal{B}^{\frac{2}{3}}}-\frac{3\mu
^{\frac{4}{3}}A^{2}\left( 1+\frac{4Q^{2}\mathcal{B}^{\frac{4}{3}}}{\upsilon
^{2}\mu ^{\frac{4}{3}}}\right) }{32\pi ^{3}\upsilon ^{4}\mathcal{B}^{\frac{4%
}{3}}\left( 1+\frac{\upsilon ^{2}A^{2}\mathcal{B}^{\frac{2}{3}}}{2\mu ^{%
\frac{2}{3}}}\right) }, \\
&& \\
\chi_{2} &=&\frac{-7A^{2}\left( 1-\frac{144Q^{2}\mathcal{B}^{\frac{2}{3}}}{%
7\mu ^{\frac{2}{3}}\upsilon ^{4}A^{2}}\right) \mathcal{B}^{\frac{2}{3}}-%
\frac{4\mu ^{\frac{2}{3}}}{\upsilon ^{2}}\left( \left( 3+\beta ^{2}\right) -%
\frac{2A^{2}Q^{2}\mathcal{B}^{2}}{\mu ^{2}}\right) }{24\pi \left( 1+\frac{%
\upsilon ^{2}A^{2}\mathcal{B}^{\frac{2}{3}}}{2\mu ^{\frac{2}{3}}}\right)
\mathcal{B}^{\frac{2}{3}}}.
\end{eqnarray*}

\begin{figure*}[tbh]
\centering
\includegraphics[width=0.28\linewidth]{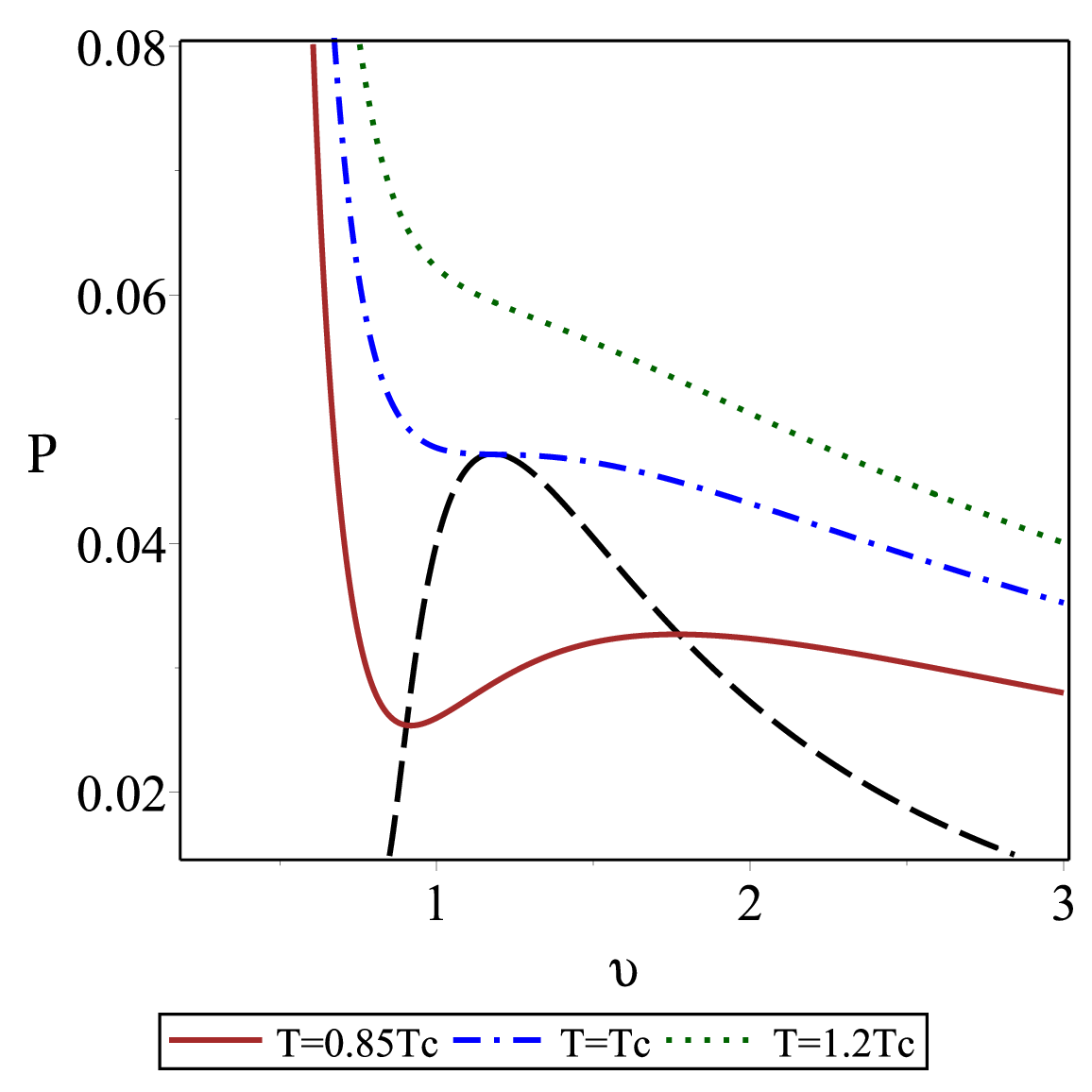}\hfil
\includegraphics[width=0.28\linewidth]{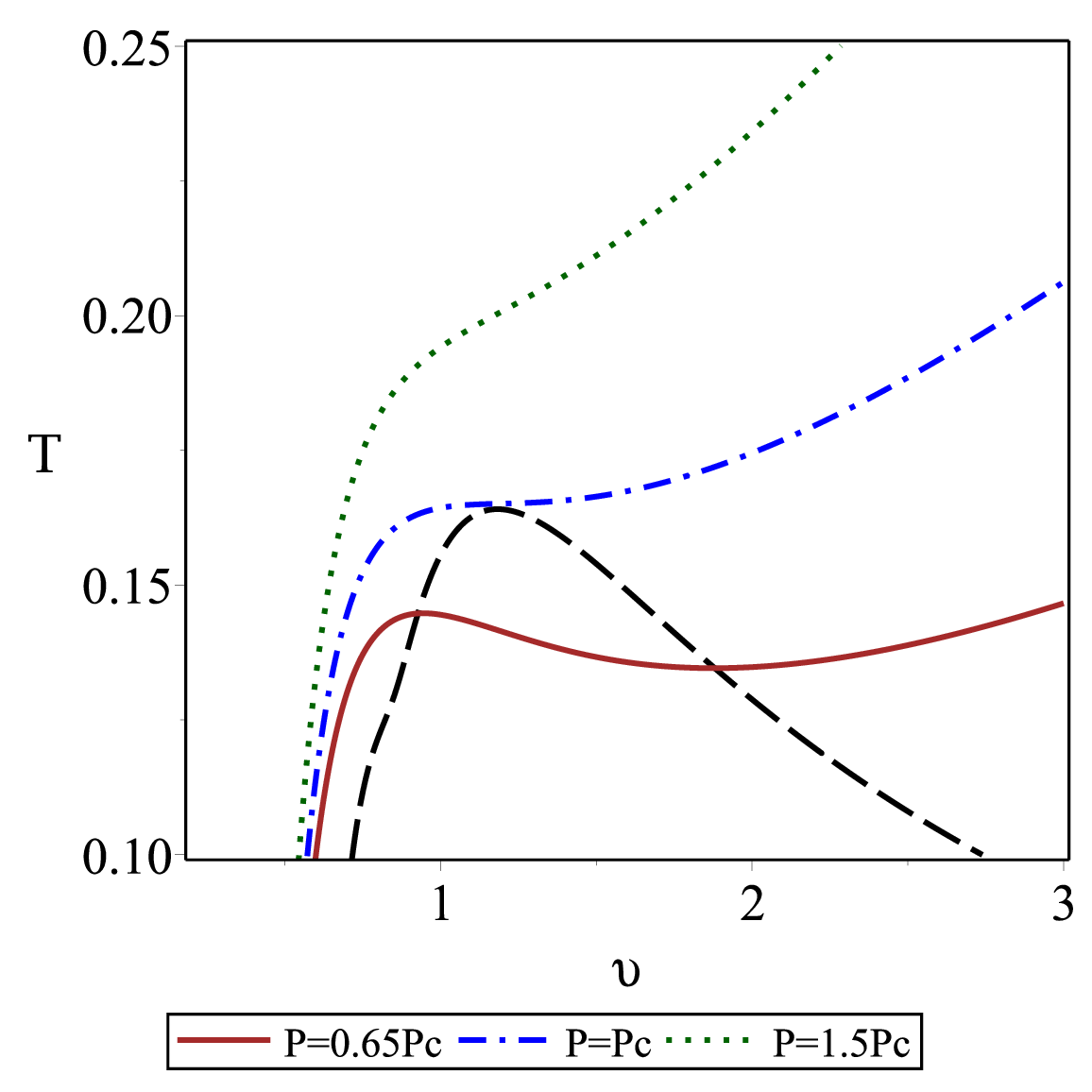}\hfil
\includegraphics[width=0.28\linewidth]{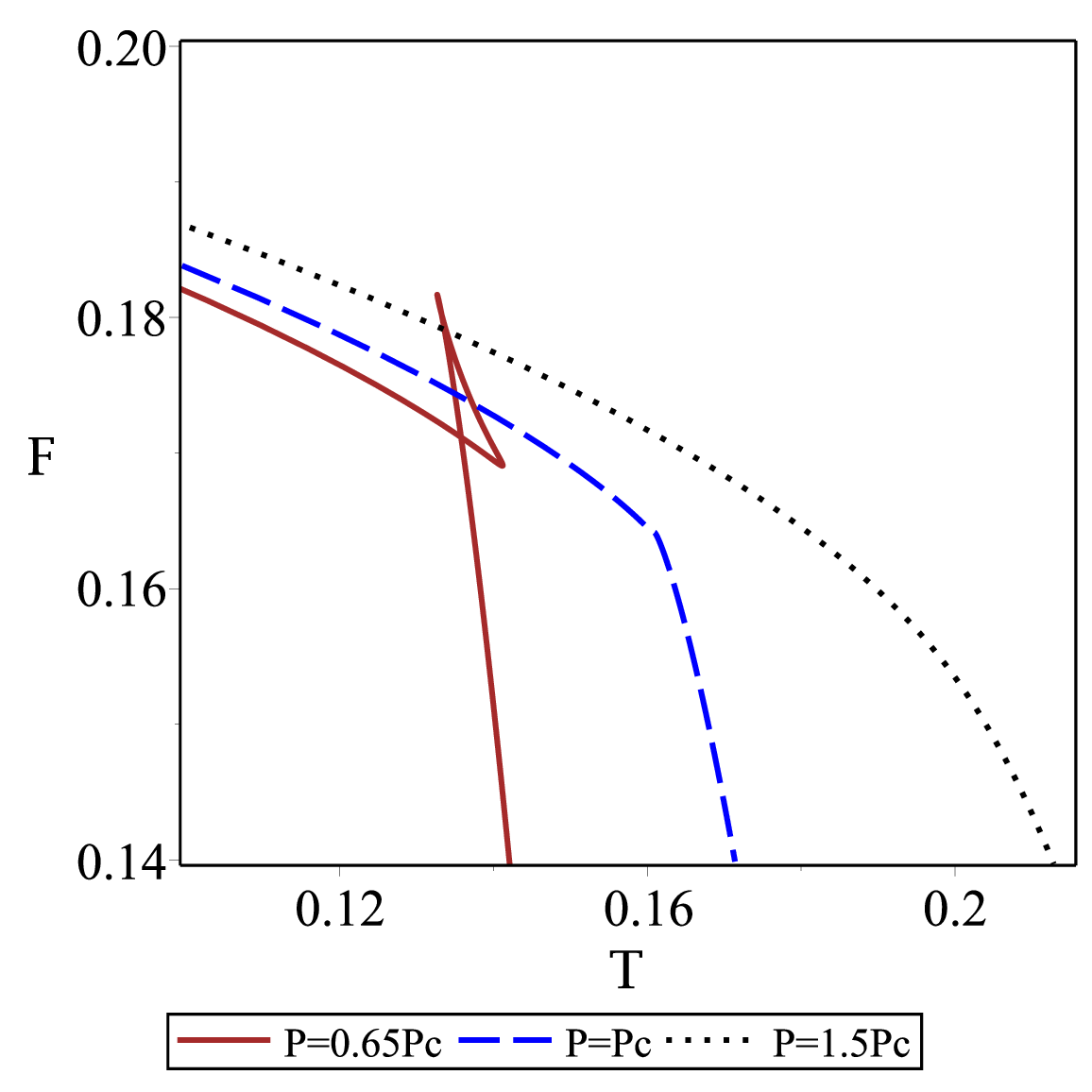}
\caption{van der Waals like phase diagrams for $\mathcal{B}=0.2$, $\protect%
\beta =0.04$, $A=0.02$, $\protect\mu =0.15$ and $Q=0.2$. Left
panel: $P - \upsilon$ diagram for $ T=0.14 $ (continuous line), $
T=0.164 $ (dash-dotted line), $ T=0.2 $ (dotted line)  and $
P_{new}$ (dashed line). Middle panel: $T - \upsilon$ diagram for $
P=0.03 $ (continuous line), $ P=0.048 $ (dash-dotted line), $
P=0.07 $ (dotted line)  and $ T_{new}$ (dashed line). Right panel:
$F - T$ diagram for $ P=0.03 $ (continuous line), $ P=0.048 $
(dash-dotted line) and $ P=0.07 $ (dotted line).} \label{Fig8}
\end{figure*}

\begin{figure*}[tbh]
\centering
\includegraphics[width=0.32\linewidth]{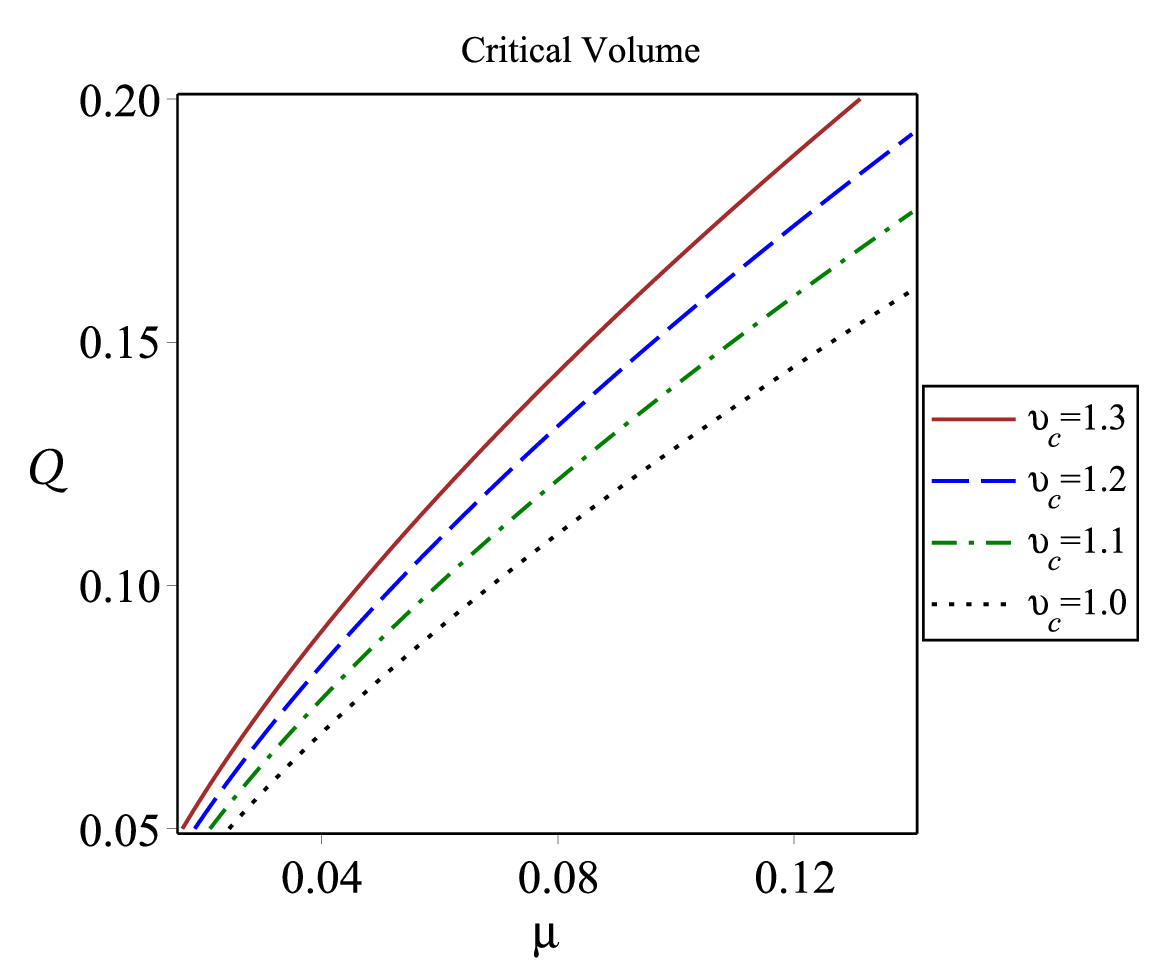}\hfil
\includegraphics[width=0.32\linewidth]{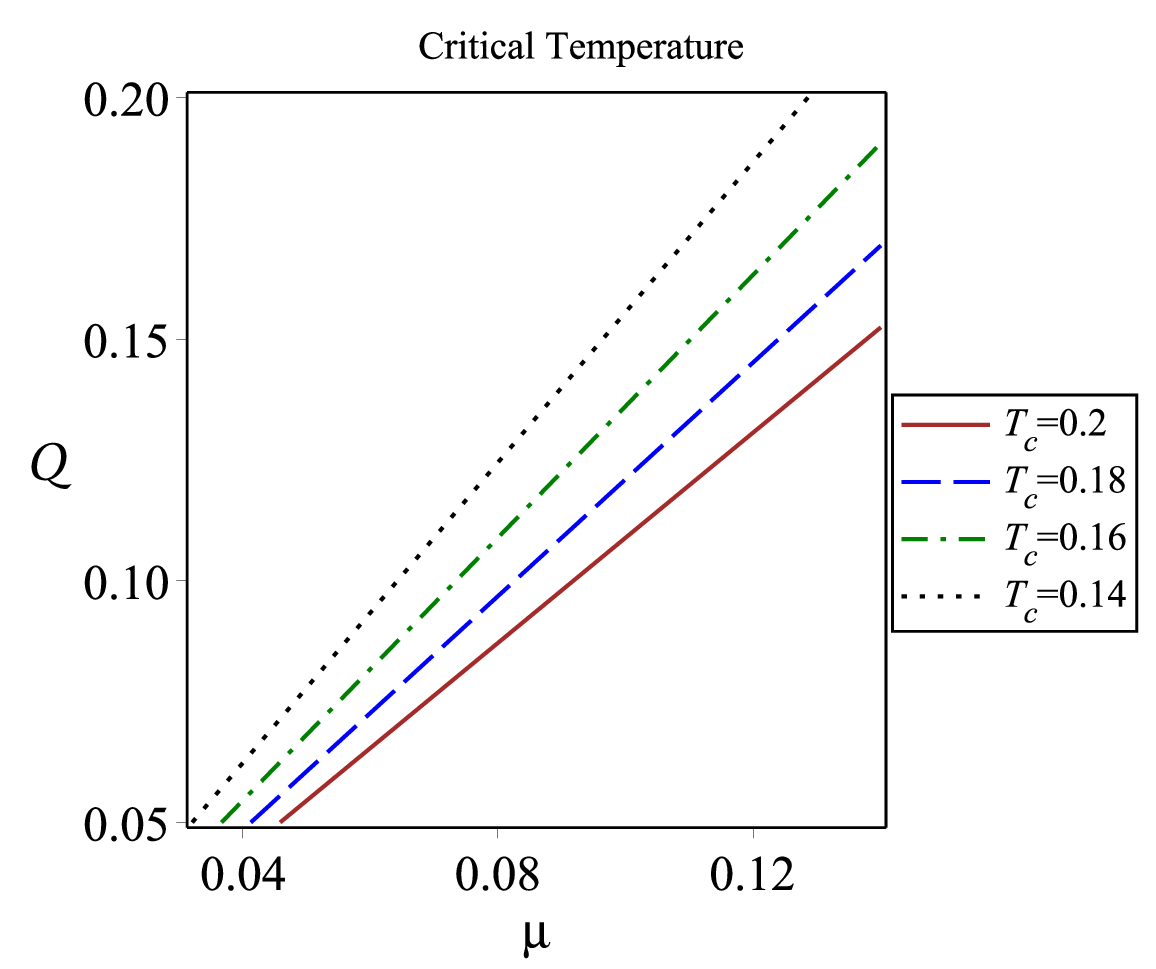}\hfil
\includegraphics[width=0.32\linewidth]{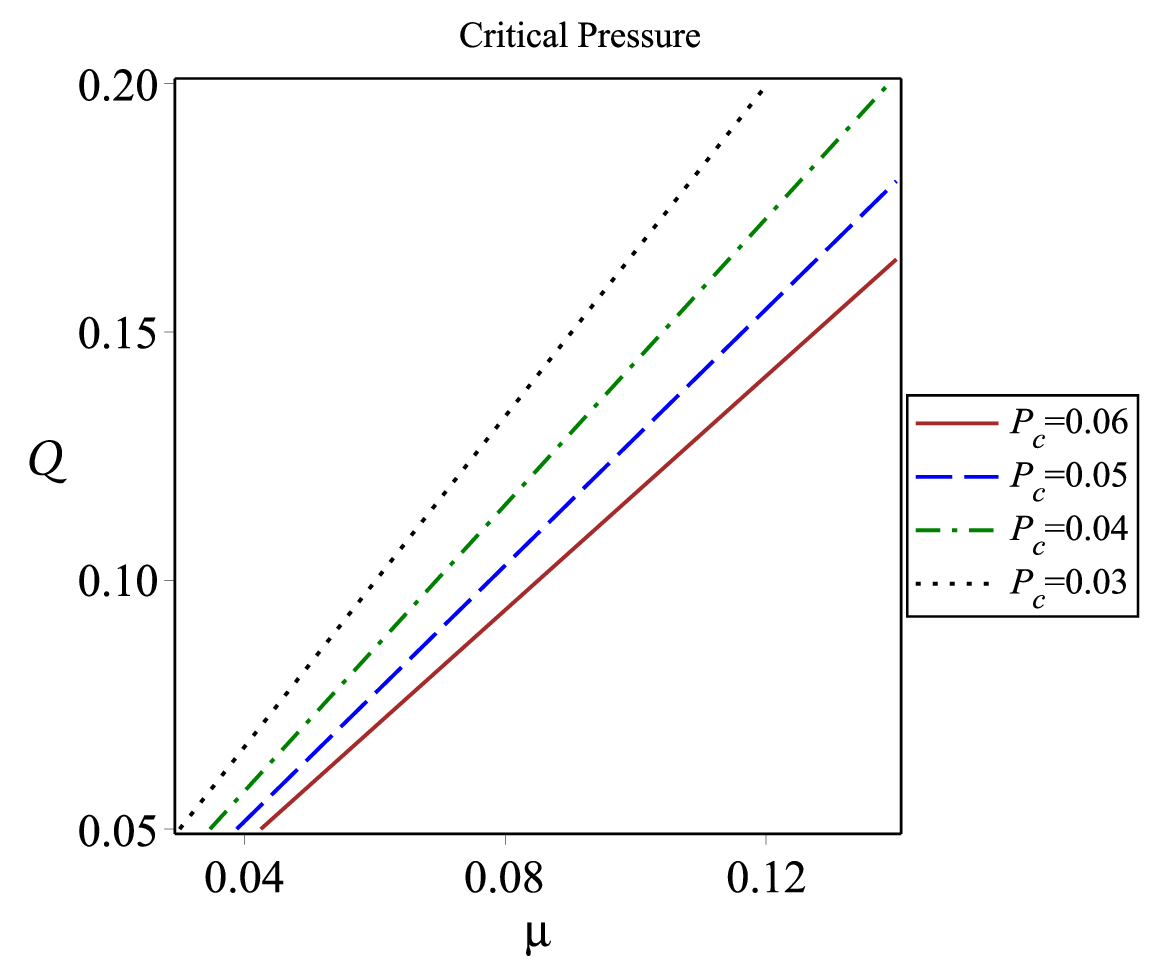}
\caption{Variation of the critical values as functions of black hole
parameters for $\mathcal{B}=0.2$, $\protect\beta =0.04$ and $A=0.02$.}
\label{Fig9}
\end{figure*}

Replacing Eqs. (\ref{Eq21}) and (\ref{Eq27}) in Eq. (\ref{Eq13}), one can
obtain a new relation for the temperature which is independent of pressure.
The new pressure and temperature have a maximum point that exactly coincides with the inflection point of $P-\upsilon $ and $T-\upsilon $ diagram. In other words, these maximum pressure and temperature are the same critical pressure and temperature and their proportional volume is critical volume (see dashed lines in left and middle panels of Fig. \ref{Fig8}).

\begin{figure*}[tbh]
\centering
\includegraphics[width=0.28\linewidth]{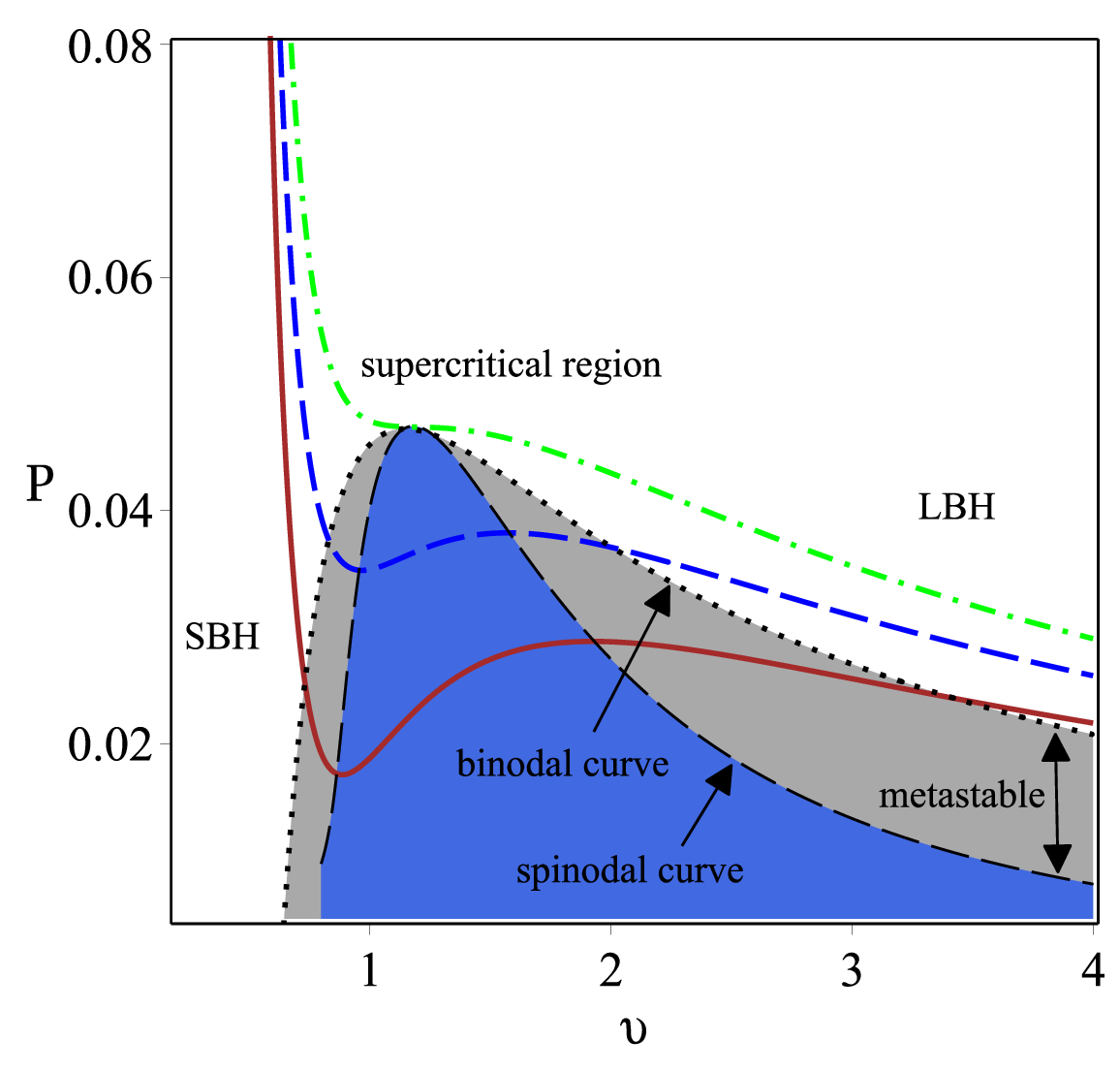}\hfil
\includegraphics[width=0.28\linewidth]{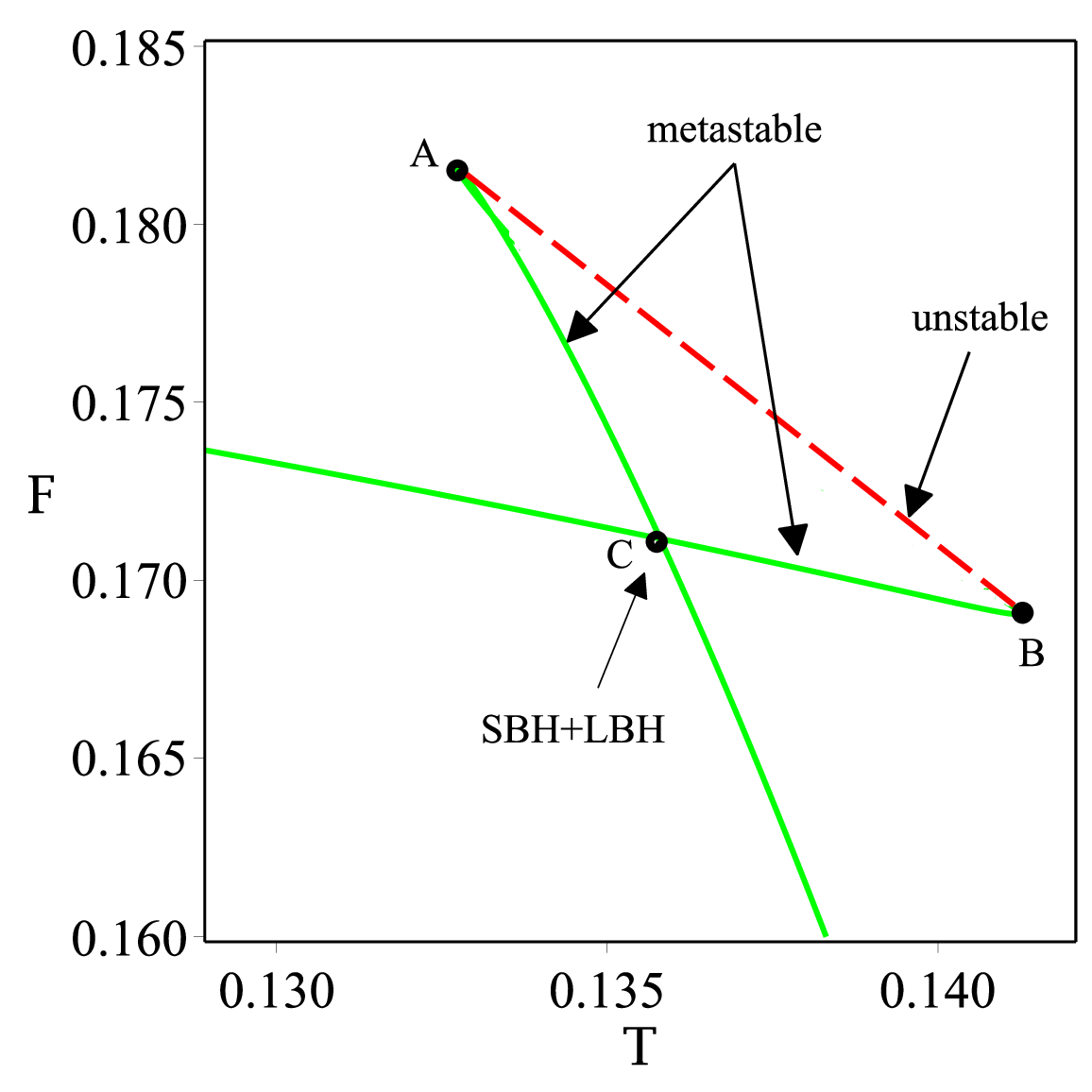}\hfil
\includegraphics[width=0.28\linewidth]{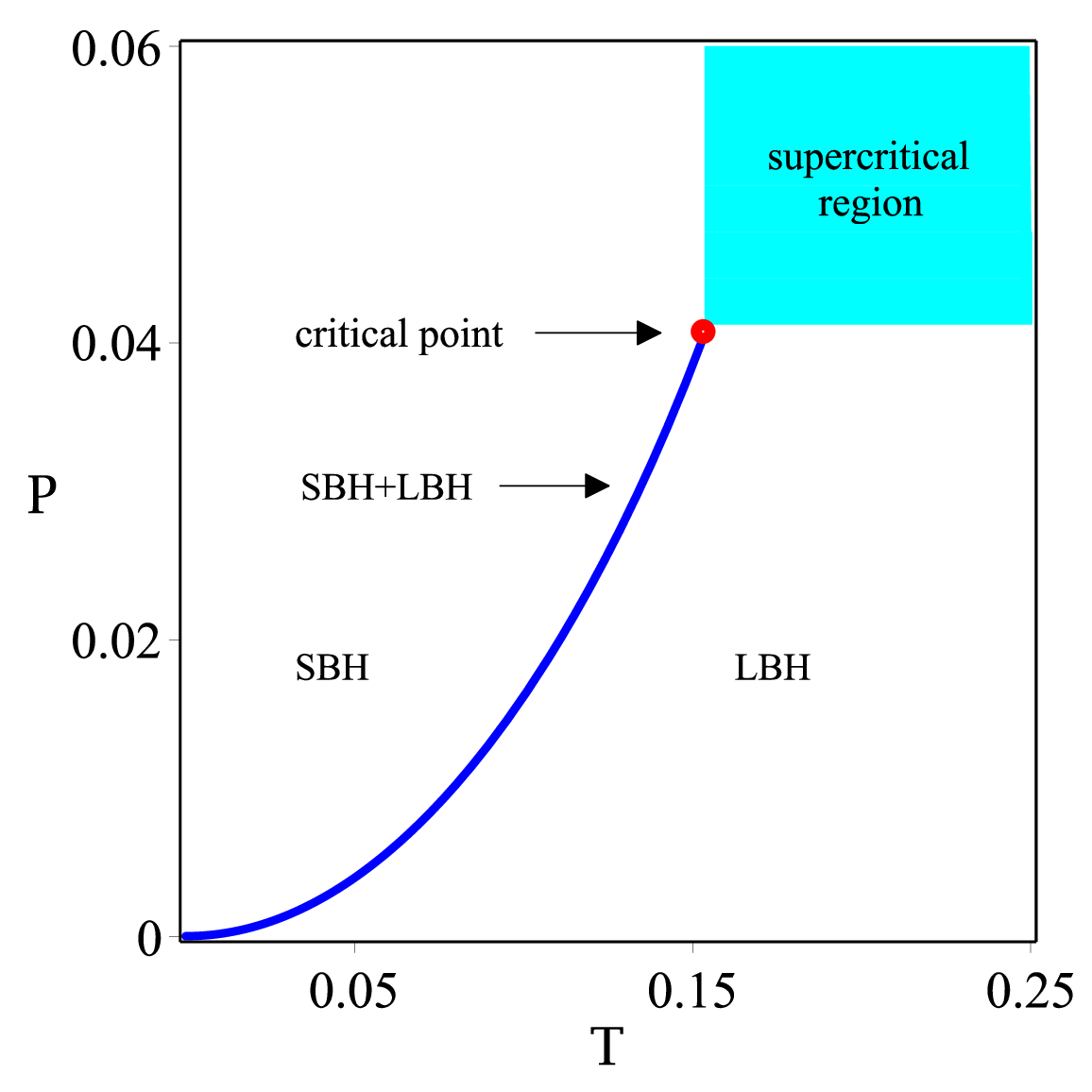}
\caption{$ P-\upsilon $, $ F-T $ and $ P-T $ diagram for $\mathcal{B}=0.2$, $\protect\beta =0.04$, $A=0.02$, $\protect\mu =0.15$ and $Q=0.2$. }
\label{Figbs}
\end{figure*}

It should be noted that  the black dashed curve in Fig.
\ref{Fig8} is not representing the boundary between small black hole and large black hole.
This curve is the same spinodal curve. In fact the first-order
phase boundary is detected by the binodal curve or coexistence
curve which has been illustrated in Fig. \ref{Figbs}. The region
between spinodal and binodal curves is related to the metastable
black holes which is equivalence to the positive heat capacity.
The region under the binodal curve indicates small black hole$+$large black hole which are thermodynamically unstable.

\begin{figure}[!htb]
\centering
\subfloat[]{
        \includegraphics[width=0.33\textwidth]{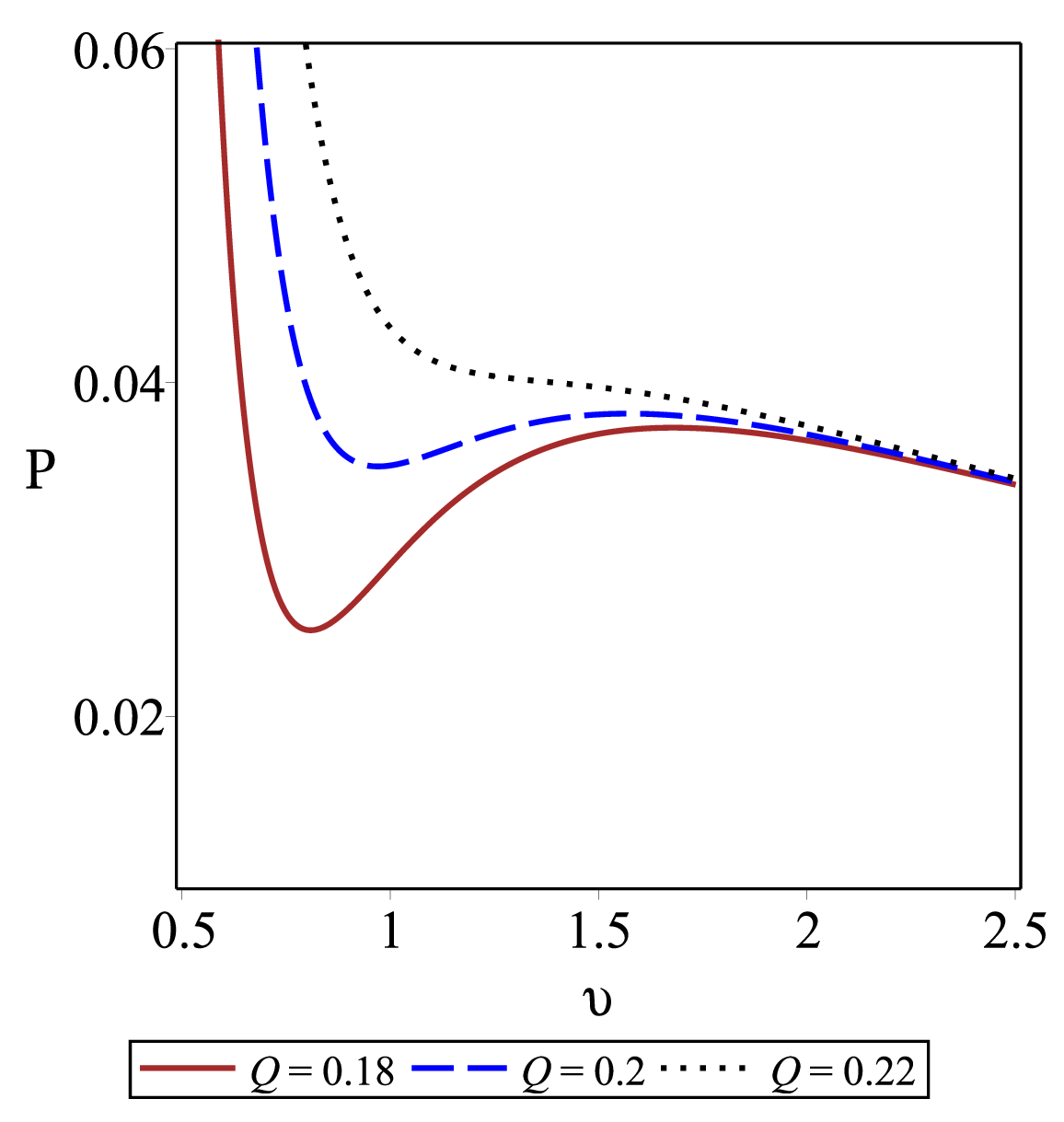}}
\subfloat[]{
        \includegraphics[width=0.33\textwidth]{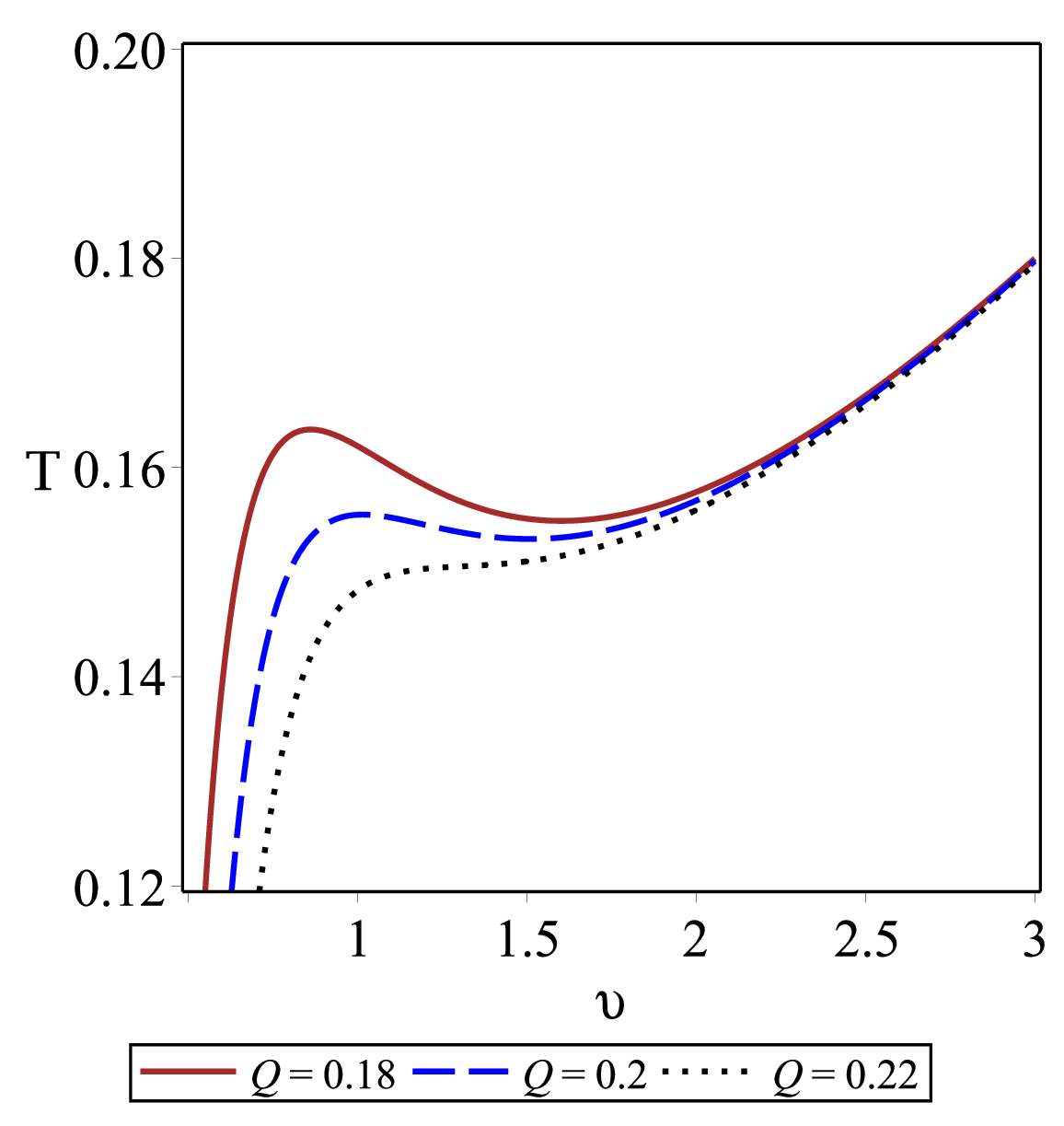}}
\subfloat[]{
        \includegraphics[width=0.35\textwidth]{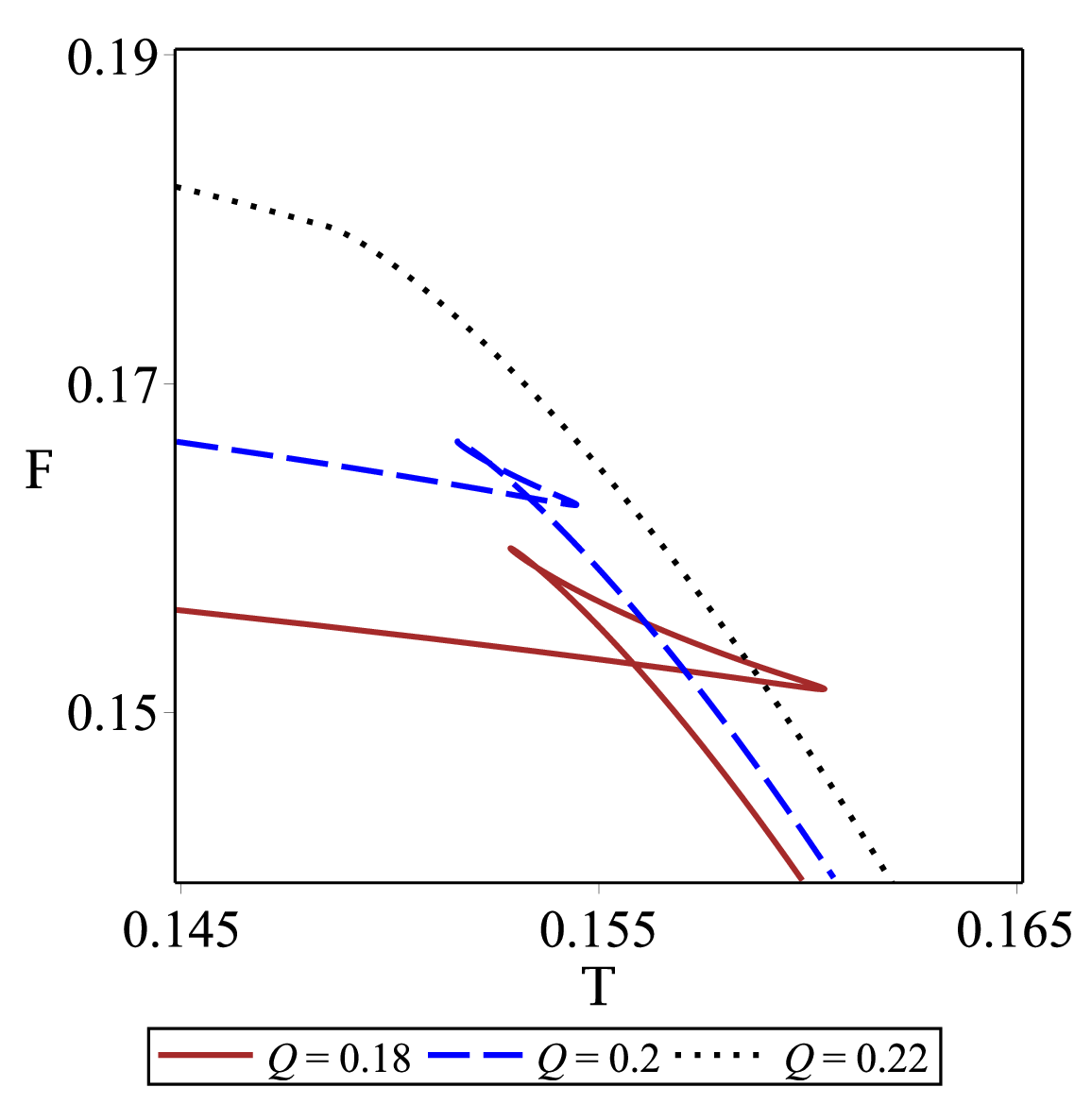}} \newline
\subfloat[]{
        \includegraphics[width=0.33\textwidth]{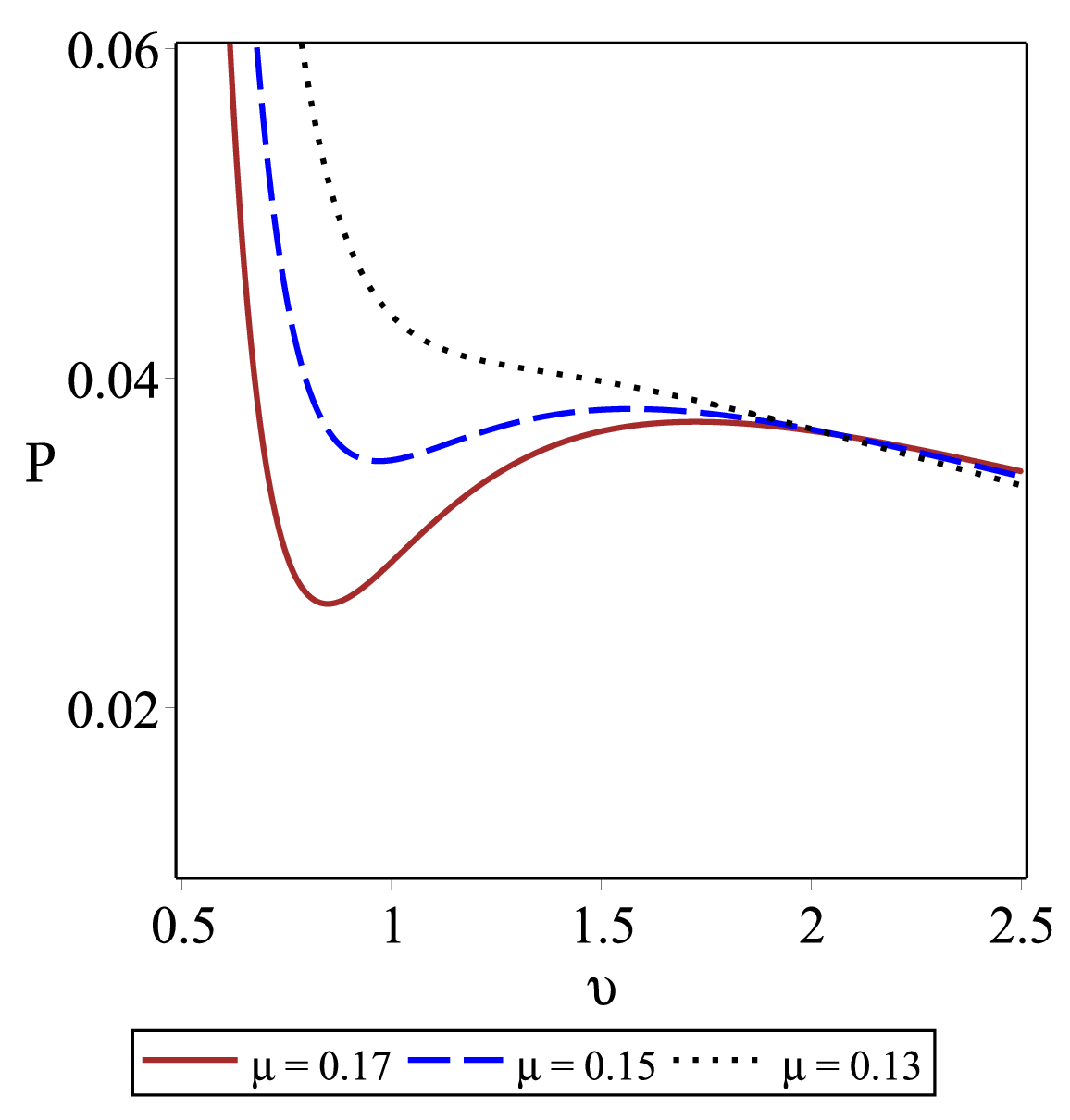}}
\subfloat[]{
        \includegraphics[width=0.33\textwidth]{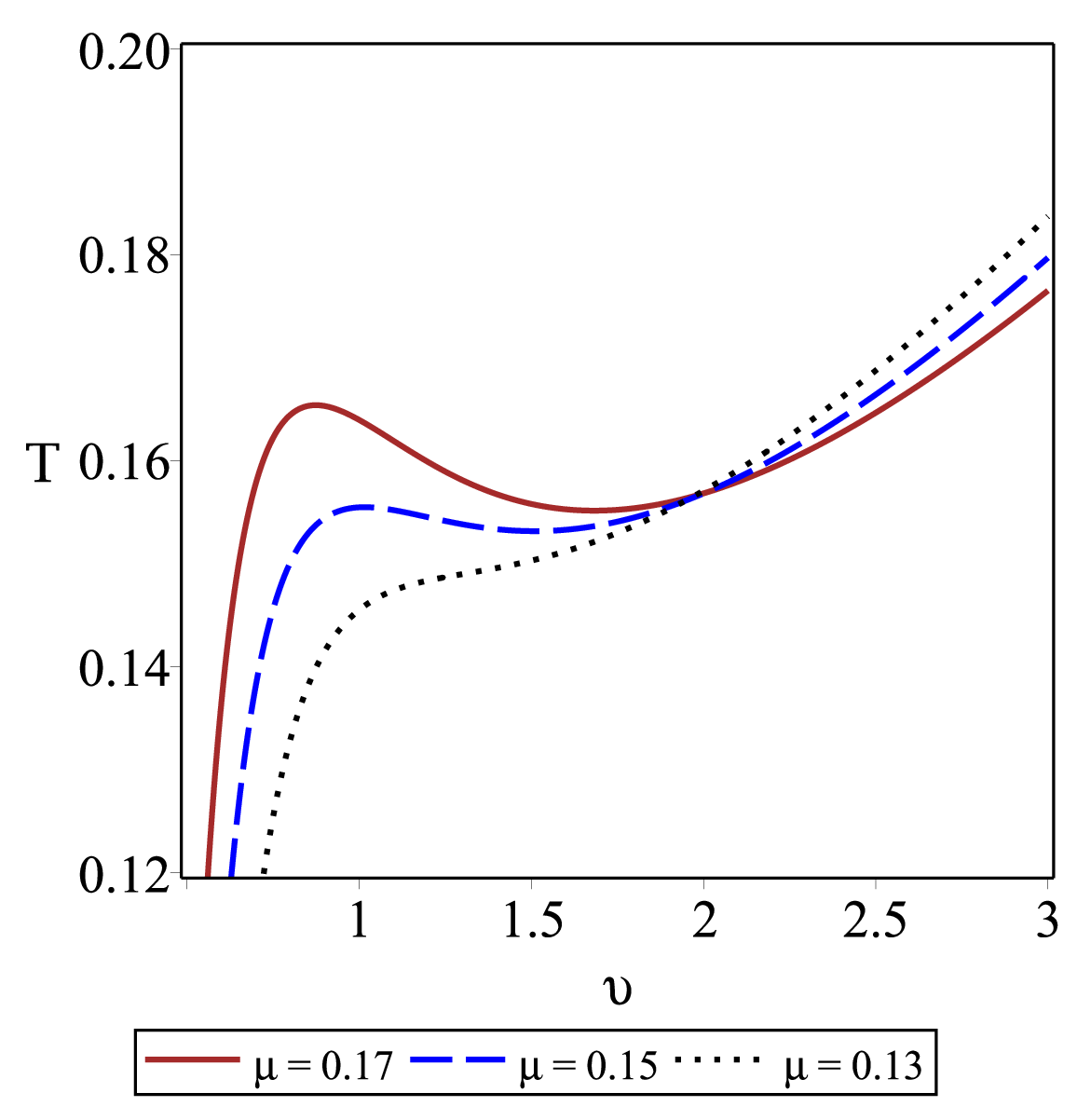}}
\subfloat[]{
        \includegraphics[width=0.33\textwidth]{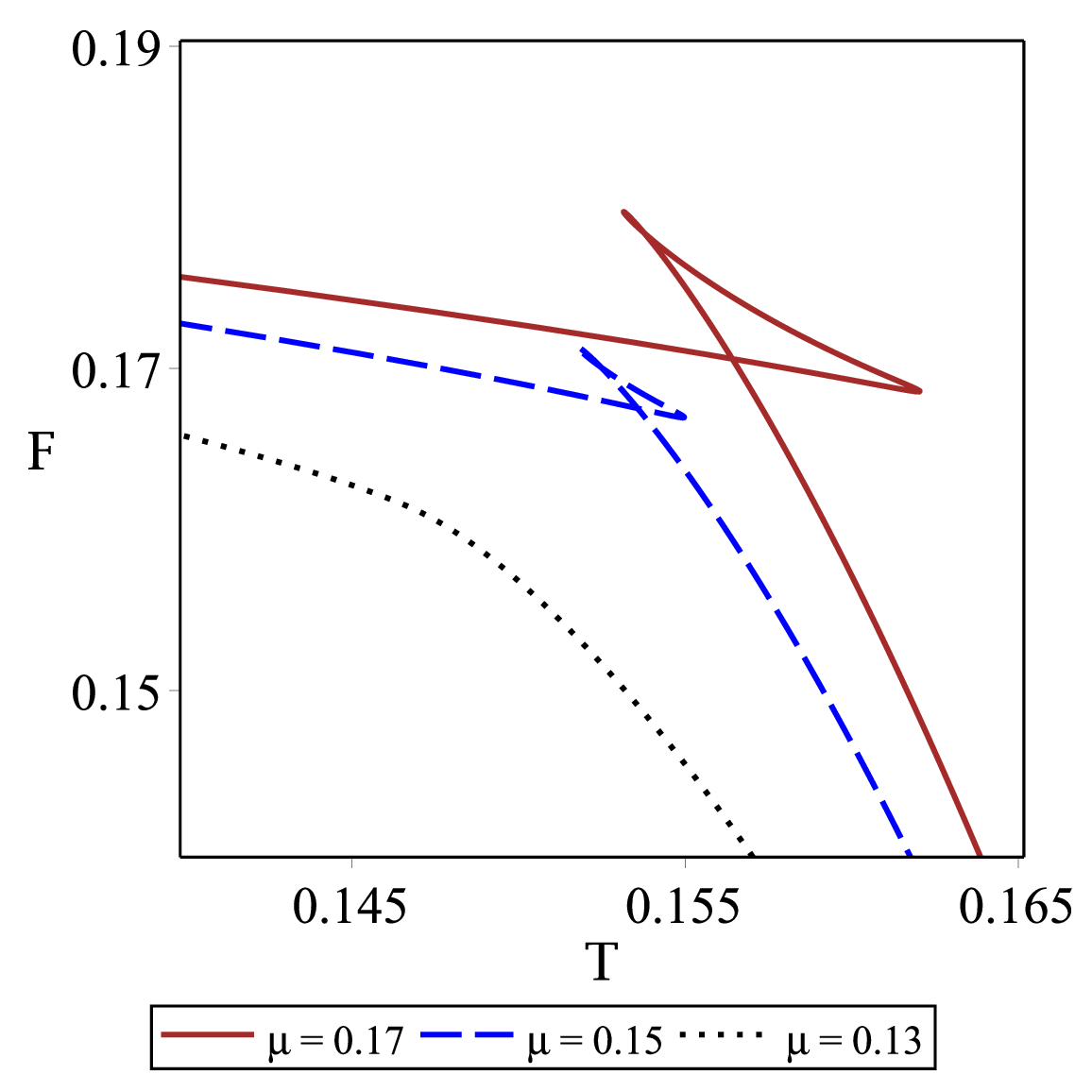}} \newline
\caption{For $\mathcal{B}=0.2$, $\protect\beta =0.04$ and $A=0.02$. $%
T=0.9T_{c}$ for $P-\protect\upsilon $ diagram and $P=0.85P_{c}$ for $T-%
\protect\upsilon $ and $F-T$ diagrams. Up panels: for $\protect\mu =0.15$
and different values of the electric charge. Down panels: for $Q=0.2$ and
different values of the string tension.}
\label{Fig10}
\end{figure}

As it was mentioned, for pressures and temperatures smaller than
the critical pressure and temperature, one can observe a first order
small/large black hole phase transition. By looking at $P-\upsilon
$ and $T-\upsilon $ diagrams in Fig. \ref{Fig10} (up panels), one
can find that for large electric charges, pressure (temperature)
is only a decreasing (an increasing) function of the volume
without any extremum. For $F-T$ diagram, evidently, free energy is
a decreasing function of the temperature and swallow-tail shape
does not appear. In this case, just a single stable phase exists
for black holes. For a specific value of $Q$, one can observe an
extremum that separates two small and large stable phases from
each other. As for small values of $Q$, two extrema are formed in
$P-\upsilon $ and $T-\upsilon $ diagrams. In this case, three
phases exist which are small, medium and large black holes. The
region between two extrema is related to medium black holes which
are unstable. Whereas, the regions before the first extremum and
after the second extremum in $P-\upsilon $ and $T-\upsilon $
diagrams are related to small and large phases respectively. Also,
$P-\upsilon $ diagram shows that as $Q$ decreases, the pressure
related to phase transition points decreases. Since, pressure is
related to the cosmological constant which is related to
asymptotical curvature of the background, one can say that as $Q$
decreases the necessity of having a background with higher
curvature decreases. Up middle and right panels of Fig.
\ref{Fig10}, show that as $Q$ decreases, the temperature related
to phase transition points increases and the difference between
free energy of different phases grows larger. So, rapidly
accelerating black holes need to absorb more mass in order to have
phase transition. In other words, by decreasing the electric
charge, the system achieves a stable state barely.

Taking a closer look at down panels of Fig. \ref{Fig10}, one can
find that the effect of string tension is the opposite of that of
the electric charge. Also, investigating the variation of string
tension on the phase structure of the system shows that as $\mu $
increases, the distance between two extrema increases. This shows
that a stable small or large charged accelerating black hole goes
to an unstable phase by increasing this parameter. Since string
tension is linearly related to the conical deficit, the larger the
tension, the steeper the cone (see Eq. \ref{Eq6}). So, it is
expected that a small/large charged accelerating black hole exits
in its stable state if it is pulled by more powerful string
tension.
\subsection{Geometrical thermodynamics in the extended phase space}

Here, we employ GTs to study phase transition in the
extended phase space. As it is known, the cosmological constant is related
to thermodynamical pressure in the extended phase space. Here, we consider
the pressure as an extensive parameter and explore its effects on different approaches of thermodynamical geometry. In this case, thermodynamical space transforms from $M(S,Q)$ to $M(S,Q,P)$. So, the Weinhold, Ruppeiner, Quevedo and HPEM metrics are modified as
\begin{equation}
ds^{2}=\left\{
\begin{array}{cc}
Mg_{ab}^{W}dX^{a}dX^{b}, & \text{Weinhold} \\
&  \\
-\frac{M}{T}g_{ab}^{R}dX^{a}dX^{b}, & \text{Ruppeiner} \\
&  \\
\left( SM_{S}+QM_{Q}+PM_{P}\right) \left(
-M_{SS}dS^{2}+M_{QQ}dQ^{2}+M_{PP}dP^{2}\right) , & \text{Quevedo I} \\
&  \\
SM_{S}\left( -M_{SS}dS^{2}+M_{QQ}dQ^{2}+M_{PP}dP^{2}\right) , & \text{%
Quevedo II} \\
&  \\
\frac{SM_{S}}{M_{QQ}^{3}M_{PP}^{3}}\left(
-M_{SS}dS^{2}+M_{QQ}dQ^{2}+M_{PP}dP^{2}\right) , & \text{HPEM}%
\end{array}%
~.\right.  \label{Eq33}
\end{equation}

By calculating the Ricci scalar of these metrics, one can obtain the
denominator of the Ricci scalar with the following forms
\begin{equation}
denom(\mathcal{R})=\left\{
\begin{array}{cc}
-2M^{3}\left(
M_{QP}^{2}M_{SS}+M_{SQ}^{2}M_{PP}+M_{SP}^{2}M_{QQ}-M_{SS}M_{QQ}M_{PP}-2M_{SQ}M_{SP}M_{QP}\right) ^{2},
& \text{Weinhold} \\
&  \\
-2M^{3}T^{3}\left(
M_{QP}^{2}M_{SS}+M_{SQ}^{2}M_{PP}+M_{SP}^{2}M_{QQ}-M_{SS}M_{QQ}M_{PP}-2M_{SQ}M_{SP}M_{QP}\right) ^{2},
& \text{Ruppeiner} \\
&  \\
2M_{SS}^{2}M_{QQ}^{2}M_{PP}^{2}\left( SM_{S}+QM_{Q}+PM_{P}\right) ^{3}, &
\text{Quevedo I} \\
&  \\
2S^{3}M_{SS}^{2}M_{QQ}^{2}M_{PP}^{2}M_{S}^{3}~, & \text{Quevedo II} \\
&  \\
S^{3}M_{S}^{3}M_{SS}^{2}, & \text{HPEM}%
\end{array}%
~.\right.  \label{Eq34}
\end{equation}

\begin{figure}[!htb]
\centering
\subfloat[]{
        \includegraphics[width=0.335\textwidth]{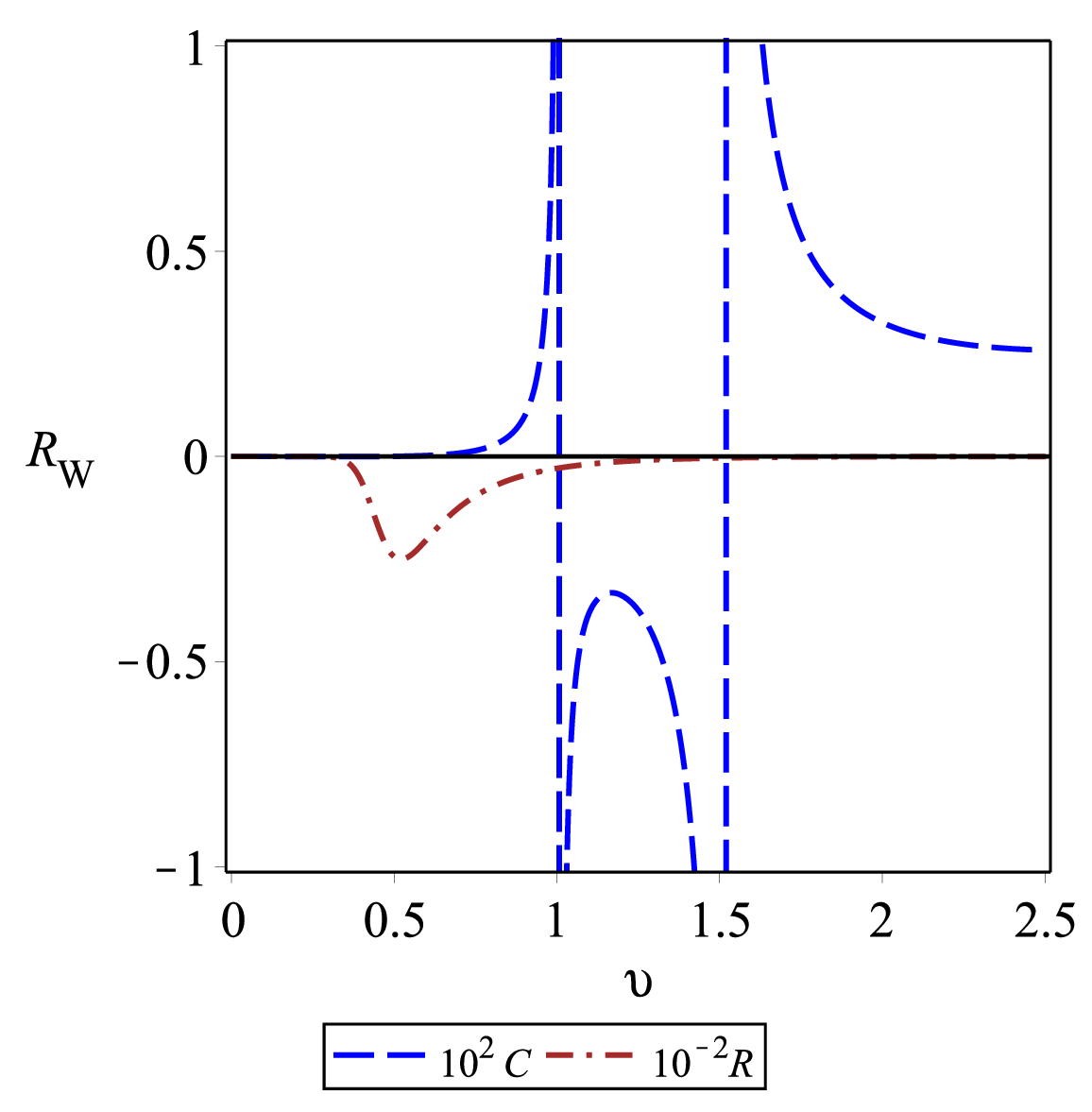}}
\subfloat[]{
        \includegraphics[width=0.33\textwidth]{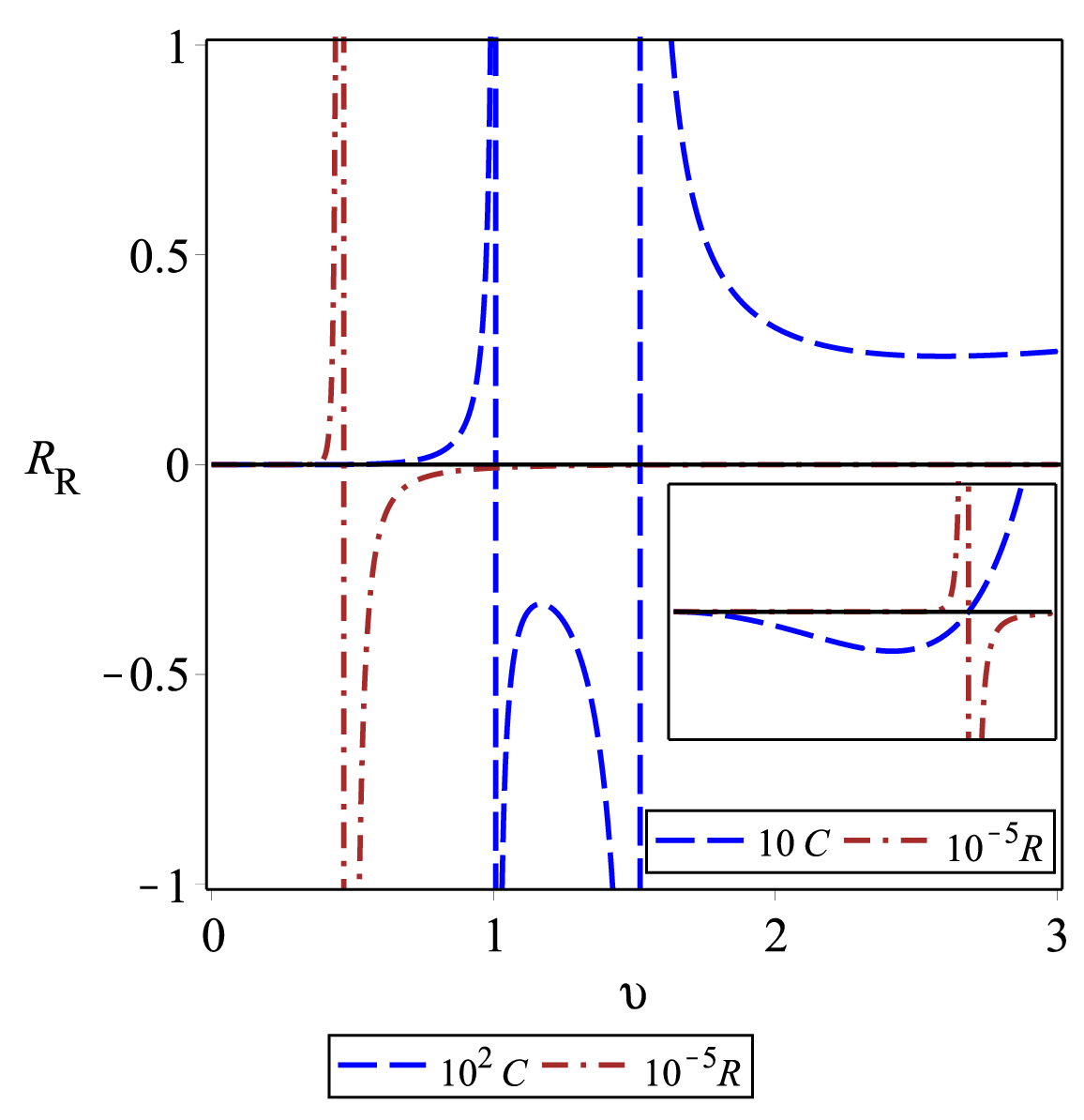}} \newline
\subfloat[]{
        \includegraphics[width=0.34\textwidth]{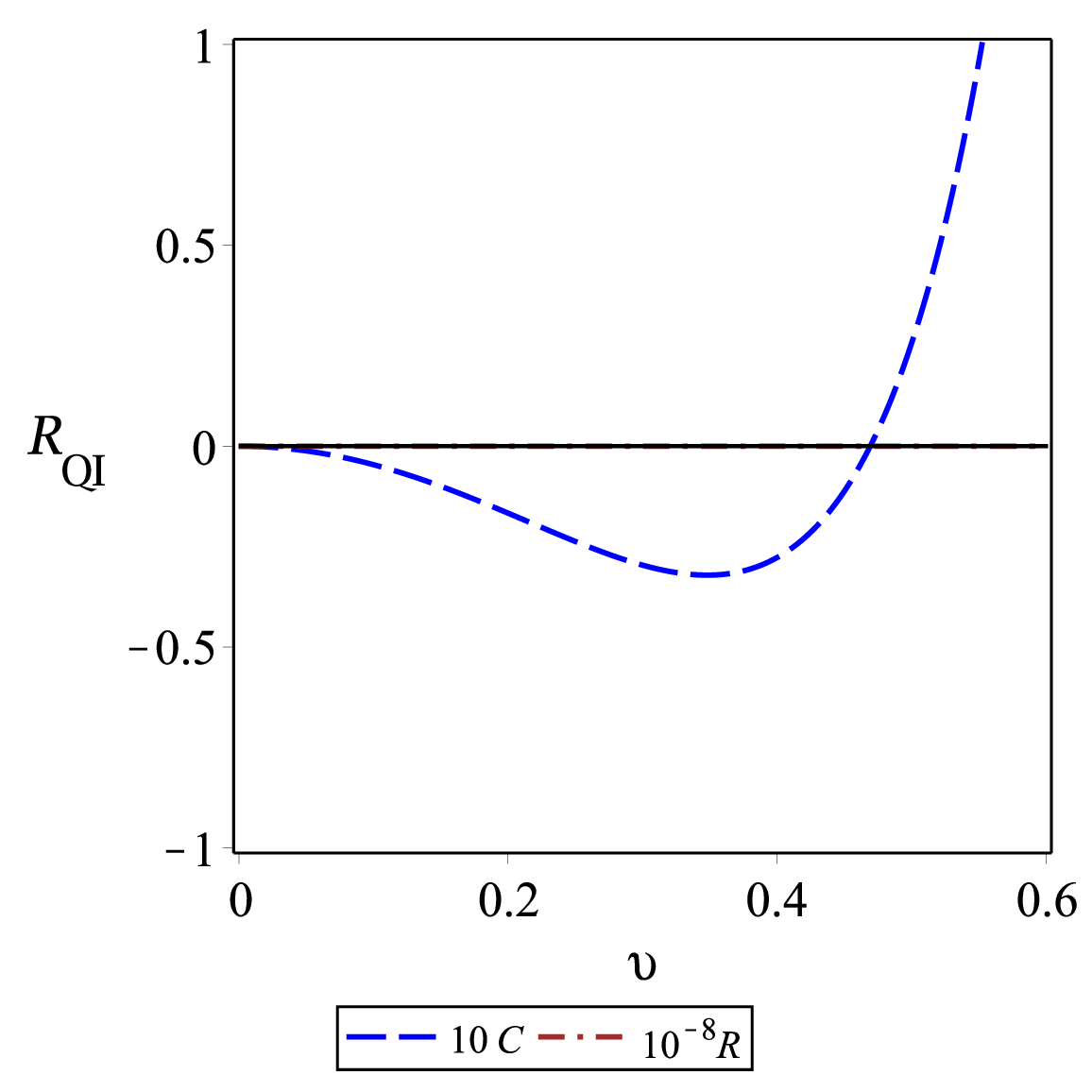}}
\subfloat[]{
        \includegraphics[width=0.33\textwidth]{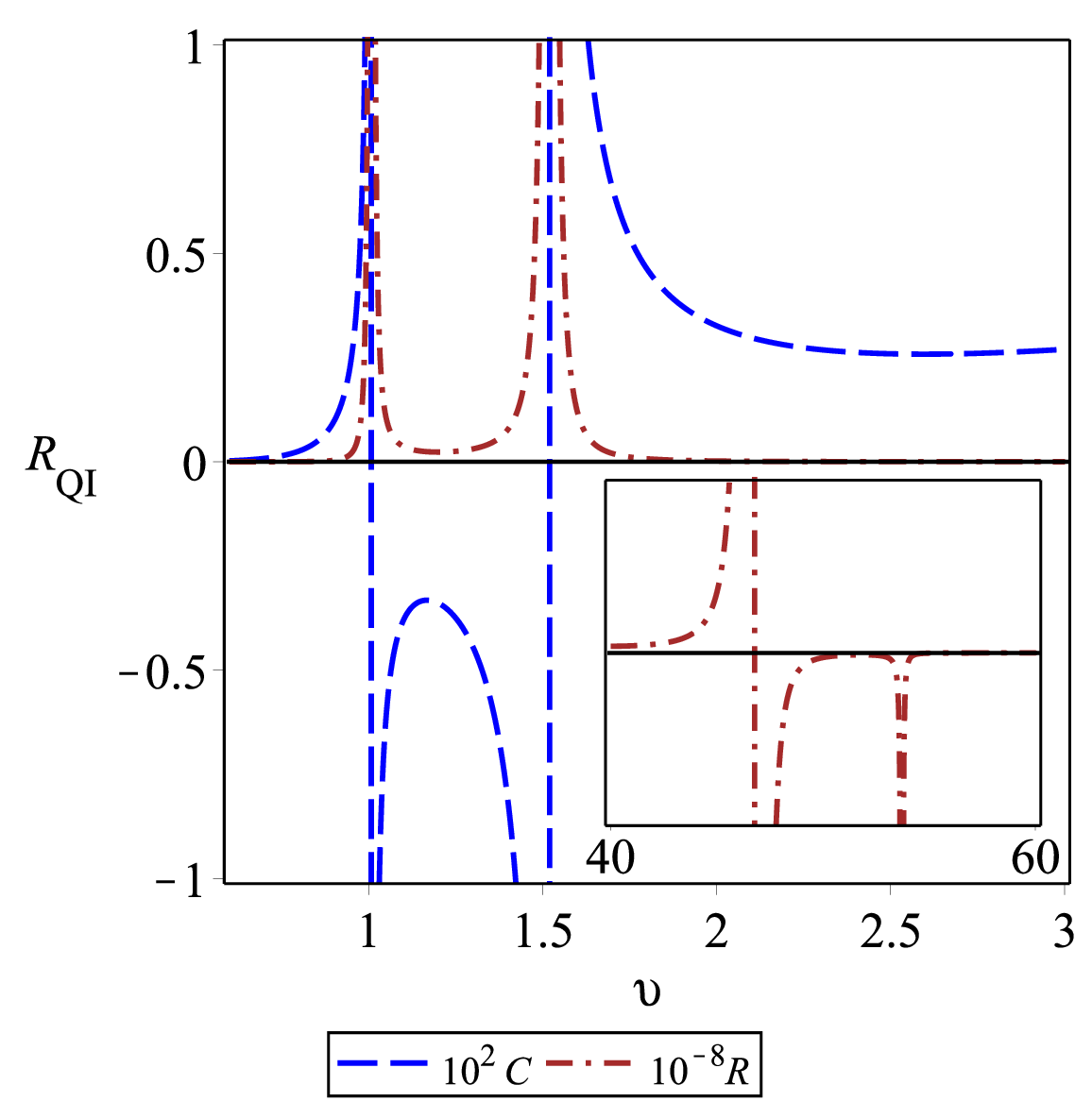}} \newline
\subfloat[]{
        \includegraphics[width=0.33\textwidth]{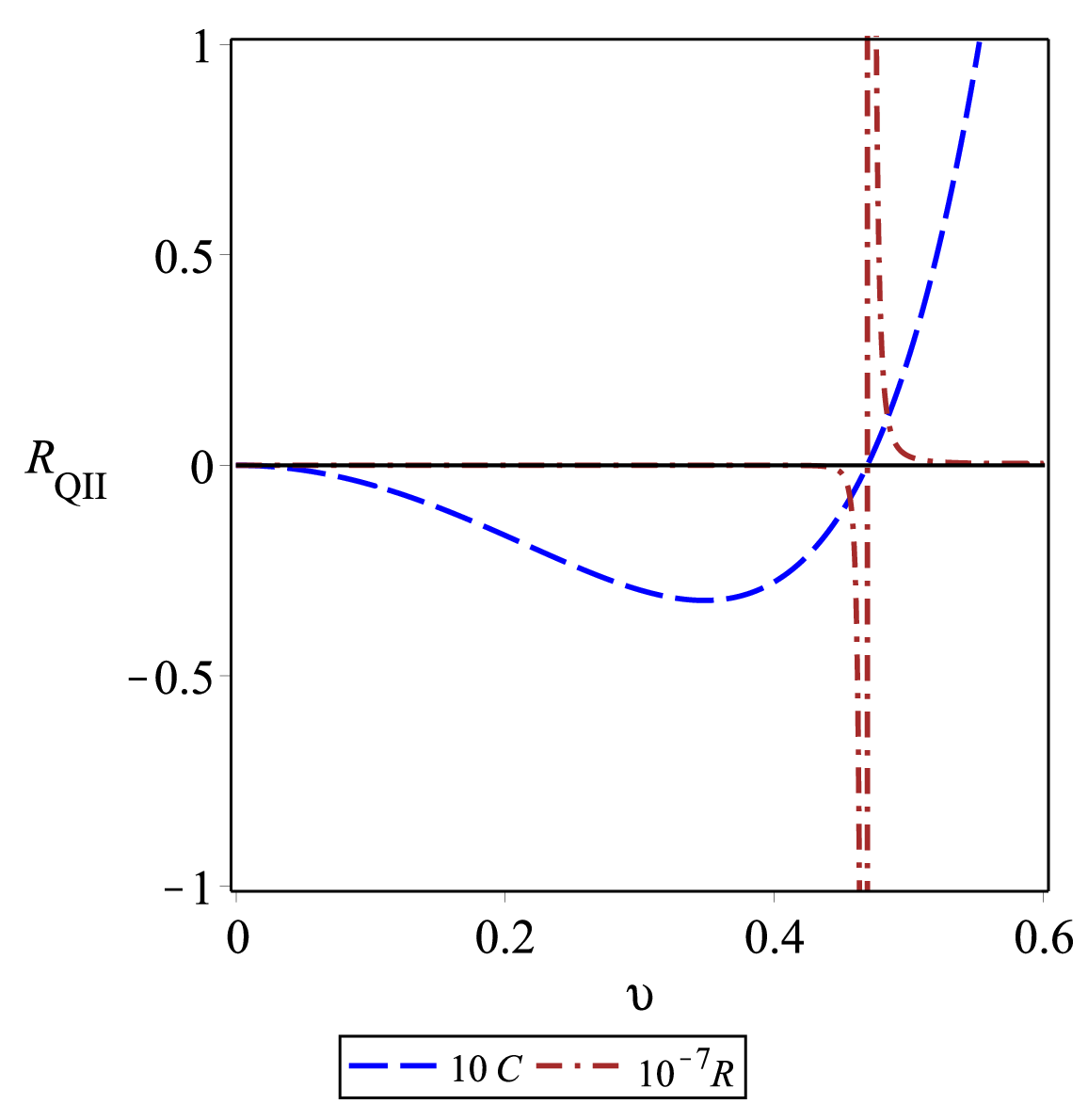}}
\subfloat[]{
        \includegraphics[width=0.33\textwidth]{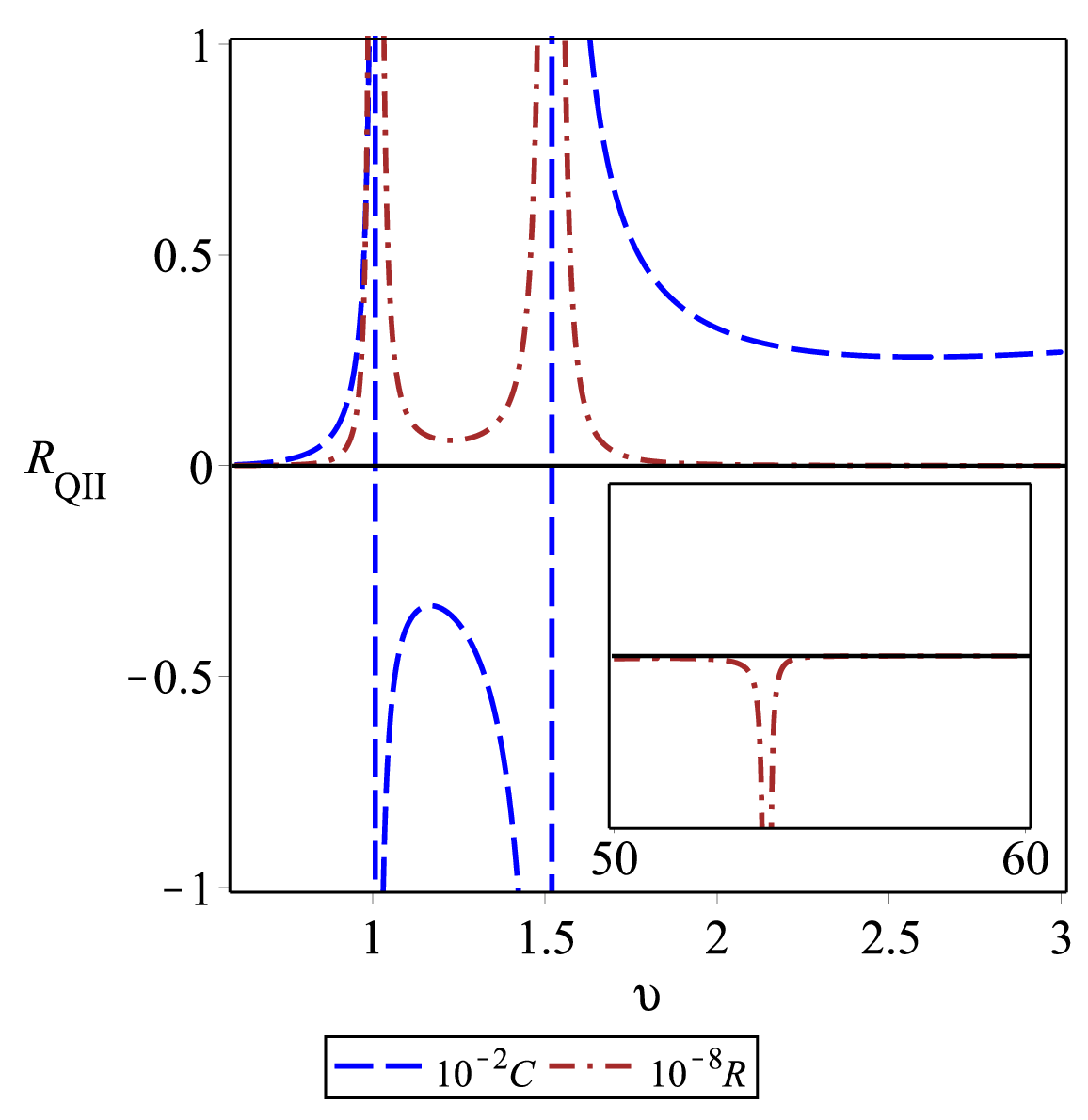}} \newline
\caption{$C_{P,Q,\mu}$ (dashed line) and Ricci scalar (dash-dotted line) versus $%
\protect\upsilon $ for $\mathcal{B}=0.2$, $\protect\beta =0.04$, $A=0.02$, $%
\protect\mu =0.15$, $Q=0.2$ and $P=0.85P_{c}$. Weinhold's Ricci scalar
(up-left panel), Ruppeiner's Ricci scalar (up-right panel), Quevedo's Ricci scalar (middle and down panels).}
\label{Fig11}
\end{figure}

\begin{figure}[!htb]
\centering
\subfloat[$P=0.85P_{c}$]{
        \includegraphics[width=0.33\textwidth]{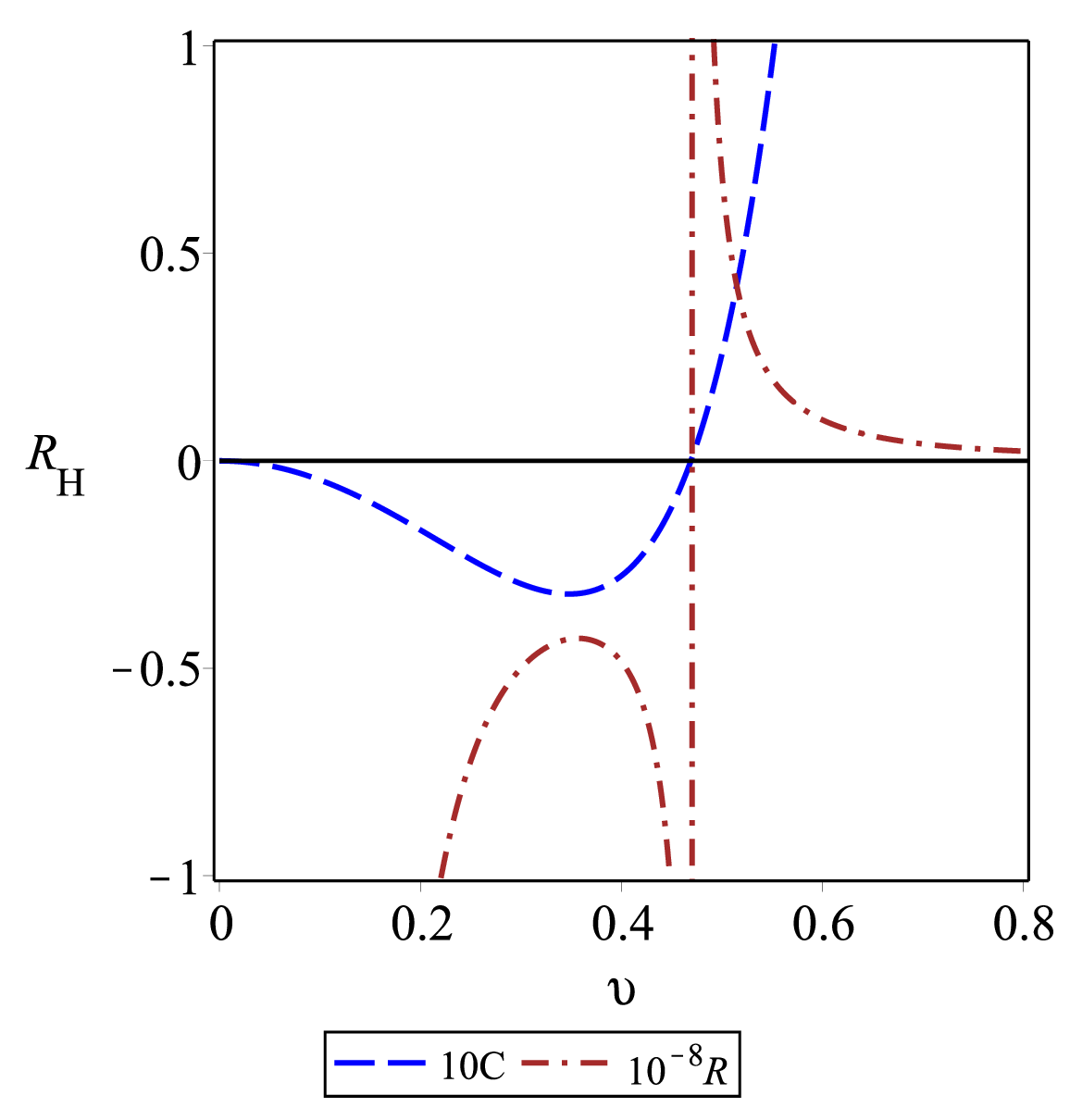}}
\subfloat[$P=0.85P_{c}$]{
        \includegraphics[width=0.33\textwidth]{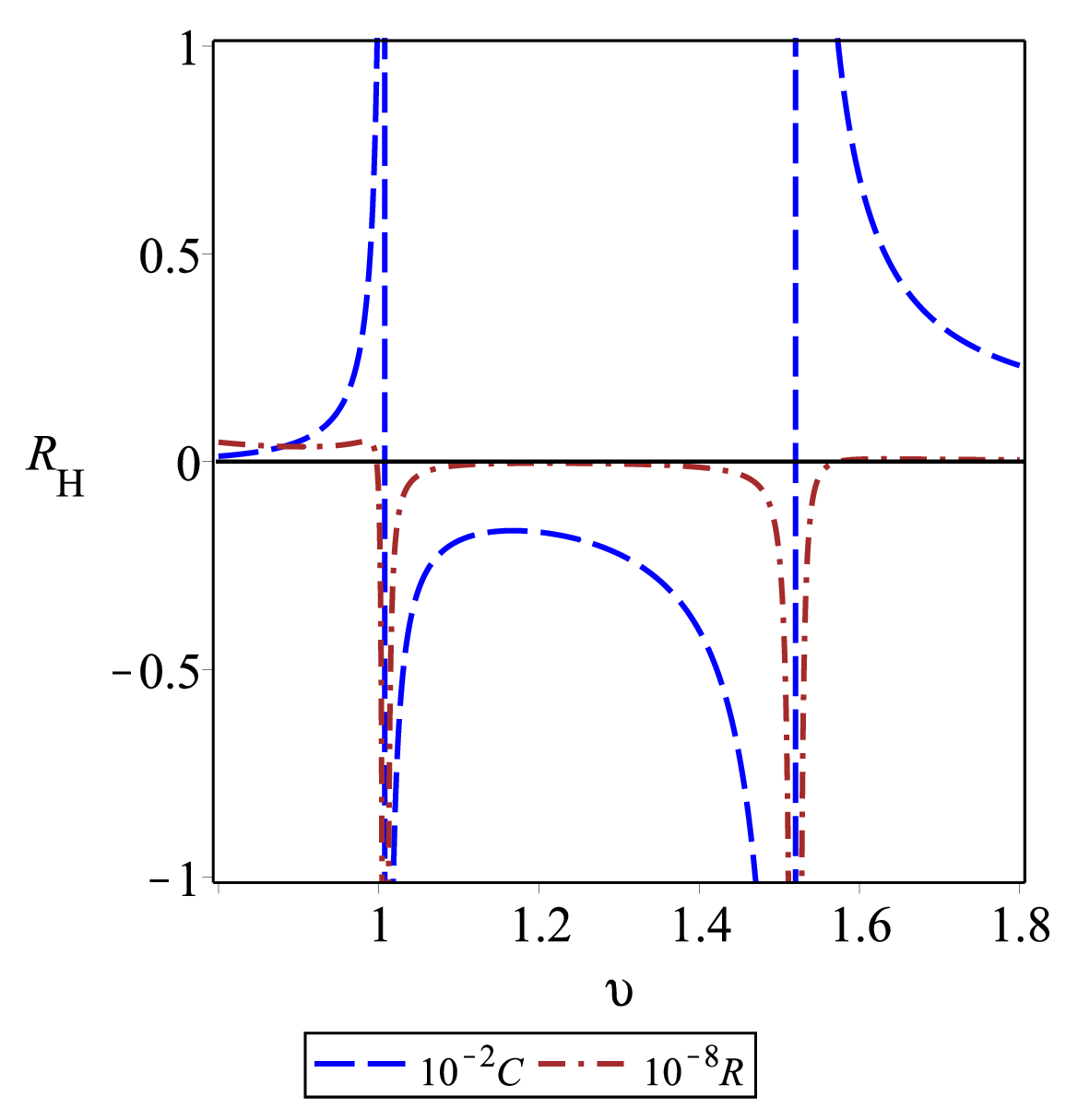}} \newline
\subfloat[$P=P_{c}$]{
        \includegraphics[width=0.33\textwidth]{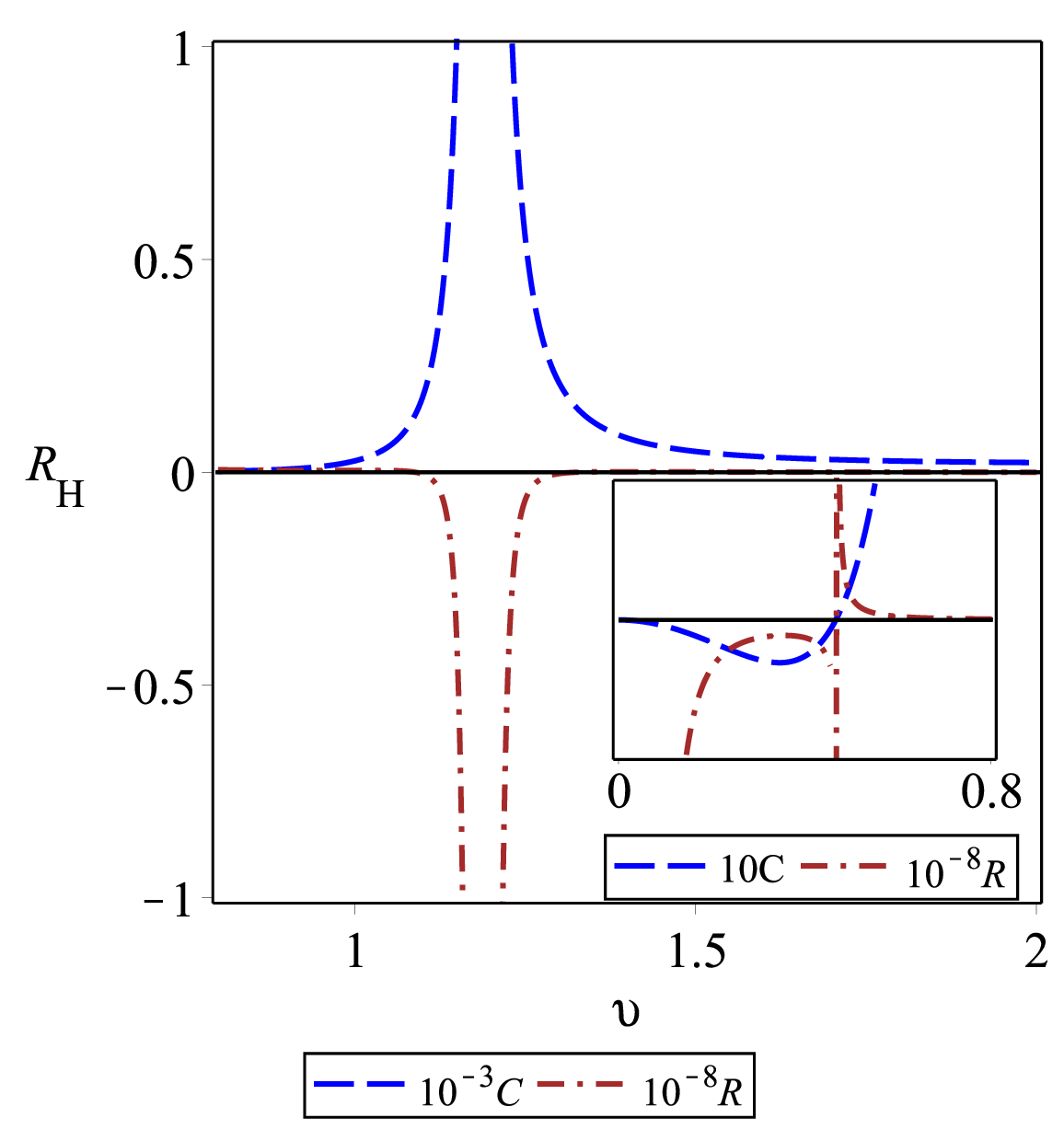}}
\subfloat[$P=1.1P_{c}$]{
        \includegraphics[width=0.33\textwidth]{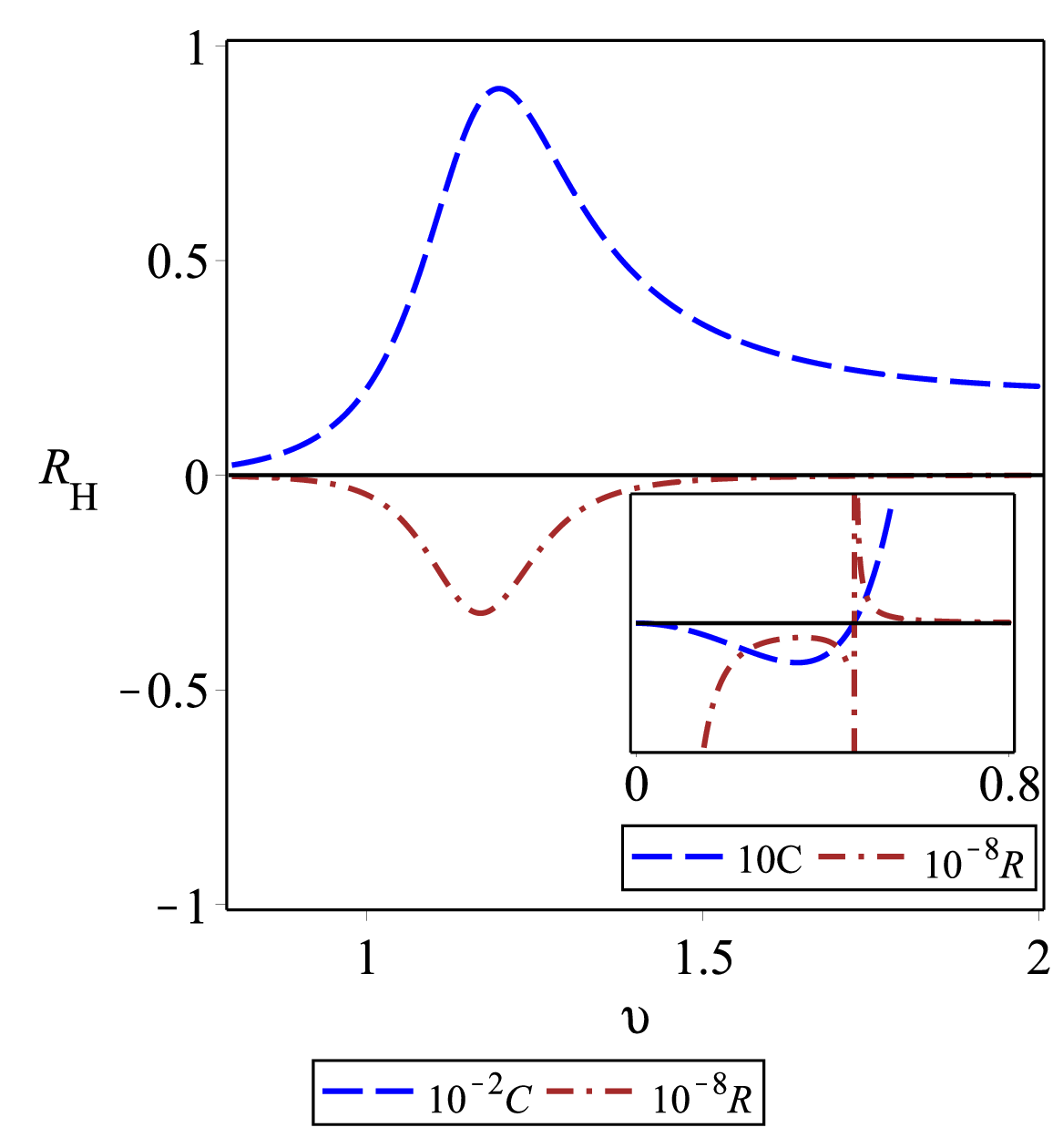}} \newline
\caption{$C_{P,Q,\mu}$ (dashed line) and $\mathcal{R}_{HPEM}$ (dash-dotted line)
versus $\protect\upsilon $ for $\mathcal{B}=0.2$, $\protect\beta =0.04$, $%
A=0.02$, $\protect\mu =0.15$ and $Q=0.2$.}
\label{Fig12}
\end{figure}

Now, we investigate thermodynamical behavior of the system through
geometrical approaches. As it is observed in Fig. \ref{Fig11},
both Weinhold and Ruppeiner metrics fail to produce consistent
results with phase transition points (see up panels of Fig.
\ref{Fig11}). In the case of Quevedo's metrics, Quevedo I does not
include bound point in divergencies of its Ricci scalar and it can
only describe phase transition points. As for Quevedo II, although
some of its divergencies coincide with bound and phase transition
points, similar to the case of Quevedo I, it has an extra
divergency that does not match with any phase transition point
(see middle and down panels of Fig. \ref{Fig11}). Regarding the
HPEM's metric, its Ricci scalar has two divergencies at $P=P_{c}$.
One divergency coincides with the root of heat capacity and the
other one matches with divergency of the heat capacity (see Fig.
\ref{Fig12}c). For the case of $P<P_{c}$, three divergencies are
observed: one of these divergencies is related to the root of heat
capacity and the other ones are coincident with divergencies of
the heat capacity (see up panels of Fig. \ref{Fig12}). As for
$P>P_{c}$, HPEM's Ricci scalar has only a divergence point which
is located at the root of heat capacity (Fig. \ref{Fig12}d). These
results show that the HPEM's metric can be considered a powerful
tool to describe the phase structures of such black holes.

As we have already mentioned, one can employ the Legendre invariant metrics for probing molecular interaction during the phase transition as well \cite{QueIII}. Since the HPEM's metric could provide a precise picture of phase transitions, we employ this method and check whether it can describe microscopic properties of the system. As we see from Fig.
\ref{Fig12}c, scalar curvature goes to negative infinity at
$P=P_{c}$, implies an attractive interaction between the microscopic molecules. Regarding Fig. \ref{Fig12}b for $P<P_{c}$, it is clear that the scalar curvature is negative around both divergence points. Taking a close look at this figure, one can find that scalar curvature becomes positive near the smaller divergence point which shows that a weak repulsive interaction dominates for small black holes. It is worthwhile to
mention that these results are similar to those obtained for van
der Waals fluid \cite{ShaoWei}. So, HPEM's metric can describe phase
structure and microstructure of the black hole, simultaneously.
Here, we should point out that we consider both the numerator
and denominator of $R_{H}$ to draw Fig. \ref{Fig12}, while in Fig. \ref%
{Fig11}, we only considered denominators of Ricci scalars. Since other
metrics have not been successful in investigating phase transitions, we have not used them for studying microscopic properties.

In 2017, two new metrics were introduced by geometric
interpretation of criticality conditions \cite{50,51}. In the next
section, we employ these two metrics and show that they are not
suitable candidates for accelerating cases in which
thermodynamical volume is not a linear function of the horizon
radius.

\subsubsection{geometrical description of critical conditions}

In this method, two independent GT metrics are defined by the
geometrical description of two critical conditions where the
defined metrics are invariant under Legendre transformations.
Ricci scalars of these metrics diverge at the critical point. For
the first condition, the Helmholtz free energy is considered as
the appropriate quantity to define GT metric. Whereas, for the
geometrical description of the second condition, the pressure is a
proper thermodynamical quantity (instead of F). Now, we employ
these two metrics for studying phase transition points of charged
accelerating AdS black hole.

In order to obtain first thermodynamic metric, one can define a
thermodynamical phase space with the coordinates $Z^{A}$ =$\{F,\xi ^{a},\rho
^{a}\}$ where $\xi ^{a}$ =$\{V,T,Q\}$ are the thermodynamical variables and $%
\rho ^{a}$ =$\{F_{V},F_{T},F_{Q}\}$ are the conjugate quantities
corresponding to these variables which are defined as follow
\begin{equation}
F_{V}=\bigg(\frac{\partial F}{\partial V}\bigg)_{T,Q}=-P~~\&~~F_{T}=\bigg(%
\frac{\partial F}{\partial T}\bigg)_{V,Q}=-S~~\&~~F_{Q}=\bigg(\frac{\partial
F}{\partial Q}\bigg)_{V,T}=\Phi ~.  \label{Eq36}
\end{equation}

In this case, the Legendre invariant metric is expressed as
\begin{equation}
g_{1}^{(PV)}=\left( -PV-ST+\Phi Q\right) \left(
-F_{VV}dV^{2}+F_{TT}dT^{2}+F_{QQ}dQ^{2}+2F_{TQ}dTdQ\right) .  \label{Eq37}
\end{equation}

The metric coefficients can be identified as \cite{51}
\begin{eqnarray}
f(V,T,Q) &=&F_{VV}(VF_{V}+TF_{T}+QF_{Q}),  \notag \\
&&  \notag \\
h(V,T,Q) &=&F_{TT}\left( VF_{V}+TF_{T}+QF_{Q}\right) ,  \notag \\
&&  \notag \\
x(V,T,Q) &=&F_{QQ}\left( VF_{V}+TF_{T}+QF_{Q}\right) ,  \notag \\
&&  \notag \\
y(V,T,Q) &=&F_{TQ}\left( VF_{V}+TF_{T}+QF_{Q}\right) .  \label{Eq38}
\end{eqnarray}

Straightforward calculation shows that the denominator of the Ricci scalar
can be written as
\begin{equation*}
denom(R_{1})=\frac{1}{2f^{2}(hx-y^{2})^{2}}.
\end{equation*}

Now, we consider the pressure as a proper thermodynamical quantity. In this
case, a thermodynamical phase space is defined with the coordinates $Z^{A}$ =%
$\{P,\xi ^{a},\rho ^{a}\}$ where $\xi ^{a}$ =$\{V,T,Q\}$ and $\rho ^{a}$ =$%
\{P_{V},P_{T},P_{Q}\}$ are the variables and conjugate quantities,
respectively. The geometrical metric in the invariant $\theta _{P}$ picture
is given by
\begin{equation}
g_{2}^{(PV)}=\left( VP_{V}+TP_{T}+QP_{Q}\right) \left(
-P_{VV}dV^{2}+P_{TT}dT^{2}+P_{QQ}dQ^{2}+2P_{TQ}dTdQ\right) ,  \label{Eq39}
\end{equation}%
with following metric coefficients
\begin{eqnarray}
\omega (V,T,Q) &=&P_{VV}\left( VP_{V}+TP_{T}+QP_{Q}\right) ,  \notag \\
&&  \notag \\
\beta (V,T,Q) &=&P_{TT}\left( VP_{V}+TP_{T}+QP_{Q}\right) ,  \notag \\
&&  \notag \\
\gamma (V,T,Q) &=&P_{QQ}\left( VP_{V}+TP_{T}+QP_{Q}\right) ,  \notag \\
&&  \notag \\
\varepsilon (V,T,Q) &=&P_{TQ}\left( VP_{V}+TP_{T}+QP_{Q}\right) .
\label{Eq40}
\end{eqnarray}

By calculating the Ricci scalar, one can obtain the following expression for
its denominator
\begin{equation*}
denom(R_{2})=\frac{1}{2\omega ^{2}\left( \beta \gamma -\varepsilon
^{2}\right) ^{2}}.
\end{equation*}

\begin{figure*}[tbh]
\centering
\includegraphics[width=0.35\linewidth]{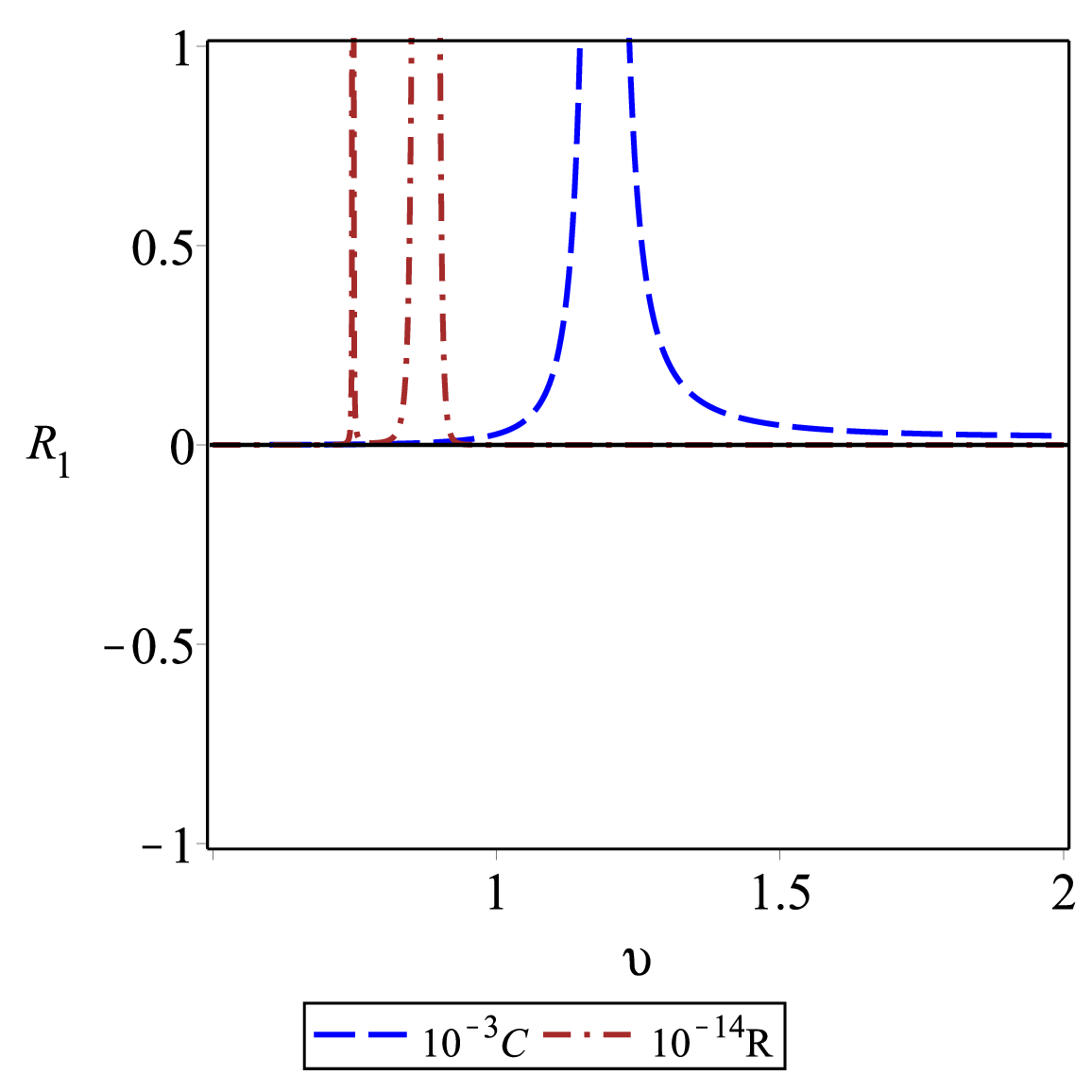} \includegraphics[width=0.35%
\linewidth]{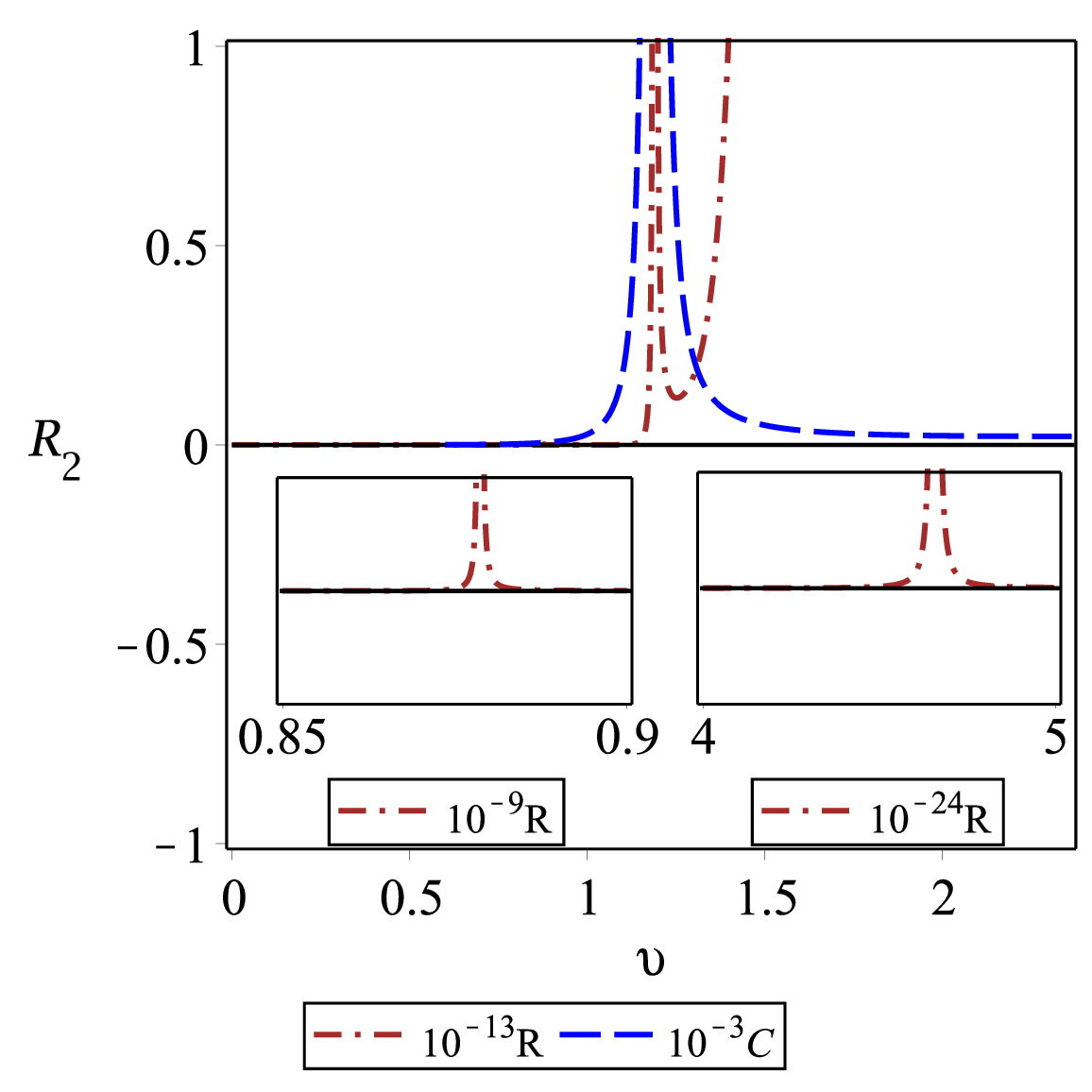}
\caption{$C_{P,Q,\mu}$ (dashed line) and $R$ (dash-dotted line) versus $\protect%
\upsilon $ for $\mathcal{B}=0.2$, $\protect\beta =0.04$, $Q=0.2$, $\protect%
\mu =0.15$, $A=0.02$ and $T=T_{c}$.}
\label{Fig13}
\end{figure*}

We draw behavior of $R_{1}$ and $R_{2}$ with respect to $\upsilon $ in Fig. %
\ref{Fig13}. One can see that $R_{1}$ and $R_{2}$ have an extra divergence
point at $T=T_{c}$. For $R_{2}$, one of these divergencies is exactly
coincident with divergency of the heat capacity at the critical temperature. But
the other ones do not\textbf{\ }match with heat capacity's divergence points
(see right panel of Fig. \ref{Fig13}). As for $R_{1}$, none of its
divergencies coincide with heat capacity's divergency (see left panel of
Fig. \ref{Fig13}). So, this method cannot provide an appropriate picture of
phase transition for charged accelerating AdS black holes.

\section{Conclusion}

In this paper, we considered the thermodynamical behavior of the charged
accelerating AdS black holes and investigated their thermal stability and
phase transition by calculating the heat capacity in canonical ensemble.
First, we studied thermodynamical structure of these black holes in the
non-extended phase space and showed that the electric charge, AdS radius and string tension are important factors for observing phase transition. We found a relation between the electric charge, string tension and AdS radius to indicate that the condition $\frac{Q\mathcal{B}}{\mu \ell }<0.1667$, should be satisfied in order to have phase transition for such black holes.
Studying the effects of electric charge and string tension on stability
conditions showed that as string tension (electric charge) increases
(decreases), the region of instability increases.

In addition, we employed GT method to investigate phase transition
of the system. We found that among well-known thermodynamical
metrics, only HPEM one provided consistent results with the heat
capacity and was able to describe both bound and phase transition
points simultaneously. Also, we saw that the behavior of HPEM's
Ricci scalar around its divergence points was different for the
corresponding root and divergence points of the heat capacity. In
other words, by (un)changing the sign of HPEM's Ricci scalar
around its divergence points, one can recognize bound and phase
transition points from each other.

Next, we extended the phase space by considering analogy between the
cosmological constant and thermodynamical pressure, and studied van der
Waals like phase transition. We also used the denominator of the heat
capacity and obtained a new relation for pressure which its maximum was the same critical pressure. Substituting the new pressure in temperature, we obtained a new relation for the temperature contains a maximum which was exactly coincident with the critical temperature. Investigating the effects of electric charge and string tension, we found that the critical pressure and temperature are decreasing (increasing) functions of electric charge (string tension). Whereas, the critical volume is an increasing (a decreasing) function of electric charge (string tension).

Studying $T-\upsilon $ and $F-T$ diagrams showed that for small
(large) values of $Q$ ($\mu $), phase transitions take place in
higher temperatures. This indicates that obtaining a stable state
becomes more difficult for these black holes and they need to
absorb more mass from surrounding in order to have a phase
transition. Also, studying $P-\upsilon $ diagram indicated that as
electric charge (string tension) increases (decreases), the
pressure related to the phase transition increases which shows
that the necessity of having a background with higher curvature increases. Also, we found that as $\mu $ increases, the distance between two extrema increases. This revealed the fact that a small/large accelerating black hole exits in its stable state by increasing this parameter.

Finally, we employed well-known thermodynamical metrics for
studying the critical behavior of the system. We noticed that
similar to the non-extended case, just divergencies of HPEM's
Ricci scalar were exactly coincident with bound and phase
transition points. We also used the geometrical approach which was
obtained by the geometrical description of critical conditions. We
saw that the Ricci scalar included an extra divergence point at
the critical temperature and was not able to describe the critical
point. So, only the HPEM's metric could provide an appropriate
picture of phase transition for accelerating AdS black holes in both extended and non-extended phase space.
\newline

\begin{acknowledgements}
SHH thanks Shiraz University Research Council. The work of BEP has been supported financially by the Research Institute for Astronomy and
Astrophysics of Maragha (RIAAM) under research project No.
1/6025-58.
\end{acknowledgements}

\begin{center}
\textbf{Appendix}

\textbf{A: normalization of the time coordinate} \label{A}
\end{center}

The action, including boundary
counterterms \cite{Balasubramanian,REmparan,RBMann}, is
\begin{equation}
I=\frac{1}{16\pi}\int_{M}d^{4}x\sqrt{g}\left[ R+\frac{6}{\ell^{2}}-F_{ab}F^{ab}%
\right] +\frac{1}{8\pi}\int_{\partial M}d^{3}x\sqrt{h}\left[ \mathcal{K}-%
\frac{2}{\ell}-\frac{\ell}{2}\mathcal{R}(h)\right] ,  \label{EqAp1}
\end{equation}
where $ F_{ab}$ is the electromagnetic field tensor  and $h_{ab}$ is the intrinsic metric on $\partial M$. $\mathcal{K}$ and $%
\mathcal{R}(h)$ are, respectively, the extrinsic curvature and Ricci scalar
of the boundary. Varying the action gives the energy momentum tensor as
\begin{equation}
8\pi \mathcal{T}_{ab}=\ell \mathcal{G}_{ab}(h)-\frac{2}{\ell}h_{ab}-\mathcal{%
K}_{ab}+h_{ab}\mathcal{K}.  \label{EqAp3}
\end{equation}

To compute the quantities appearing in this expression, we require
new coordinates near the boundary of AdS, typically parameterized
by Fefferman-Graham coordinates, in which the geometry takes a
standard format \cite{AccV,AccIX}
\begin{equation}
ds^{2}=\frac{\ell ^{2}}{z^{2}}dz^{2}+z^{-2}\left( \gamma
_{ab}^{0}+z^{2}\gamma _{ab}^{2}+...\right) dx^{a}dx^{b}.  \label{EqAp4}
\end{equation}

One can perform an asymptotic expansion for the coordinate transformation as
\begin{equation}
\frac{1}{r}=-A\xi -\sum_{n=1}^{4}X_{n}(\xi )z^{n}~,~~~\&~~~\cos \theta =\xi
+\sum_{n=1}^{4}Y_{n}(\xi )z^{n}.  \label{EqAp5}
\end{equation}

The functions $X_{n}$ and $Y_{n}$ are fixed by requiring the metric (\ref%
{Eq1}) be of the form Eq. (\ref{EqAp4}). By a systematic expansion, one can
determine the $Y_{n}$ in terms of the $X_{n}$. For example, at leading
order, $\mathcal{O}(z^{-2})$
\begin{equation}
Y_{1}(\xi )=-\frac{A\ell ^{2}X_{1}(\xi )G(\xi )}{F^{2}(\xi )}%
,~~~\&~~~X_{1}(\xi )=-\frac{F(\xi )^{3}}{\alpha \omega (\xi )},
\label{EqAp6}
\end{equation}%
which elucidates the conformal degree of freedom in the boundary metric, $%
\omega $, with
\begin{equation}
G(x)=\left( 1-x^{2}\right) \left( 1+2mAx+e^{2}A^{2}x^{2}\right)
~,~~~\&~~~F(x)=\sqrt{1-A^{2}\ell ^{2}G(x)}.  \label{EqAp7}
\end{equation}

The expression (\ref{EqAp6}) ensures $g_{zz}=\frac{\ell^{2}}{z^{2}} $ to
this order, and gives the boundary metric as \cite{AccV,AccIX}
\begin{equation}
ds^{2}(0)=-\frac{\omega^{2}d\tau^{2}}{\ell^{2}}+\frac{\alpha^{2}\omega^{2}d%
\xi^{2}}{GF^{4}}+\frac{\alpha^{2}\omega^{2}Gd\varphi^{2}}{F^{2}K^{2}}.
\label{EqAp2}
\end{equation}

The expectation value of the energy momentum of the $CFT_{3}$ is calculated
by the following relation \cite{AccV,AccIX,Bernardi}
\begin{equation}
<\mathcal{T}_{ab}>=\lim_{z\longrightarrow 0}\frac{1}{\ell z}\mathcal{T}_{ab}=%
\frac{3}{2}\rho _{E}U_{a}U_{b}+\frac{\rho _{E}}{2}\ell ^{2}\gamma
_{ab}^{0}+\pi _{ab},  \label{EqAp8}
\end{equation}%
with $U=\omega ^{-1}\partial \tau $. The energy density is

\begin{equation}
\rho _{E}=\frac{\left( m+2e^{2}A\xi \right) }{8\pi \ell ^{2}\alpha
^{3}\omega ^{3}}F^{3}\left( \xi \right) \left( 2-3A^{2}\ell ^{2}G\left( \xi
\right) \right) ,  \label{EqAp9}
\end{equation}%
yielding the mass
\begin{equation}
M=\int \rho _{E}\ell ^{3}\sqrt{-\gamma ^{0}}dxd\varphi =\frac{m\left(
1-A^{2}\ell ^{2}-A^{4}e^{2}\ell ^{2}\right) }{K\alpha }.  \label{EqAp10}
\end{equation}

The variation of the boundary metric with respect to the parameters results
to
\begin{equation}
\delta\gamma^{ab}=\frac{\partial \gamma^{ab}}{\partial K}\delta K+\frac{%
\partial \gamma^{ab}}{\partial A}\delta A+\frac{\partial \gamma^{ab}}{%
\partial m}\delta m +\frac{\partial \gamma^{ab}}{\partial e}\delta e ,
\label{EqAp11}
\end{equation}
considering $\ell $ and $\mu $ as constant parameters, one can calculate
\begin{equation}
\delta I=\int_{\partial M}\sqrt{-\gamma} \tau_{ab}\delta\gamma^{ab}d^{3}x.
\label{EqAp12}
\end{equation}

Imposing that the variation vanishes, the parameter $\alpha$ is obtained as
\cite{AccIX}
\begin{equation}
\alpha =\sqrt{\left( 1+A^{2}e^{2}\right) \left( 1-A^{2}\ell ^{2}-A^{4}\ell
^{2}e^{2}\right) }.
\end{equation}

\begin{center}
\textbf{B: thermodynamic mass} \label{Appendix B}
\end{center}

Setting $m=e=0$ in Eq. (\ref{Eq1}), we can eliminate the conical
singularity \cite{AccII}. Using the mentioned adjustment, one
finds
\begin{equation}
ds^{2}=\frac{1}{\Omega ^{2}}\left[ \mathop{f}\limits^{\sim }dt^{2}-\frac{%
dr^{2}}{\mathop{f}\limits^{\sim }}-r^{2}d\theta ^{2}-r^{2}sin^{2}\theta
d\phi ^{2}\right] ,
\end{equation}%
where
\begin{equation}
\mathop{f}\limits^{\sim }=1+\frac{r^{2}}{\ell ^{2}}\left( 1-A^{2}\ell
^{2}\right) .
\end{equation}

This spacetime no longer has a conical singularity and is locally pure AdS
at $r=-\frac{1}{Acos\theta }$ (not $r\rightarrow \infty $). To transform
Rindler coordinate to global AdS coordinates $\{R,\Theta \}$, one takes \cite{AccV,AccIX,52}
\begin{equation}
Rsin\Theta =\frac{rsin\theta }{\Omega },~~~\&~~~R=\ell \sqrt{\frac{\mathop{f}%
\limits^{\sim }}{\Omega ^{2}(1-A^{2}\ell ^{2})}-1},
\end{equation}%
which results to the metric of AdS space in global coordinates:
\begin{equation}
ds^{2}=\left( 1+\frac{R^{2}}{\ell ^{2}}\right) dt^{2}-\frac{dR^{2}}{1+\frac{%
R^{2}}{\ell ^{2}}}-R^{2}\left( d\Theta ^{2}+sin^{2}\Theta d\phi ^{2}\right) .
\end{equation}


\begin{thebibliography}{999}
\bibitem{Bardeen} J. M. Bardeen, B. Carter, and S. W. Hawking, Commun. Math. Phys. \textbf{31}, 161 (1973).

\bibitem{Hawking1a} S. W. Hawking, Commun. Math. Phys. \textbf{43}, 199  (1975).

\bibitem{Bekenstein1a} J. D. Bekenstein, Phys. Rev. D \textbf{7}, 2333 (1973).

\bibitem{Bekenstein1b} J. D. Bekenstein, Phys. Rev. D \textbf{9}, 3292 (1974).

\bibitem{2} J. Maldacena, Adv. Theor. Math. Phys. \textbf{2}, 231 (1998).

\bibitem{3} E. Witten, Adv. Theor. Math. Phys. \textbf{2}, 253 (1998).

\bibitem{4} S. S. Gubser, I. R. Klebanov, and A. M. Polyakov, Phys. Lett. B \textbf{428}, 105 (1998).

\bibitem{5} O. Aharony, S. S. Gubser, J. Maldacena, H. Ooguri, and Y. Oz,
Phys. Rept. \textbf{323}, 183 (2000).

\bibitem{6} D. Bazeia, L. Losano, G. J. Olmo, and D. Rubiera-Garcia, Phys.
Rev. D \textbf{90}, 044011 (2014).

\bibitem{7} S. W. Hawking, and D. N. Page, Commun. Math. Phys. \textbf{87}, 577 (1983).

\bibitem{8} M. Cvetic, and S. S. Gubser, JHEP \textbf{04}, 024 (1999).

\bibitem{9} M. Cvetic, and S. S. Gubser, JHEP \textbf{07}, 010 (1999).

\bibitem{10} A. Chamblin, R. Emparan, C. Johnson, and R. Myers, Phys. Rev. D \textbf{60}, 064018 (1999).

\bibitem{11} A. Chamblin, R. Emparan, C. Johnson, and R. Myers, Phys. Rev. D \textbf{60}, 104026 (1999).

\bibitem{12} D. Kubiznak, and R. B. Mann, JHEP \textbf{07}, 033 (2012).

\bibitem{13} S. Gunasekaran, D. Kubiznak, and R. B. Mann, JHEP \textbf{11}, 110 (2012).

\bibitem{GB} R. G. Cai, L. M. Cao, L. Li, and R. Q. Yang, JHEP \textbf{09}, 005 (2013).

\bibitem{GBI} J. X. Mo, and W. B. Liu, Phys. Rev. D \textbf{89}, 084057
(2014).

\bibitem{GBII} M. S. Ma, L. C. Zhang, H. H. Zhao, and R. Zhao, Adv. High
Energy Phys. \textbf{2015}, 134815 (2015).

\bibitem{GBIII} Y. G. Miao, and Z. M. Xu, Phys. Rev. D \textbf{98}, 084051
(2018).

\bibitem{Dilaton} R. Zhao, H. Zhao, M. S. Ma, and L. C. Zhang , Eur. Phys.
J. C \textbf{73}, 2645 (2013).

\bibitem{DilatonII} M. H. Dehghani, S. Kamrani, and A. Sheykhi, Phys. Rev. D \textbf{90}, 104020 (2014).

\bibitem{LoveI} J. X. Mo, and W. B. Liu, Eur. Phys. J. C \textbf{74}, 2836
(2014).

\bibitem{LoveII} S. H. Hendi, S. Panahiyan, and B. Eslam Panah, Prog. Theor. Exp. Phys. \textbf{2015}, 103E01 (2015).

\bibitem{LoveIII} A. Haldar, and R. Biswas, Gen. Relativ. Gravit. \textbf{50}, 69 (2018).

\bibitem{HL0} B. R. Majhi, and D. Roychowdhury, Class. Quantum Gravit.
\textbf{29}, 245012 (2012).

\bibitem{HLI} J. X. Mo, Astrophys. Space Sci. \textbf{356}, 319 (2015).

\bibitem{HLII} Kh. Jafarzade, and J. Sadeghi, Int. J. Mod. Phys. D \textbf{26}, 1750138 (2017).

\bibitem{HLIII} M. S. Ma, and R. H. Wang, Phys. Rev. D \textbf{96}, 024052
(2017).

\bibitem{HLIV} Y. Z. Du, R. Zhao, and L. C. Zhang, [arXiv:1909.09968].

\bibitem{massiveI} B. Mirza, and Z. Sherkatghanad, Phys. Rev. D \textbf{90}, 084006 (2014).

\bibitem{massiveII} J. Xu, L. M. Cao, and Y. P. Hu, Phys. Rev. D \textbf{91}, 124033 (2015).

\bibitem{massiveIII} S. H. Hendi, B. Eslam Panah, and S. Panahiyan, Class.
Quantum Gravit. \textbf{33}, 235007 (2016).

\bibitem{massiveIV} S. Fernando, Phys. Rev. D \textbf{94}, 124049 (2016).

\bibitem{massiveV} S. H. Hendi, S. Panahiyan, B. Eslam Panah, and M.
Momennia, Ann. Phys. \textbf{528}, 819 (2016).

\bibitem{massiveVI} S. H. Hendi, R. B. Mann, S. Panahiyan, and B. Eslam
Panah, Phys. Rev. D \textbf{95}, 021501(R) (2017).

\bibitem{massiveVII} D. C. Zou, R. Yue, and M. Zhang, Eur. Phys. J. C
\textbf{77}, 256 (2017).

\bibitem{massiveVIII} D. C. Zou, Y. Liu, and R. Yue, Eur. Phys. J. C \textbf{77}, 365 (2017).

\bibitem{massiveIX} S. Upadhyay, B. Pourhassan, and H. Farahani, Phys. Rev. D \textbf{95}, 106014 (2017).

\bibitem{massiveX} M. Chabab, H. El Moumni, S. Iraoui, and K. Masmar, Eur.
Phys. J. C \textbf{79}, 342 (2019).

\bibitem{F(R)I} J. X. Mo, G. Q. Li, and Y. C. Wu, JCAP \textbf{04}, 045
(2016).

\bibitem{F(R)II} A. Ovg\"{u}n, Adv. High Energy Phys. \textbf{2018}, 8153721 (2018).

\bibitem{RainI} S. H. Hendi, S. Panahiyan, B. Eslam Panah, M. Faizal, and M. Momennia, Phys. Rev. D \textbf{94}, 024028 (2016).

\bibitem{RainII} Z. W. Feng, and S. Z. Yang, Phys. Lett. B \textbf{772}, 737 (2017).

\bibitem{MassRain} S. H. Hendi, B. Eslam Panah, and S. Panahiyan, Phys.
Lett. B \textbf{769}, 191 (2017).

\bibitem{NED} Y. S. Myung, Y. W. Kim, and Y. J. Park, Phys. Rev. D \textbf{78}, 084002 (2008).

\bibitem{NEDI} S. H. Hendi, and M. H. Vahidinia, Phys. Rev. D \textbf{88},
084045 (2013).

\bibitem{NEDIII} S. H. Hendi, G. Q. Li, J. X. Mo, S. Panahiyan, and B. Eslam Panah, Eur. Phys. J. C \textbf{76}, 571 (2016).

\bibitem{NEDIV} J. X. Mo, G. Q. Li, and X. B. Xu, Phys. Rev. D \textbf{93}, 084041 (2016).

\bibitem{NEDVI} H. F. Li, H. H. Zhao, L. C. Zhang, and R. Zhao, Eur. Phys.
J. C \textbf{77}, 295 (2017).

\bibitem{NEDVII} M. Zhang, D. C. Zou, and R. H. Yue, Adv. High Energy Phys. \textbf{2017}, 3819246 (2017).

\bibitem{NEDVIII} Z. Dayyani, A. Sheykhi, M. H. Dehghani, and S. Hajkhalili, Eur. Phys. J. C \textbf{78}, 152 (2018).

\bibitem{NEDIX} P. Wang, H. Wu, and H. Yang, JHEP \textbf{07}, 002 (2019).

\bibitem{NEDX} P. Wang, H. Wu, and H. Yang, JCAP \textbf{04}, 052 (2019).

\bibitem{NEDXI} Y. M. Xu, H. M. Wang, Y. X. Liu, and S. W. Wei, Phys. Rev. D \textbf{100}, 104044 (2019).

\bibitem{NEDXII} H. Li, Y. Chen, and S. J. Zhang, Nucl. Phys. B \textbf{954}, 114975 (2020).

\bibitem{HeatI} S. Grunau, and H. Neumann, Class. Quantum Gravit. \textbf{32}, 175004 (2015).

\bibitem{HeatII} B. P. Dolan, Class. Quantum Gravit. \textbf{31}, 165011
(2014).

\bibitem{HeatIII} S. H. Hendi, and S. Panahiyan, Phys. Rev. D \textbf{90},
124008 (2014).

\bibitem{HeatIV} B. Eslam Panah, Phys. Lett. B \textbf{787}, 45 (2018).

\bibitem{HeatV} B. Eslam Panah, S. H. Hendi, S. Panahiyan, and M. Hassaine, Phys. Rev. D \textbf{98}, 084006 (2018).

\bibitem{WeinI} F. Weinhold, J. Chem. Phys. \textbf{63}, 2479 (1975).

\bibitem{WeinII} F. Weinhold, J. Chem. Phys. \textbf{63}, 2484 (1975).

\bibitem{RupI} G. Ruppeiner, Phys. Rev. A \textbf{20}, 1608 (1979).

\bibitem{RupII} G. Ruppeiner, Rev. Mod. Phys. \textbf{67}, 605 (1995).

\bibitem{Salamon} P. Salamon, J. Nulton, and E. Ihrig, J. Chem. Phys.
\textbf{80}, 436 (1984).

\bibitem{QueI} H. Quevedo, J. Math. Phys. \textbf{48}, 013506 (2007).

\bibitem{QueII} H. Quevedo, and A. Sanchez, JHEP \textbf{09}, 034 (2008).

\bibitem{QueIII} H. Quevedo and D. Tapias, J. Math. Chem. \textbf{52}, 141
(2014).

\bibitem{MansooriI} S. A. H. Mansoori, and B. Mirza, Eur. Phys. J. C \textbf{74}, 2681 (2014).

\bibitem{MansooriII} S. A. H. Mansoori, B. Mirza, and M. Fazel, JHEP \textbf{04}, 115 (2015).

\bibitem{HPEMI} S. H. Hendi, S. Panahiyan, B. Eslam Panah, and M. Momennia, Eur. Phys. J. C \textbf{75}, 507 (2015).

\bibitem{HPEMII} S. H. Hendi, S. Panahiyan, and B. Eslam Panah, Adv. High
Energy Phys. \textbf{2015}, 743086 (2015).

\bibitem{HPEMIII} S. H. Hendi, A. Sheykhi, S. Panahiyan, and B. Eslam Panah, Phys. Rev. D \textbf{92}, 064028 (2015).

\bibitem{HPEMV} S. H. Hendi, B. Eslam Panah, and S. Panahiyan, JHEP \textbf{11}, 157 (2015).

\bibitem{HPEMVI} S. H. Hendi, S. Panahiyan, and B. Eslam Panah, JHEP \textbf{01}, 129 (2016).

\bibitem{50} R. Banerjee, B. R. Majhi, and S. Samanta, Phys. Lett. B \textbf{767}, 25 (2017).

\bibitem{51} K. Bhattacharya, and B. R. Majhi, Phys. Rev. D \textbf{95},
104024 (2017).

\bibitem{Pineda} V. Pineda, H. Quevedo, M. N. Quevedo, A. Sanchez, and E.
Valdes, Int. J. Geom. Meth. Mod. Phys. \textbf{11}, 1950168 (2019).

\bibitem{AccI} M. Appels, R. Gregory, and D. Kubiznak, Phys. Rev. Lett.
\textbf{117}, 131303 (2016).

\bibitem{AccIII} R. Gregory, J. Phys. Conf. Ser. \textbf{942}, 012002 (2017).

\bibitem{AccIV} M. Astorino, Phys. Rev. D \textbf{95}, 064007 (2017).

\bibitem{AccV} A. Anabalon, M. Appels, R. Gregory, D. Kubiznak, R. B. Mann, and A. Ovg\"{u}n, Phys. Rev. D \textbf{98}, 104038 (2018).

\bibitem{AccVI} M. Appels, Thermodynamics of Accelerating Black Holes, Ph.D. thesis, Durham University (2018).

\bibitem{AccVII} J. Zhang, Y. Li, and H. Yu, Eur. Phys. J. C \textbf{78},
645 (2018).

\bibitem{AccVIII} J. Zhang, Y. Li, and H. Yu, JHEP \textbf{02}, 144 (2019).

\bibitem{AccIX} A. Anabalon, F. Gray, R. Gregory, D. Kubiznak, and R. B.
Mann, JHEP \textbf{04}, 096 (2019).

\bibitem{AccX} N. Abbasvandi, W. Cong, D. Kubiznak, and R. B. Mann, Class.
Quantum Gravit. \textbf{36}, 104001 (2019).

\bibitem{AccXI} N. Abbasvandi, W. Ahmed, W. Cong, D. Kubiznak, and R. B.
Mann, Phys. Rev. D\textbf{\ 100}, 064027 (2019).

\bibitem{AccXII} S. Gregory, and A. Scoins, Phys. Lett. B \textbf{796}, 191 (2019).

\bibitem{AccXIII} B. Eslam Panah, and Kh. Jafarzade, [arXiv:1906.09478].

\bibitem{AccXIV} W. Ahmed, H. Z. Chen, E. Gesteau, R. Gregory, and A.
Scoins, Class. Quantum Gravit. \textbf{36} 214001 (2019).

\bibitem{AccII} M. Appels, R. Gregory, and D. Kubiznak, JHEP \textbf{05},
116 (2017).

\bibitem{CMI} W. Kinnersley, and M. Walker, Phys. Rev. D \textbf{2}, 1359
(1970).

\bibitem{CMII} J. F. Plebanski, and M. Demianski, Ann. Phys. \textbf{98}, 98 (1976).

\bibitem{CMIII} O. J. C. Dias, and J. P. S. Lemos, Phys. Rev. D \textbf{67}, 064001 (2003).

\bibitem{CMIV} J. B. Griffiths, and J. Podolsky, Int. J. Mod. Phys. D
\textbf{15}, 335 (2006).

\bibitem{52} J. Podolsky, Czech. J. Phys. \textbf{52}, 1 (2002).

\bibitem{Zhang1kj} M. Zhang, and J. Jiang, Phys. Rev. D \textbf{103}, 025005 (2021).

\bibitem{Ashtekar1abc} A. Ashtekar, and T. Dray, Comm. Math. Phys. \textbf{79}, 581 (1981).

\bibitem{Podolsky2bc} J. Podolsky, M. Ortaggio, and P. Krtouss, Phys. Rev. D \textbf{68}, 124004 (2003).

\bibitem{Horowitz3abc} S. W. Hawking, G. T. Horowitz, and S. F. Ross, Phys. Rev. D \textbf{51}, 4302 (1995).

\bibitem{Emparan4abc} R. Emparan, and H. S. Reall, Phys. Rev. Lett. \textbf{88}, 101101 (2002).

\bibitem{Dowker1ab} F. Dowker, J. P. Gauntlett, D. A. Kastor, and J. H. Traschen, Phys. Rev. D \textbf{49}, 2909 (1994).

\bibitem{Eardley1ab} D. M. Eardley, G. T. Horowitz, D. A. Kastor, and J. H. Traschen, Phys. Rev. Lett \textbf{75}, 3390 (1995).

\bibitem{Ross1mn} S. W. Hawking, and S. F. Ross, Phys. Rev. D \textbf{56}, 6403 (1997).

\bibitem{Zhang2kjl} M. Zhang, and J. Jiang, Phys. Rev. D \textbf{101}, 104012 (2020).

\bibitem{Frost1kjl} T. C. Frost, and V. Perlick, Class. Quantum Grav. \textbf{38}, 085016 (2021).

\bibitem{Destounis2kn} K, Destounis, R. D. B. Fontana, and F. C. Mena, Phys. Rev. D \textbf{102}, 044005 (2020).

\bibitem{Shun1sd} S. Jiang, and J. Jiang, [arXiv:2106.09371].

\bibitem{Destounis2sd} K. Destounis, R. D. B. Fontana, and F. C. Mena, 	Phys. Rev. D \textbf{102}, 104037 (2020).

\bibitem{Guha2sd} S. Guha, and S. Chakraborty, Int. J. Mod. Phys. D \textbf{29}, 5 (2020).

\bibitem{Ferrero1kj} P. Ferrero, J. P. Gauntlett, J. M. Perez Ipina, D. Martelli, and J. Sparks,  [arXiv:2012.08530].

\bibitem{Tavakoli2bkj} M. Tavakoli, B. Mirza, and Z. Sherkatghanad, Nucl. Phys. B \textbf{943},  114620 (2019).

\bibitem{Huang2br} Y. Huang, and S. Guo, [arXiv:2009.09401].

\bibitem{Ball2bkj} A. Ball, [arXiv:2103.07521].

\bibitem{Ball2kj} A. Ball, and N. Miller, Class. Quantum Gravit. \textbf{38}, 145031 (2021).

\bibitem{Zhang2akj} M. Zhang, and R. B. Mann, Phys. Rev. D \textbf{100}, 084061 (2019).

\bibitem{Belhaj2fd} A. Belhaj, H. El Moumni, and K. Masmar, Adv. High Energy Phys. \textbf{2020},  4092730 (2020).

\bibitem{Pourhassan} M. Rostami, J. Sadeghi, S. Miraboutalebi, A. A. Masoudi, and B. Pourhassan, Int. J. Geome. Meth. Mod. Phys. \textbf{17}, 2050136 (2020).

\bibitem{Podolsky2lkb} J. Podolsky, and A. Vratny, Phys. Rev. D \textbf{102}, 084024 (2020).

\bibitem{Ashtekar} A. Ashtekar, and S. Das, Class. Quantum Gravit. \textbf{17}, L17 (2000).

\bibitem{Das} S. Das, and R. B. Mann, JHEP \textbf{08}, 033 (2000).

\bibitem{Gibbons} G. W. Gibbons, M. J. Perry, and C. N. Pope, Class. Quantum Gravit. \textbf{22}, 1503 (2005).

\bibitem{Dolana1008} B. P. Dolan, Class. Quantum Gravit. \textbf{28}, 125020 (2011).

\bibitem{HendiarXiv} S. H. Hendi, S. Panahiyan, B. Eslam Panah, and M.
Jamil, Chin. Phys. C \textbf{43}, 113106 (2019).

\bibitem{54} S. H. Hendi, S. Panahiyan, and B. Eslam Panah, Int. J. Mod.
Phys. D \textbf{25}, 1650010 (2016).

\bibitem{53} H. Liu, and X. Meng, Mod. Phys. Lett. A \textbf{37}, 1650199
(2016).

\bibitem{55} J. X. Mo, and W. B. Liu, Phys. Lett. B \textbf{727}, 336 (2013).

\bibitem{ShaoWei} S. W. Wei, Y. X. Liu, and R. B. Mann, Phys. Rev. D \textbf{100}, 124033 (2019).

\bibitem{Balasubramanian} V. Balasubramanian, and P. Kraus, Commun. Math.
Phys. \textbf{208}, 413 (1999).

\bibitem{REmparan} R. Emparan, C. V. Johnson, and R. C. Myers, Phys. Rev. D \textbf{60}, 104001 (1999).

\bibitem{RBMann} R. B. Mann, Phys. Rev. D \textbf{60}, 104047 (1999).

\bibitem{Bernardi} G. Bernardi de Freitas, and H. S. Reall, JHEP \textbf{06}, 148 (2014).

\end{thebibliography}
\end{document}